\documentclass[a4paper,11pt]{article}
\pdfoutput=1 

\usepackage{jheppub} 

\usepackage[T1]{fontenc} 
\usepackage[table]{xcolor}
\usepackage[caption=false]{subfig}
\usepackage{epsfig}
\usepackage{colortbl}
\usepackage[T1]{fontenc} 
\usepackage[compat=1.1.0]{tikz-feynman}
\usepackage{adjustbox}
\usepackage{multirow}
\usepackage{slashed}
\usepackage{lipsum}
\usepackage{float}
\usepackage{chngcntr}
\usepackage{bigints}
\usepackage[normalem]{ulem}
\usepackage{amsmath}
\usepackage{textcomp}
\usepackage{cleveref}
\usepackage{lscape}

\usepackage{float}
\usepackage{pstricks}
\usepackage[toc,page]{appendix}
\usepackage{pifont}
\usepackage{braket}
\usepackage{ragged2e}
\usepackage{slashed}
\usepackage{rotating}
\usepackage{tablefootnote}

\usepackage{hyperref}
\usepackage{tikz}
\usepackage{tikz-feynman}
\usepackage{enumitem}
\usetikzlibrary{arrows}
\usepackage{caption} 
\crefname{figure}{fig\,.}{figs\,.} 
\crefname{equation}{eq\,.}{eqs\,.} 
\newcommand{\stkout}[1]{\ifmmode\text{\sout{\ensuremath{#1}}}\else\sout{#1}\fi}
\definecolor{calpolypomonagreen}{rgb}{0.12, 0.3, 0.17}
\newcommand{\bea}{\begin{eqnarray}}
	\newcommand{\eea}{\end{eqnarray}}

\usepackage{xcolor}
   
\allowdisplaybreaks
\title{Study of Form Factors and Observables in $B_c^- \rightarrow \bar{D}^{(*)0}\ell^-\bar{\nu}_{\ell}$ and $B_c^- \rightarrow D^{(*)-}\ell^+\ell^-$ decays}
	
	\author[a]{Utsab Dey,}
	\author[a]{Soumitra Nandi.}

\affiliation[a]{Department of Physics, Indian Institute of Technology Guwahati,\\North Guwahati, Assam-781039, India,}

\emailAdd{utsab\_dey@iitg.ac.in}
\emailAdd{soumitra.nandi@iitg.ac.in}

\abstract{We investigate the decays $B_c^- \rightarrow \bar{D}^{(*)0}\ell^-\bar{\nu}_{\ell}$ and $B_c^- \rightarrow D^{(*)-}\ell^+\ell^-$ within the Standard Model (SM), employing perturbative QCD (pQCD) form factors that are sensitive to the wave functions of $B_c$ and $D^{(*)}$ mesons. Using lattice QCD inputs for $B \to D^{(*)}$ and $B_c \to D$ transitions, we extract the shape parameters of the meson wave functions and estimate the $q^2$-dependence of $B_c \to D^*$ form factors via heavy quark spin symmetry. We present predictions for branching fractions, lepton flavor violating observables, and perform a detailed angular analysis of the cascade decay $B_{c}^{-} \to D^{*-}(\to D^{0}\pi^{-}) \ell^+\ell^-$, providing SM expectations for several angular observables.}


\begin{document}

		\maketitle

\section{Introduction}
\label{section:Introduction}

The study of the decay modes of these $B$ mesons play a crucial role in probing the competence of the Standard Model (SM) and also probing for the possibilities of any New Physics (NP) scenarios. For example, these related semileptonic and leptonic decay modes are useful in extracting the Cabibbo-Kobayashi-Maskawa (CKM) matrix elements. In addition, there are a number of physical observables, an examination of which might hint towards the desired inferences, out of which we confine ourselves within the analysis of the more prominent ones, namely the lepton flavor universality violating (LFUV) observables. Lepton flavor universality (LFU) is a fundamental principle in the SM that asserts the interactions involving different flavors of charged leptons should be the same, provided identical kinematic conditions are met for all three cases, and if any violation arises, it should be solely due to the mass difference of the leptons.

Recently, a different member of the $B$ meson family has gained much more interest in the phenomenological arena. It is the bound state of a $b$ and $c$ quark, also known as the $B_{c}$ meson. Being first discovered by the CDF collaboration \cite{CDF:1998ihx} at Tevatron in 1998 through the semileptonic modes $B_{c}\rightarrow J/\psi(\mu^{+}\mu^{-})\ell^{+}X$, it paved the way and gathered enough interest for further experimental studies to be conducted in full swing. At present, LHCb is expected to produce about $5\times 10^{10}$ $B_{c}$ events each year \cite{PepeAltarelli:2008yyl}, making it possible to venture further into the physics concerning the decay modes of the $B_{c}$ mesons, with precision that had not been possible to attain before.

Compared to the other members of the $B$ family, the $B_{c}$ meson, being the bound state of two heavy quarks, decays via weak interaction only, with the strong and electromagnetic annihilation processes being forbidden. Since both quarks are heavy, each of them can decay independently, with the other playing the role of a spectator quark. The weak transitions can occur through the bottom quark decay with the charm quark acting as the spectator quark, or through the charm quark decay with the bottom quark acting as the spectator quark, or through the pure weak annihilation channel. Thus, it is quite evident that there is a plethora of possible channels that the $B_{c}$ meson can decay into. A thorough and extensive study of these channels can shed light on a number of aspects of weak interactions, either in SM or possible NP scenarios. In a recent study, we have analysed the semileptonic and the non-leptonic decay modes of the $B_c$ meson to $S$ and $P$ wave charmonium states \cite{Dey:2025xdx}. In this work, we focus on the semileptonic $B_{c}^{-}\rightarrow D^{(*)0}\ell^{-}\bar{\nu}_{\ell}$ and rare $B_{c}^{-}\rightarrow D^{(*)-}\ell^{+}\ell^{-}$ channels with $\ell$ being either $e$, $\mu$ or $\tau$ lepton. These channels gather interest as a detailed analysis can contribute to a precise determination of the relevant CKM matrix elements $V_{ub}$, $V_{tb}$, and $V_{td}$. Currently, sufficient information is not available on these modes to make them useful for the precise extraction of the CKM elements. Hence, in this work, we have taken their PDG values as inputs. In addition, studying these modes also allows us to probe into the relevant form factors, which encode information about the internal structure of hadrons and the underlying dynamics of weak interactions. Precise determination of these form factors, either experimentally or through theoretical predictions, is crucial for achieving an in-depth knowledge about the matrix elements and other relevant observables that will be discussed in this work.

The rare $b\rightarrow s\ell^{+}\ell^{-}$ and $b\rightarrow d\ell^{+}\ell^{-}$ decays, however, are much more interesting in today's research scenario. With NP analysis being at the core of most of the phenomenological analysis of heavy quark decays, these decays tend to be highly sensitive to physics beyond the Standard Model (BSM). Being governed by flavor-changing neutral current (FCNC) processes, these are forbidden at the tree level and can only operate through loop diagrams in the SM. At the lowest order, there are basically three diagrams that contribute to the decay width, the $Z$ and photo-penguin diagrams shown in Fig. \ref{fig:FCNC feynman diagram BcDStar} and the $W^{+}W^{-}$ box diagrams shown in Fig. \ref{fig:FCNC feynman diagram BcDStar2}. Experimental search for these FCNC processes started through the $B\rightarrow K^{(*)}\ell^{+}\ell^{-}$ channels by CDF collaboration in 1998 \cite{CDF:1999uew}. Further studies on these channels were presented by BELLE \cite{Belle:2001oey,Belle:2003ivt,Belle:2009zue,Belle:2016fev,BELLE:2019xld,Belle:2019oag}, BABAR \cite{BaBar:2003szi,BaBar:2008jdv,BaBar:2012mrf}, CMS\cite{CMS:2015bcy} and LHCb \cite{LHCb:2012juf,LHCb:2013ghj,LHCb:2014vgu,LHCb:2017avl,LHCb:2021trn}, thus kickstarting a significant interest in these channels. Unlike the $b\rightarrow s\ell^{+}\ell^{-}$ modes, however, the $b\rightarrow d\ell^{+}\ell^{-}$ modes have less information available from the experimental arena. This can be attributed to the low branching fractions for the latter case, which in turn is primarily due to the CKM suppression induced by $V_{td}$ in the expression for decay width, which is much smaller than $V_{ts}$ for the former case. The related decay modes associated with the $B_c$ meson will be the rare processes such as the semileptonic decays $B_{c}^{-}\rightarrow D^{(*)-}\ell^{+}\ell^{-}$. These modes will be equally important in testing new physics and will provide complementary information. This is the rare decay mode we will be mainly focusing on in this work. Our intention is to obtain $q^{2}$ distribution of the $B_{c}\rightarrow D^{*}$ form factors by adopting certain parametrization technique and then ultimately present predictions on branching fractions for the different lepton modes, and also perform a full angular analysis by predicting the various angular observables and other observables like the forward backward asymmetry, longitudinal and transverse polarization fractions of the $D^{*}$ meson and certain form factor independent clean observables in the SM framework.  

 The contents of the paper are organized as follows: In section \ref{section:Theoretical background}, we describe analytic expressions of the various physical observables we intend to predict and analyze in this work, along with brief discussions on the relevant form factors in the modified pQCD framework. In addition, we also discuss the form of the light cone distribution amplitudes (LCDAs) of the participating mesons. In section \ref{section:Extraction of LCDA parameters} we extract the LCDA shape parameters of the participating mesons, and present predictions of the relevant form factors at $q^{2}=0$, calculated using the extracted parameters as inputs. In section \ref{section:Extrapolation to full physical range} we extract information on $B_{c}\rightarrow D^{*}$ form factors over the full physical $q^{2}$ region, using some suitable symmetry relations and appropriate form factor parametrization. In section \ref{section:Prediction of physical observables}, we present our prediction of some physical observables, involving branching fractions and a number of angular observables, along with observables like forward-backward asymmetry and longitudinal and transverse polarization fractions. Finally, in section \ref{section:Summary and conclusions} we briefly summarize our work.

\section{Theoretical Background}
\label{section:Theoretical background}
We begin by briefly discussing the theoretical background required for this work. Like the $B$ meson decays, studying the rare decay mode $B_{c}^{-}\rightarrow D^{*-}\ell^{+}\ell^{-}$ of the $B_{c}$ meson also poses a challenge due to the unavailability of concrete information of the form factors over the full physical $q^{2}$ region. 
From a theoretical perspective, there has been a considerable amount of work in recent years that has attempted to predict the form factors governing these decay modes. These include the QCD sum rules (QCDSR) \cite{Azizi:2008vv,Kiselev:2002vz}, light front quark model (LFQM) \cite{Geng:2001vy}, constituent quark model (CQM) \cite{Geng:2001vy}, to name a few. In this work, however, we will be performing the analysis in the perturbative QCD (pQCD) framework. There are a few reasons why we chose pQCD as the framework of choice for our analysis. First, this approach renders high-precision predictions in the low $q^{2}$ region, compared to other approaches, which, although have a broader range of applicability, do so at the cost of relatively lower precision. Second, the calculations of pQCD are based on fundamental principles of QFT and perturbation theory, making fewer model assumptions in its region of applicability, compared to QCDSR or LFQM, which rely on specific model assumptions that might, in turn, lead to loss of generality in their predictions. 

The pQCD approach has already been widely adopted to calculate the form factors of various decay modes of heavy-light B meson systems \cite{Kurimoto:2002sb,Li:2008ts,Hu:2019bdf}. Moreover, a number of works have also been done to study the same for decay modes involving $B_{c}$ mesons \cite{Rui:2014tpa,Rui:2017pre,Rui:2018kqr,Wang:2012lrc,PhysRevD.90.094018,Hu:2019qcn}, during which, a modified pQCD approach was also formulated in \cite{PhysRevD.97.113001}. The form factors are generally expressed in terms of a hard part encoding all the perturbative contributions, and light cone distribution amplitudes (LCDAs) of the participating mesons. A quick inspection of these papers would reveal that the shape of these LCDAs are defined by certain shape parameters, which in all existing works has been considered as numerical inputs. The novelty of our work lies in the fact that we have taken a data-driven approach to constrain these shape parameters through statistical methods. This enables us to predict the form factors and, hence, a number of physical observables to a degree of precision that is in greater agreement with existing experimental observations and also paves the way for predictions whose observations are yet to be obtained in future experiments. The details of the approach and methodology will be discussed in the later sections of this paper.

In subsection \ref{subsection:physical observables}, we introduce all the physical observables that we will be predicting and mention the relevant expressions. From there, we move forward, simplifying each observable in terms of more fundamental quantities, ultimately highlighting the primary motivation for this work. In subsection \ref{subsection:form factors}, we introduce the form factors and briefly discuss the modified perturbative QCD framework. Finally, in subsection \ref{subsection:LCDAs}, we introduce the light cone distribution amplitudes of the participating mesons. 

\subsection{Physical Observables}
\label{subsection:physical observables}
In this subsection, we briefly discuss the various physical observables that we will be predicting in this work. 

\subsubsection{Decay widths and branching fractions}
\label{subsubsection:decay width and branching fractions}

\begin{itemize}
		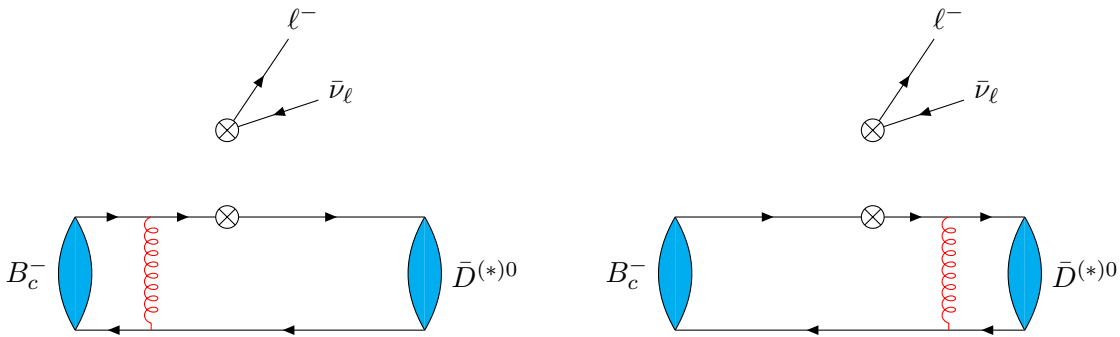
\begin{figure}[htb!]
		\centering
		\begin{tikzpicture}
			\begin{feynman}
				\vertex[crossed dot](a){};
				\vertex[left=1.0cm of a](a1);
				\vertex[left=1.0cm of a1](a2);
				\vertex[right=2.6cm of a](a3);
				\vertex[below=1.5cm of a1](b1);
				\vertex[left=1.0cm of b1](b2);
				\vertex[right=3.6cm of b1](b3);
				\vertex[crossed dot][above=1.0cm](c){};
				\vertex[above right=1.5cm and 1.0cm of c](c1){$\ell^{-}$};
				\vertex[above right=0.5cm and 1.5cm of c](c2){$\bar{\nu}_{\ell}$};
				\diagram*{(a2)--[arrow size=1pt,fermion](a1)--[arrow size=0pt,fermion](a)--[arrow size=1pt,fermion](a3),(b3)--[arrow size=1pt,fermion](b1)--[arrow size=1pt,fermion](b2),(c)--[arrow size=1pt,fermion](c1),(c2)--[arrow size=1pt,fermion](c),(a1)--[style=red,gluon](b1),(a2)--[fill=cyan,bend right, plain,edge label'={\(B_c^-\)}](b2),(a2)--[fill=cyan,bend left, plain](b2),(a3)--[fill=cyan,bend left,plain,edge label={\(\bar{D}^{(*)0}\)}](b3),(a3)--[fill=cyan,bend right,plain](b3)};
			\end{feynman}
			{}		\end{tikzpicture}
		\qquad
		\begin{tikzpicture}
			\begin{feynman}
				\vertex[crossed dot](a){};
				\vertex[left=2.6cm of a](a1);
				\vertex[right=1.0cm of a](a2);
				\vertex[right=1.0cm of a2](a3);
				\vertex[below=1.5cm of a2](b1);
				\vertex[left=3.6cm of b1](b2);
				\vertex[right=1.0cm of b1](b3);
				\vertex[crossed dot][above=1.0cm](c){};
				\vertex[above right=1.5cm and 1.0cm of c](c1){$\ell^{-}$};
				\vertex[above right=0.5cm and 1.5cm of c](c2){$\bar{\nu}_{\ell}$};
				\diagram*{(a1)--[arrow size=1pt,fermion](a)--[arrow size=0pt,fermion](a2)--[arrow size=1pt,fermion](a3),(b3)--[arrow size=1pt,fermion](b1)--[arrow size=1pt,fermion](b2),(c)--[arrow size=1pt,fermion](c1),(c2)--[arrow size=1pt,fermion](c),(a2)--[style=red,gluon](b1),(a1)--[fill=cyan,bend right,plain,edge label'={\(B_c^-\)}](b2),(a1)--[fill=cyan,bend left,plain](b2),(a3)--[fill=cyan,bend right,plain](b3),(a3)--[fill=cyan,bend left,plain,edge label={\(\bar{D}^{(*)0}\)}](b3)};
			\end{feynman}
		\end{tikzpicture}
		\caption{Leading order Feynman diagram for $B_{c}^{-}\rightarrow \bar{D}^{(*)0}\ell^{-}\bar{\nu}_{\ell}$ charged current semileptonic decays with $\ell=(e,\mu,\tau)$.}
		\label{fig:charged current feynman diagram BcDStar}
	\end{figure}
	\item The effective Hamiltonian for $B_{c}\rightarrow D^{(*)}$ semileptonic decays shown in Fig. \ref{fig:charged current feynman diagram BcDStar}, involving $b\rightarrow u$ charged current interactions, can be expressed as
	\begin{equation}
		\mathcal{H}_{eff}(b\rightarrow u\ell^{-}\bar{\nu}_{\ell})=\frac{G_{F}}{\sqrt{2}}V_{ub}\bar{u}\gamma_{\mu}(1-\gamma_{5})b.\bar{\ell}\gamma^{\mu}(1-\gamma_{5})\nu_{\ell},
	\end{equation}

	has the same form as that for $B\rightarrow \pi(\rho) \ell\nu_{\ell}$ transitions \cite{Wang:2012ab}. Owing to the fact that the underlying quark level transition are the same for both cases, we are going to draw an analogy between $B_{c}^{-}\rightarrow \bar{D}^{0}\ell^{-}\bar{\nu}_{\ell}$ and $B\rightarrow \pi \ell\nu_{\ell}$, and $B_{c}^{-}\rightarrow \bar{D}^{*0}\ell^{-}\bar{\nu}_{\ell}$ and $B\rightarrow \rho \ell\nu_{\ell}$, the only difference between their expressions of semileptonic decay width being the masses of the initial and the final mesons. The differential decay widths for $B_{c}^{-}\rightarrow \bar{D}^{0}\ell^{-}\bar{\nu}_{\ell}$ channels \cite{Wang:2012ab,Sakaki:2013bfa} are expressed as
	\begin{equation}
	\begin{split}
	\frac{d\Gamma(B_{c}^{-}\rightarrow \bar{D}^{0}\ell^{-}\bar{\nu}_{\ell})}{dq^{2}}=&\frac{G_{F}^{2}|V_{ub}|^{2}}{192\pi^{3}m_{B_{c}}^{3}}q^{2}\sqrt{\lambda(q^{2})}\left(1-\frac{m_{\ell}^{2}}{q^{2}}\right)^{2}\\&\biggl[\left(1+\frac{m_{\ell}^{2}}{2q^{2}}\right)H_{V,0}^{s\hspace{0.1cm}2}+\frac{3}{2}\frac{m_{\ell}^{2}}{q^{2}}H_{V,t}^{s\hspace{0.2cm}2}\biggr],
	\end{split}
	\label{eqn:differential decay width Bc_D}
	\end{equation}
	and for $B_{c}^{-}\rightarrow \bar{D}^{*0}\ell^{-}\bar{\nu}_{\ell}$ channels  \cite{Biswas:2021cyd,Sakaki:2013bfa} are expressed as
	\begin{equation}
	\begin{split}
	\frac{d\Gamma(B_{c}^{-}\rightarrow \bar{D}^{*0}\ell^{-}\bar{\nu}_{\ell})}{dq^{2}}=&\frac{G_{F}^{2}|V_{ub}|^{2}}{192\pi^{3}m_{B_{c}}^{3}}q^{2}\sqrt{\lambda(q^{2})}\left(1-\frac{m_{\ell}^{2}}{q^{2}}\right)^{2}\\&\biggl[\left(1+\frac{m_{\ell}^{2}}{2q^{2}}\right)(H_{V,+}^{2}+H_{V,-}^{2}+H_{V,0}^{2})+\frac{3}{2}\frac{m_{\ell}^{2}}{q^{2}}H_{V,t}^{2}\biggr],
	\end{split}
	\label{eqn:differential decay width Bc_DStar}
	\end{equation}
	extracting out only the Standard Model contribution from the references and with 
	\begin{equation}
	\begin{split}
	H_{V,0}^{s}(q^{2})&=\sqrt{\frac{\lambda(q^{2})}{q^{2}}}F_{+}(q^{2}),\\
	H_{V,t}^{s}(q^{2})&=\frac{m_{B_{c}}^{2}-m_{D}^{2}}{\sqrt{q^{2}}}F_{0}(q^{2}),\\
	H_{V,\pm}(q^{2})&=(m_{B_{c}}+m_{D^{*}})A_{1}(q^{2})\mp \frac{\sqrt{\lambda(q^{2})}}{m_{B_{c}}+m_{D^{*}}}V(q^{2}),\\
	H_{V,0}(q^{2})&=\frac{m_{B_{c}}+m_{D^{*}}}{2m_{D^{*}}\sqrt{q^{2}}}\Biggl[-(m_{B_{c}}^{2}-m_{D^{*}}^{2}-q^{2})A_{1}(q^{2})+\frac{\lambda(q^{2})}{(m_{B_{c}}+m_{D^{*}})^{2}}A_{2}(q^{2})\Biggr],\\
	H_{V,t}(q^{2})&=-\sqrt{\frac{\lambda(q^{2})}{q^{2}}}A_{0}(q^{2}),
	\end{split}
	\end{equation}
	being the expressions for the Helicity amplitudes \cite{Sakaki:2013bfa} expressed as functions of the form factors.
		\begin{figure}[htb!]
		\begin{tikzpicture}
		\begin{feynman}
		\vertex[crossed dot](a){};
		\vertex[crossed dot][right=1.7cm of a](a1){};
		\vertex[left=0.85cm of a](a2);
		\vertex[left=0.85cm of a2](a4);
		\vertex[right=1.7cm of a1](a3);
		\vertex[above=1.0cm of a](d);
		\vertex[right=1.7cm of d](d1);
		\vertex[above right=0.85cm and 0.85cm of d](c);
		\vertex[above=1.0cm of c](c1);
		\vertex[above left=0.85cm and 0.85cm of c1](c2){$\ell^{-}$};
		\vertex[above right=0.85cm and 0.85cm of c1](c3){$\ell^{+}$};
		\vertex[below=1.25cm of a](b);
		\vertex[left=0.85cm of b](b1);
		\vertex[left=0.85cm of b1](b2);
		\vertex[right=3.4cm of b](b3);
		\vertex[right=0.1cm of d1](d2){$V_{i}$};
		\diagram*{(a4)--[arrow size=1pt,fermion](a2)--[arrow size=0pt,fermion](a)--[arrow size=1pt,fermion](d)--[arrow size=1pt,fermion,edge label=$t$](c)--[arrow size=1pt,fermion,edge label=$t$](d1)--[arrow size=1pt,fermion](a1)--[arrow size=1pt,fermion](a3),(b3)--[arrow size=1pt,fermion](b)--[arrow size=1pt,plain](b1)--[arrow size=1pt,fermion](b2),(c)--[style=blue,photon,edge label={$Z^{0}$, $\gamma$}](c1),(c1)--[arrow size=1pt,fermion](c2),(c3)--[arrow size=1pt,fermion](c1),(a2)--[style=red,gluon](b1),(d)--[fill=gray!20,bend left,plain](d1),(d)--[fill=gray!20,bend right,plain](d1),(a4)--[fill=cyan,bend left,plain](b2),(a4)--[fill=cyan,bend right,plain,edge label'=\(B_{c}^{-}\)](b2),(a3)--[fill=cyan,bend left,plain,edge label={\(D^{(*)-}\)}](b3),(a3)--[fill=cyan,bend right,plain](b3)};
		\end{feynman}
		\end{tikzpicture}
		\qquad
		\begin{tikzpicture}
		\begin{feynman}
		\vertex[crossed dot](a){};
		\vertex[crossed dot][right=1.7cm of a](a1){};
		\vertex[left=1.7cm of a](a2);
		\vertex[right=0.85cm of a1](a3);
		\vertex[right=0.85cm of a3](a4);
		\vertex[above=1.0cm of a](d);
		\vertex[right=1.7cm of d](d1);
		\vertex[above right=0.85cm and 0.85cm of d](c);
		\vertex[above=0.85cm of c](c1);
		\vertex[above left=0.85cm and 0.85cm of c1](c2){$\ell^{-}$};
		\vertex[above right=0.85cm and 0.85cm of c1](c3){$\ell^{+}$};
		\vertex[below=1.25cm of a](b);
		\vertex[left=1.7cm of b](b1);
		\vertex[right=2.55cm of b](b2);
		\vertex[right=0.85cm of b2](b3);
		\vertex[right=0.1cm of d1](d2){$V_{i}$};
		\diagram*{(a2)--[arrow size=1pt,fermion](a)--[arrow size=1pt,fermion](d)--[arrow size=1pt,fermion,edge label=$t$](c)--[arrow size=1pt,fermion,edge label=$t$](d1)--[arrow size=1pt,fermion](a1)--[arrow size=0pt,fermion](a3)--[arrow size=1pt,fermion](a4),(b3)--[arrow size=1pt,fermion](b2)--[plain](b)--[arrow size=1pt,fermion](b1),(c)--[style=blue,photon,edge label={$Z^{0}$, $\gamma$}](c1),(c1)--[arrow size=1pt,fermion](c2),(c3)--[arrow size=1pt,fermion](c1),(a3)--[style=red,gluon](b2),(d)--[fill=gray!20,bend left,plain](d1),(d)--[fill=gray!20,bend right,plain](d1),(a2)--[fill=cyan,bend left,plain](b1),(a2)--[fill=cyan,bend right,plain,edge label'=\(B_{c}^{-}\)](b1),(a4)--[fill=cyan,bend left,plain,edge label={\(D^{(*)-}\)}](b3),(a4)--[fill=cyan,bend right,plain](b3)};
		\end{feynman}
		\end{tikzpicture}
		\caption{$Z^{0}$ and $\gamma$ penguin diagrams in effective theory for $B_{c}^{-}\rightarrow D^{(*)-}\ell^{+}\ell^{-}$ channel with $\ell=(e,\mu,\tau)$. $V_{i}$ denotes the intermediate resonance states $\rho$, $\omega$, $\phi$, $J/\psi$ and $\psi(2S)$.}
		\label{fig:FCNC feynman diagram BcDStar}
	\end{figure}
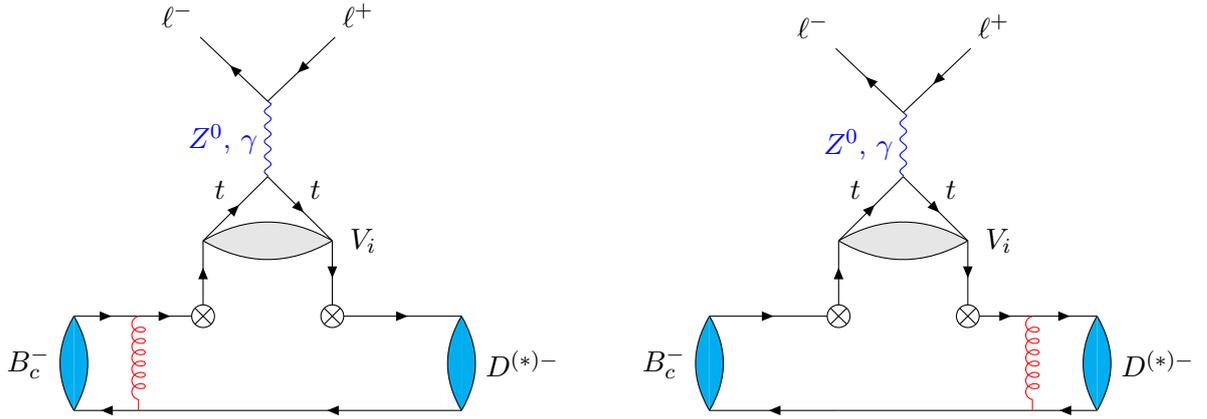
	\begin{figure}[htb!]
		\begin{tikzpicture}
		\begin{feynman}
		\vertex[crossed dot](a){};
		\vertex[crossed dot][right=1.7cm of a](a1){};
		\vertex[right=1.5cm of a1](a2);
		\vertex[left=1.7cm of a](a3);
		\vertex[right=0.85cm of a3](a4);
		\vertex[crossed dot][above=1.0cm of a](b){};
		\vertex[crossed dot][right=1.7cm of b](b1){};
		\vertex[above right=0.85cm and 0.85cm of b1](b2){$\ell^{-}$};
		\vertex[above left=0.85cm and 0.85cm of b](b3){$\ell^{+}$};
		\vertex[below left=1.3cm and 0.85cm of a](c);
		\vertex[right=4.05cm of c](c1);
		\vertex[left=0.85cm of c](c2);
		\diagram*{(a3)--[arrow size=1pt,fermion](a4)--[plain](a)--[arrow size=1pt,fermion, edge label=$t$](a1)--[arrow size=1pt,fermion](a2),(c1)--[arrow size=1pt,fermion](c)--[arrow size=1pt,fermion](c2),(b3)--[arrow size=1pt,fermion](b)--[arrow size=1pt,fermion, edge label=$\nu_{l}$](b1)--[arrow size=1pt,fermion](b2),(a4)--[style=red,gluon](c),(a3)--[fill=cyan,bend left,plain](c2),(a3)--[fill=cyan,bend right,plain,edge label'=\(B_{c}^{-}\)](c2),(a2)--[fill=cyan,bend left,plain,edge label={\(D^{(*)-}\)}](c1),(a2)--[fill=cyan,bend right,plain](c1)};
		\end{feynman}
		\end{tikzpicture}	
		\qquad
		\begin{tikzpicture}
		\begin{feynman}
		\vertex[crossed dot](a){};
		\vertex[crossed dot][right=1.7cm of a](a1){};
		\vertex[right=1.7cm of a1](a2);
		\vertex[left=1.7cm of a](a3);
		\vertex[right=0.85cm of a1](a4);
		\vertex[crossed dot][above=1.0cm of a](b){};
		\vertex[crossed dot][right=1.7cm of b](b1){};
		\vertex[above right=0.85cm and 0.85cm of b1](b2){$\ell^{-}$};
		\vertex[above left=0.85cm and 0.85cm of b](b3){$\ell^{+}$};
		\vertex[below right=1.3cm and 0.85cm of a1](c);
		\vertex[right=0.85cm of c](c1);
		\vertex[left=4.25cm of c](c2);
		\diagram*{(a3)--[arrow size=1pt,fermion](a)--[arrow size=1pt,fermion, edge label=$t$](a1)--[plain](a4)--[arrow size=1pt,fermion](a2),(c1)--[arrow size=1pt,fermion](c)--[arrow size=1pt,fermion](c2),(b3)--[arrow size=1pt,fermion](b)--[arrow size=1pt,fermion, edge label=$\nu_{l}$](b1)--[arrow size=1pt,fermion](b2),(a4)--[style=red,gluon](c),(a3)--[fill=cyan,bend left,plain](c2),(a3)--[fill=cyan,bend right,plain,edge label'=\(B_{c}^{-}\)](c2),(a2)--[fill=cyan,bend left,plain,edge label={\(D^{(*)-}\)}](c1),(a2)--[fill=cyan,bend right,plain](c1)};
		\end{feynman}
		\end{tikzpicture}	
		\caption{$W^{+}W^{-}$ box diagrams in effective theory for $B_{c}^{-}\rightarrow D^{(*)-} \ell^{+}\ell^{-}$ channel with $\ell=(e,\mu,\tau)$.}
		\label{fig:FCNC feynman diagram BcDStar2}
	\end{figure}
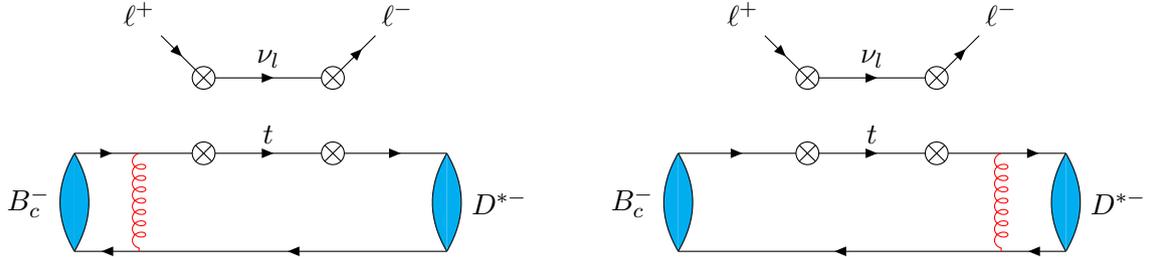
	\item Effective Hamiltonian in SM for the FCNC decay mode $B_{c}^{-}\rightarrow D^{(*)-}~\ell^{+}\ell^{-}$ shown in Figs \ref{fig:FCNC feynman diagram BcDStar} and \ref{fig:FCNC feynman diagram BcDStar2}, with the underlying quark-level transition being $b\rightarrow d \ell^{+}\ell^{-}$, can be expressed as
	\begin{equation}
	\mathcal{H}_{eff}=-\frac{G_{F}}{\sqrt{2}}V_{tb}V_{td}^{*}(C_{1}(\mu)O_{1}(\mu)+C_{2}(\mu)O_{2}(\mu)),
	\end{equation}	
	where $C_{1,2}(\mu)$ are the Wilson coefficients and the local operators $O_{1,2}(\mu)$ are expressed as
	\begin{equation}
	\begin{split}
	O_{1}=&(\bar{d_{\alpha}}c_{\alpha})_{V-A}(\bar{d_{\beta}}c_{\beta})_{V-A},\\
	O_{2}=&(\bar{d_{\alpha}}c_{\beta})_{V-A}(\bar{d_{\beta}}c_{\alpha})_{V-A}.
	\end{split}
	\end{equation}	
	 The expressions for the decay width are pretty much complicated and involved to be mentioned here. Hence we refrain from discussing it in details in this work and refer the readers to the references \cite{Colangelo:2010bg,Ali:1999mm,Wang:2012ab} for their convenience.
	\item For the rare decay modes $B_{c}^{-}\rightarrow D^{(*)-}\nu\bar{\nu}$ the effective Hamiltonian is
	\begin{equation}
	\begin{split}
	\mathcal{H}_{eff}(b\rightarrow d \nu\bar{\nu})=&\frac{G_{F}}{\sqrt{2}}\frac{\alpha_{EM}}{2\pi \sin^{2}(\theta_{W})}V_{tb}V_{td}^{*}\eta_{X}X(x_{t})\left[\bar{d}\gamma^{\mu}(1-\gamma_{5})b\right]\left[\bar{\nu}\gamma_{\mu}(1-\gamma_{5})\nu\right],\\[1em]
    =&C_{L}^{b\to d}\mathcal{O}_{L}^{b\to d},
	\end{split}
	\end{equation}
	where $\theta_{W}$ is the Weinberg angle with $\sin^{2}(\theta_{W})=0.231$. $V_{tb}$ and $V_{td}$ are the CKM matrix elements. The function $X(x_{t})$ has been taken from \cite{Buchalla:1995vs}, and $\eta_{X}\approx 1$ represents the QCD correction factor. The differential decay width for $B_{c}^{-}\rightarrow D^{-}\nu\bar{\nu}$ mode is expressed as \cite{Wang:2012ab}
    \begin{equation}
          \frac{d\Gamma(B_{c}^{-}\rightarrow D^{-}\nu\bar{\nu}))}{dq^2} = 3 \frac{|C_L^{b \to d}|^2 \lambda(q^{2})^{3/2}(M^{2}, m^{2}, q^{2})}{96 M^{3} \pi^3} |F_+(q^2)|^2 \,.  
    \end{equation}
    and that for $B_{c}^{-}\rightarrow D^{*-}\nu\bar{\nu}$ mode is expressed as \cite{Barakat:2001ef,Colangelo:2010bg,Aliev:2007gr}
\begin{equation}
\resizebox{0.95\hsize}{!}{%
$
\begin{aligned}
\frac{d\Gamma(B_{c}^{-}\rightarrow D^{*-}\nu\bar{\nu})}{dq^{2}}=&\frac{G_{F}^{2}\alpha_{EM}^{2}}{2^{10}\pi^{5}M^{3}}\cdot\Biggl|\frac{X(x_{t})}{\sin^{2}(\theta_{W})}\Biggr|^{2}\cdot \eta_{X}^{2}\cdot |V_{tb}V_{td}^{*}|^{2}\sqrt{\lambda(q^{2})}
\Biggl\{8\lambda(q^{2})q^{2}\frac{V(q^{2})^{2}}{(M+m)^{2}}\\[10pt]&+\frac{\lambda(q^{2})^{2}}{m^{2}}\cdot\frac{A_{2}(q^{2})^{2}}{(M+m)^{2}}+\frac{1}{m^{2}}(M+m)^{2}(\lambda(q^{2})+12m^{2}q^{2})\cdot A_{1}(q^{2})^{2}\\[10pt]&-\frac{2\lambda(q^{2})}{m^{2}}(M^{2}-m^{2}-q^{2})\cdot Re[A_{1}(q^{2})^{*}A_{2}(q^{2})]\Biggr\},
\end{aligned}
$}
\end{equation}	
where $\lambda(q^{2})$, the phase space factor is expressed as 
\begin{equation}
	\label{eqn:phasefactor}
	\lambda(q^{2})=(m_{B_{c}}^{2}+m_{D^{(*)}}^{2}-q^{2})^{2}-4m_{B_{c}}^{2}m_{D^{(*)}}^{2}.
\end{equation}
\end{itemize}
The total decay width can be obtained by simply integrating the expressions for the differential decay widths over $m_{l}^{2}$ to $(m_{B_{c}}-m_{D^{(*)}})^{2}$ and 4$m_{l}^{2}$ to $(m_{B_{c}}-m_{D^{(*)}})^{2}$ as the physical $q^{2}$ region for charged current and neutral processes respectively. For $B_{c}^{-}\to D^{(*)-}\nu\bar{\nu}$ modes the integration is done over $0\text{ to }(m_{B_{c}}-m_{D^{(*)}})^{2}$. Once the total decay width is obtained, the branching fractions can be calculated by a straightforward division of the decay width with the total width of $B_{c}$ meson.

\subsubsection{Angular analysis of rare $B_{c}^{-}\rightarrow D^{(*)-}\ell^{+}\ell^{-}$ channel}
\label{subsubsection:Angular analysis}
\begin{figure}[htb!]
	\centering
	\includegraphics[width=7.0cm]{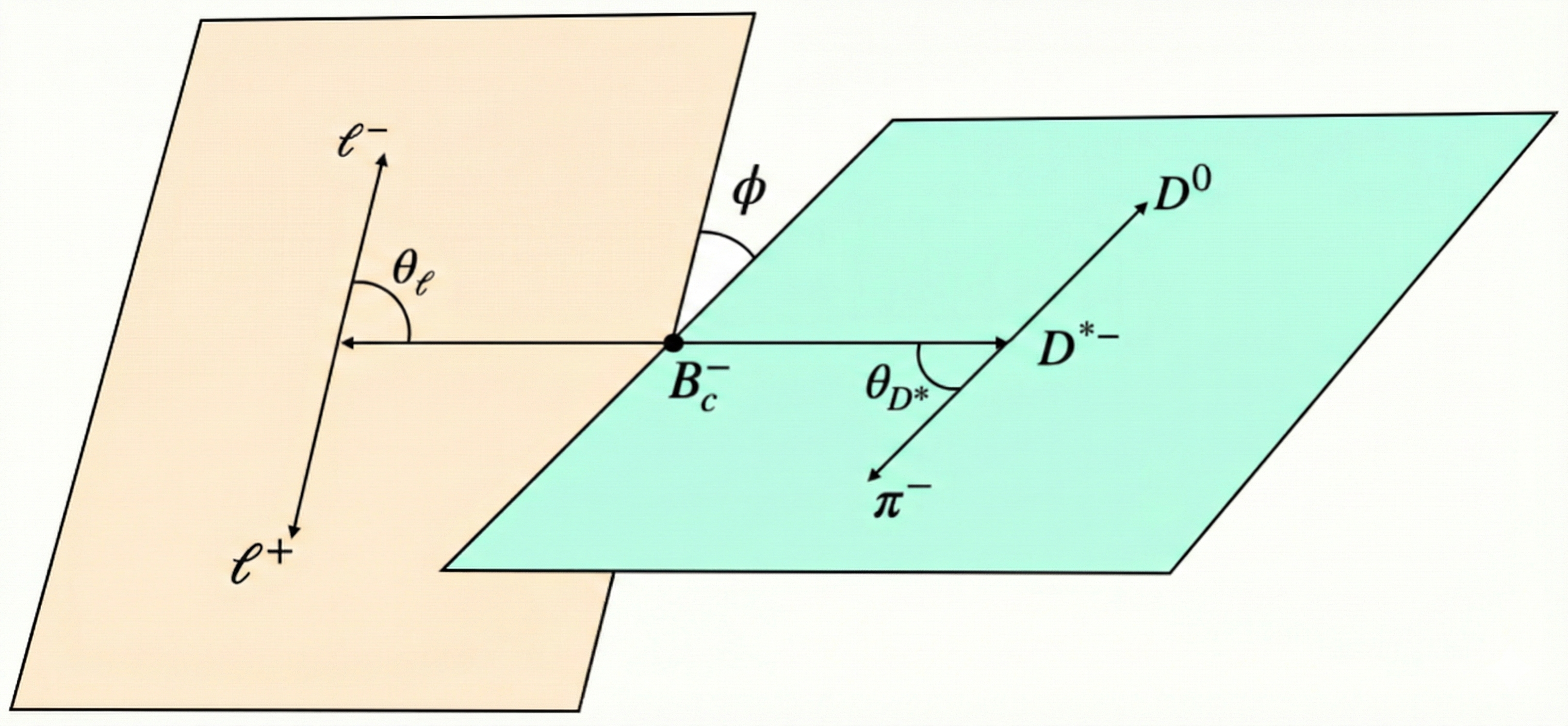}
	\caption{Kinematics of $B_{c}^{-}\rightarrow D^{*-}(\rightarrow D^{0}\pi^{-}) \ell^{+}\ell^{-}$ four body decay.}
	\label{fig: Angular observables decay planes}
\end{figure}
In addition to the branching fractions, there are a multitude of observables that come up through a full angular analysis of the decay channel. Angular analysis of $B_{c}^{-}\rightarrow D^{(*)-}\ell^{+}\ell^{-}$ channel particularly gathers more interest due to the underlying $b\rightarrow dl^{+}l^{-}$ quark level transition being an FCNC process, which is forbidden in tree level and can only operate through loop diagrams in SM. Moreover, the angular observables extracted from these processes being highly sensitive to NP effects can also serve as a probe to explore possible NP scenarios. The effective Hamiltonian for $b\rightarrow d\ell^{+}\ell^{-}$ channel has the form as \cite{Altmannshofer:2008dz}
\begin{equation}
\mathcal{H}_{eff}=-\frac{4G_{F}}{\sqrt{2}}\left(\lambda_{t}\mathcal{H}_{eff}^{t}+\lambda_{u}\mathcal{H}_{eff}^{u}\right),
\end{equation}
where $\lambda_{i}=V_{ib}V_{id}^{*}$ represents the CKM combination, and 
\begin{equation}
		H_{eff}^{t}=C_{1}O_{1}^{c}+C_{2}O_{2}^{c}+\sum_{i=3}^{6}C_{i}O_{i}+\sum_{i=7}^{10}\left(C_{i}O_{i}+C_{i}^{'}O_{i}^{'}\right),
	\end{equation}	 
	and
	\begin{equation}
	H_{eff}^{u}=C_{1}\left(O_{1}^{c}-O_{1}^{u}\right)+C_{2}\left(O_{2}^{c}-O_{2}^{u}\right),
	\end{equation}
	in the SM framework, $O_{i}\equiv O_{i}(\mu)$ represent the four Fermi operators, and $C_{i}\equiv C_{i}(\mu)$ represent the corresponding Wilson coefficients, both being functions of the re-normalization scale $\mu$. The operators $O_{1,2}^{c,u}$ are the current-current operators, $O_{3-6}$ the QCD penguin operators, $O_{7,8}$ the electromagnetic and chromomagnetic operators, and $O_{9,10}$ the semileptonic operators, respectively. 
	With the effective Hamiltonian of the $b\rightarrow d \ell^{+}\ell^{-}$ decay, the differential decay width of the four-body decay $B_{c}^{-}\rightarrow D^{*-}(\rightarrow D^{0}\pi^{-}) \ell^{+}\ell^{-}$ can be written, where, since we are more interested in probing the angular dependence of the decay, will have the relevant decay angles along with $q^{2}$ as the variables. The differential decay width can be expressed as
\begin{equation}
\frac{d^{4}\Gamma}{dq^{2}d\cos\theta_{D^{*}}d\cos\theta_{l}d\phi}=\frac{9}{32\pi}\sum_{i}I_{i}(q^{2})f_{i}(\theta_{D^{*}},\theta_{l},\phi),
\label{eqn:decay width angular}
\end{equation}
where $I_{i}(q^{2})$ are the various angular coefficients which encode all the $q^{2}$ dependence and can be further expressed in terms of various transversity amplitudes $\mathcal{A}_{\perp}^{L,R}$, $\mathcal{A}_{\parallel}^{L,R}$ and $\mathcal{A}_{0}^{L,R}$ explicitly as shown later, and $f_{i}(\theta_{{D}^{*}},\theta_{l},\phi)$ are the various angular coefficients which encode the necessary angular information. Their expressions has been shown in Table \ref{table:anglularf}.

\begin{table}[htb!]
\centering
\renewcommand{\arraystretch}{1.3}
\begin{tabular}{|cc|cc|}
\hline
Angular Observables&$f_{i}(\theta_{D^{*}},\theta_{l},\phi)$&Angular Observables&$f_{i}(\theta_{{D}^{*}},\theta_{l},\phi)$\\
\hline
$I_{1s}(q^{2})$&$\sin^{2}\theta_{D^{*}}$&$I_{5}(q^{2})$&$\sin 2\theta_{D^{*}}\sin\theta_{l}\cos \phi$\\
$I_{1c}(q^{2})$&$\cos^{2}\theta_{D^{*}}$&$I_{6s}(q^{2})$&$\sin^{2}\theta_{D^{*}}\cos\theta_{l}$\\
$I_{2s}(q^{2})$&$\sin^{2}\theta_{D^{*}}\cos2\theta_{l}$&$I_{7}(q^{2})$&$\sin 2\theta_{D^{*}}\sin\theta_{l}\sin\phi$\\
$I_{2c}(q^{2})$&$\cos^{2}\theta_{D^{*}}\cos2\theta_{l}$&$I_{8}(q^{2})$&$\sin 2\theta_{D^{*}}\sin 2\theta_{l}\sin\phi$\\
$I_{3}(q^{2})$&$\sin^{2}\theta_{D^{*}}\sin^{2}\theta_{l}\cos 2\phi$&$I_{9}(q^{2})$&$\sin^{2}\theta_{D^{*}}\sin^{2}\theta_{l}\sin 2\phi$\\
$I_{4}(q^{2})$&$\sin 2\theta_{D^{*}}\sin 2\theta_{l}\cos\phi$&&\\
\hline
\end{tabular}
\caption{Expressions for $f_{i}(\theta_{D^{*}},\theta_{l},\phi)$.}
\label{table:anglularf}
\end{table}
Regarding the angles $\theta_{D^{*}}$, $\theta_{\ell}$ and $\phi$ are shown in Fig. \ref{fig: Angular observables decay planes}, where
\begin{itemize}
	\item $\theta_{D^{*}}$ denotes the angle made by $\pi$ in the centre of mass system of $D^{0}$ and $\pi^{-}$ with respect to the direction of flight of $D^{*-}$,
	\item $\theta_{\ell}$ denotes the angle made by $\ell^{-}$ in the centre of mass system of $\ell^{+}$ and $\ell^{-}$ with respect to the direction of flight of the lepton pair, and
	\item $\phi$ denotes the angle between the decay planes formed by the $(D^{0}\pi^{-})$ and $(\ell^{+},\ell^{-})$ pairs.
\end{itemize}
 Since the primary observables that we will be predicting in this work are the CP-conserving ones, which consists of contributions from the CP conjugate mode as well, in addition to the above differential decay width, we also need to mention the differential decay width of the CP conjugate decay mode $B_{c}^{+}\rightarrow D^{*+}(\rightarrow D^{0}\pi^{+}) \ell^{+}\ell^{-}$ which can be expressed as
\begin{equation}
\frac{d^{4}\bar{\Gamma}}{dq^{2}d\cos\theta_{D^{*}}d\cos\theta_{\ell}d\phi}=\frac{9}{32\pi}\sum_{i}\bar{I_{i}}(q^{2})f_{i}(\theta_{D^{*}},\theta_{l},\phi),
\label{eqn:decay width angular CP}
\end{equation}
where $f_{i}(\theta_{D^{*}},\theta_{l},\phi)$ has the same functional form as in Table \ref{table:anglularf}, while the angular observables $\bar{I_{i}}$ of the CP conjugate mode are related to corresponding $I_{i}$ as
\begin{equation}
\begin{split}
&I_{1s,1c,2s,2c,3,4,7}\rightarrow \bar{I}_{1s,1c,2s,2c,3,4,7},\\
&I_{5,6s,8,9}\rightarrow -\bar{I}_{5,6s,8,9}.
\end{split}
\end{equation}
The observables $I_{1s,1c,2s,2c,3,4,7}$ do not change sign upon writing the CP conjugate mode, while, on the contrary, the observables $I_{5,6s,8,9}$ do change sign. The reason behind this lies in the convention of the angles that we have mentioned previously. For the angle $\theta_{l}$, a CP transformation interchanges the lepton and the anti-lepton pair, which leads to the transformation $\theta_{l}\rightarrow \theta_{l}-\pi$. For the angle $\phi$, the same change in the lepton and anti-lepton upon CP transformation leads to the reversal of the angle between the decay planes, leading to the transformation $\phi\rightarrow -\phi$.

The angular coefficients are explicitly defined in terms of the transversity amplitudes and the relevant form factors as \cite{Li:2023mrj}
\begin{align}
		I_{1s}=&\left(\frac{3}{4}-\hat{m_{\ell}}^{2}\right)\left(|\mathcal{A}_{\parallel}^{L}|^{2}+|\mathcal{A}_{\perp}^{L}|^{2}+|\mathcal{A}_{\parallel}^{R}|^{2}+|\mathcal{A}_{\perp}^{R}|^{2}\right)+4\hat{m_{\ell}}^{2}\text{Re}\left[\mathcal{A}_{\perp}^{L}\mathcal{A}_{\perp}^{R*}+\mathcal{A}_{\parallel}^{L}\mathcal{A}_{\parallel}^{R*}\right],\notag\\[1em]
		I_{1c}=&|\mathcal{A}_{0}^{L}|^{2}+|\mathcal{A}_{0}^{R}|^{2}+4\hat{m_{\ell}}^{2}\left(|\mathcal{A}_{t}|^{2}+2\text{Re}[\mathcal{A}_{0}^{L}\mathcal{A}_{0}^{R*}]\right),\notag\\[1em]
		I_{2s}=&\beta_{\ell}^{2}\frac{|\mathcal{A}_{\parallel}^{L}|^{2}+|\mathcal{A}_{\parallel}^{R}|^{2}+|\mathcal{A}_{\perp}^{L}|^{2}+|\mathcal{A}_{\perp}^{R}|^{2}}{4},\notag\\[1em]
		I_{2c}=&-\beta_{\ell}^{2}\left(|\mathcal{A}_{0}^{L}|^{2}+|\mathcal{A}_{0}^{R}|^{2}\right),\notag\\[1em]
		I_3=&\beta_{\ell}^{2}\frac{|\mathcal{A}_{\perp}^{L}|^{2}+|\mathcal{A}_{\perp}^{R}|^{2}-|\mathcal{A}_{\parallel}^{L}|^{2}-|\mathcal{A}_{\parallel}^{R}|^{2}}{2},\notag\\[1em]
		I_4=&\beta_{\ell}^{2}\frac{\text{Re}\left[\mathcal{A}_{0}^{L}\mathcal{A}_{\parallel}^{L*}+\mathcal{A}_{0}^{R}\mathcal{A}_{\parallel}^{R*}\right]}{\sqrt{2}},\label{eqn:angular observables}\\[1em]
		I_5=&\sqrt{2}\beta_{\ell}\text{Re}\left[\mathcal{A}_{0}^{L}\mathcal{A}_{\perp}^{L*}-\mathcal{A}_{0}^{R}\mathcal{A}_{\perp}^{R*}\right],\notag\\[1em]
		I_{6s}=&2\beta_{\ell}\text{Re}\left[\mathcal{A}_{\parallel}^{L}\mathcal{A}_{\perp}^{L*}-\mathcal{A}_{\parallel}^{R}\mathcal{A}_{\perp}^{R*}\right],\notag\\[1em]
		I_7=&\sqrt{2}\beta_{\ell}\text{Im}\left[\mathcal{A}_{0}^{L}\mathcal{A}_{\parallel}^{L*}-\mathcal{A}_{0}^{R}\mathcal{A}_{\parallel}^{R*}\right],\notag\\[1em]
		I_8=&\beta_{\ell}^{2}\frac{\text{Im}\left[\mathcal{A}_{0}^{L}\mathcal{A}_{\perp}^{L*}+\mathcal{A}_{0}^{R}\mathcal{A}_{\perp}^{R*}\right]}{\sqrt{2}},\notag\\[1em]
		I_9=&\beta_{\ell}^{2}\text{Im}\left[\mathcal{A}_{\parallel}^{L*}\mathcal{A}_{\perp}^{L}-\mathcal{A}_{\parallel}^{R*}\mathcal{A}_{\perp}^{R}\right],\notag\notag
\end{align}
where the transverse amplitudes thus shown can further be expressed in terms of Wilson coefficients $C_{7,9}^{eff}(\mu)$ and form factors $A_{0,1,2}(q^{2})$, $V(q^{2})$ and $T_{1,2,3}(q^{2})$ as
\begin{equation}
\scriptsize
\begin{split}
\mathcal{A}_{\perp}^{L,R}=&-N_{\ell}\sqrt{2N_{D^{*}}}\sqrt{\lambda(m_{B_{c}}^{2},m_{D^{*}}^{2},q^{2})}\left[(C_{9}^{eff}\mp C_{10})\frac{V(q^{2})}{m_{B_{c}}+m_{D^{*}}}+2\hat{m_{b}}C_{7}^{eff}T_{1}(q^{2})\right],\\[1em]
\mathcal{A}_{\parallel}^{L,R}=&N_{l}\sqrt{2N_{D^{*}}}\left[(C_{9}^{eff}\mp C_{10})(m_{B_{c}}+m_{D^{*}})A_{1}(q^{2})+2\hat{m_{b}}C_{7}^{eff}(m_{B_{c}}^{2}-m_{D^{*}}^{2})T_{2}(q^{2})\right],\\[1em]
\mathcal{A}_{0}^{L,R}=&\frac{N_{\ell}\sqrt{N_{D^{*}}}}{2m_{D^{*}}\sqrt{q^{2}}}\biggl[(C_{9}^{eff}\mp C_{10})\left\lbrace (m_{B_{c}}^{2}-m_{D^{*}}^{2}-q^{2})(m_{B_{c}}+m_{D^{*}})A_{1}(q^{2})-\frac{\lambda(m_{B_{c}}^{2},m_{D^{*}}^{2},q^{2})}{m_{B_{c}}+m_{D^{*}}}A_{2}(q^{2})\right\rbrace\\&+2m_{b}C_{7}^{eff}\left\lbrace (m_{B_{c}}^{2}+3m_{D^{*}}^{2}-q^{2})T_{2}(q^{2})-\frac{\lambda(m_{B_{c}}^{2},m_{D^{*}}^{2},q^{2})}{m_{B_{c}}^{2}-m_{D^{*}}^{2}}T_{3}(q^{2})\right\rbrace\biggr],\\[1em]
\mathcal{A}_{t}(q^{2})=&2N_{\ell}\sqrt{N_{D^{*}}}\sqrt{\frac{\lambda(m_{B_{c}}^{2},m_{D^{*}}^{2},q^{2})}{q^{2}}}C_{10}A_{0}(q^{2}),
\label{eqn:transversity amplitudes}
\end{split}
\end{equation}
with $\hat{m_{b}}=m_{b}/q^{2}$ and the normalization factors are expressed as
\begin{equation}
\begin{split}
&N_{\ell}=\frac{i\alpha_{e}G_{F}}{4\sqrt{2}\pi}V_{tb}V_{td}^{*},\\[1em]
&N_{D^{*}}=\frac{8\sqrt{\lambda}q^{2}}{3\times 256 \pi^{3}m_{B_{c}}^{3}}\sqrt{1-\frac{4m_{\ell}^{2}}{q^{2}}}\mathcal{B}(D^{*-}\rightarrow D^{0}\pi^{-}),
\end{split}
\end{equation}
with $\mathcal{B}(D^{*-}\rightarrow D^{0}\pi^{-})=67.7(5)\%$ \cite{CLEO:1997rew}. The Wilson coefficients $C_{7,9}^{eff}(\mu)$ in Eqn.\eqref{eqn:transversity amplitudes} have the form \cite{Chen:2001zc,Soni:2020bvu}
	\begin{equation}
	\begin{split}
	&C_{7}^{eff}(\mu)=C_{7}(\mu)+i \alpha_{s}(\mu)\left[\frac{2}{9}\eta^{14/23}\left\lbrace\frac{x_{t}(x_{t}^{2}-5x_{t}-2)}{8(x_{t}-1)^{3}}+\frac{3x_{t}^{2}\ln(x_{t})}{4(x_{t}-1)^{4}}-0.1687\right\rbrace-0.03C_{2}(\mu)\right],\\[1em]
	&C_{9}^{eff}(q^{2},\mu)=C_{9}(\mu)+Y_{pert}(q^{2},\mu)+Y_{res}(q^{2},\mu),
	\end{split}
	\label{eqn:Wilson Coefficients}
	\end{equation}	 
with $x_{t}=m_{t}^{2}/m_{W}^{2}$ and $\eta=\alpha_{s}(m_{W})/\alpha_{s}(\mu)$ with $\alpha_{s}(\mu)$ being calculated at $\mu=m_{b}$. The short-distance contributions from the soft gluon emission and the one-loop contribution from the four Fermi operators $O_{1-6}$ are collected in the $Y_{pert}(q^{2},\mu)$ part of $C_{9}^{eff}(\mu)$ and can be written as \cite{Jin:2020jtu}
\begin{equation}
\begin{split}
Y_{pert}(\hat{s},\mu)=&0.124\omega(\hat{s})+g(\hat{m_{c}},\hat{s})C(\mu)+\lambda_{u}\left[g(\hat{m_{c}},\hat{s})-g(0,\hat{s})\right](3C_{1}(\mu)+C_{2}(\mu))\\
&-\frac{1}{2}g(0,\hat{s})(C_{3}(\mu)+3C_{4}(\mu))-\frac{1}{2}g(1,\hat{s})(4C_{3}(\mu)+4C_{4}(\mu)+3C_{5}(\mu)+C_{6}(\mu))\\
&+\frac{2}{9}(3C_{3}(\mu)+C_{4}(\mu)+3C_{5}(\mu)+C_{6}(\mu)),
\end{split}
\label{eqn:Ypert}
\end{equation}
where $\hat{s}=q^{2}/m_{b}^{2}$, $\hat{m_{c}}=m_{c}/m_{b}$ and $C(\mu)=3C_{1}(\mu)+C_{2}(\mu)+3C_{3}(\mu)+C_{4}(\mu)+3C_{5}(\mu)+C_{6}(\mu)$, with the Wilson Coefficients $C_{1-10}(\mu)$ being evaluated up to the next-to-leading order correction at $\mu=m_{b}$ scale. The numerical values of the Wilson coefficients at $m_{b}$ scale \cite{Buchalla:1995vs,Soni:2020bvu} that have been used in this work are tabulated in Table \ref{table:tabWilson}.

\begin{table}[htb!]
\centering
\begin{tabular}{|cccccccccc|}
\hline
$C_{1}$&$C_{2}$&$C_{3}$&$C_{4}$&$C_{5}$&$C_{6}$&$C_{7}$&$C_{8}$&$C_{9}$&$C_{10}$\\
\hline
-0.175&1.076&0.01258&-0.03279&0.01112&-0.03634&-0.302&-0.148&4.232&-4.410\\
\hline
\end{tabular}
\caption{Numerical values of Wilson coefficients $C_{1-10}$ evaluated at $m_{b}$ scale \cite{Buchalla:1995vs}.}
\label{table:tabWilson}
\end{table}

In addition, in Eqn \eqref{eqn:Ypert}, the term $\omega(\hat{s})$ representing the one-gluon correction to the matrix element of the operator $O_{9}$ is represented as \cite{Jin:2020jtu,Soni:2020bvu,Wang:2012ab}
\begin{equation}
\begin{split}
\omega({\hat{s}})=&-\frac{2}{9}\pi^{2}+\frac{4}{3}\int_{0}^{\hat{s}}\frac{\ln(1-u)}{u}du-\frac{2}{3}\ln(\hat{s})\ln(1-\hat{s})-\frac{5+4\hat{s}}{3(1+2\hat{s})}\ln(1-\hat{s})\\
&-\frac{2\hat{s}(1+\hat{s})(1-2\hat{s})}{3(1-\hat{s})^{2}(1+2\hat{s})}\ln(\hat{s})+\frac{5+9\hat{s}-6\hat{s}^{2}}{6(1-\hat{s})(1+2\hat{s})},
\end{split}
\end{equation}
the functions $g(z,\hat{s})$ and $g(0,\hat{s})$ in Eqn.\eqref{eqn:Ypert} representing the one-loop contributions of the $O_{1-6}$ are represented as \cite{Jin:2020jtu}

\begin{equation}
\begin{split}
g(z,\hat{s})=-\frac{8}{9}\ln(z)+\frac{8}{27}+\frac{4}{9}x-\frac{2}{9}(2+x)\sqrt{|1-x|}
\times \begin{cases}
\ln|\frac{1+\sqrt{1-x}}{1-\sqrt{1-x}}|-i\pi \quad \text{for }x\equiv \frac{4z^{2}}{\hat{s}}< 1\\
2 \arctan(\frac{1}{\sqrt{x-1}}) \quad \text{for }x\equiv \frac{4z^{2}}{\hat{s}}> 1,
\end{cases}
\end{split}
\end{equation}
and
\begin{equation}
g(0,\hat{s})=\frac{8}{27}-\frac{8}{9}\ln(\frac{m_{b}}{\mu})-\frac{4}{9}\ln(\hat{s})+\frac{4}{9}i\pi,
\end{equation}
and $\lambda_{u}=V_{ub}V_{ud}^{*}/V_{tb}V_{td}^{*}$. In addition to $Y_{pert}(q^{2},\mu)$ and the third term $Y_{res}(q^{2},\mu)$ in Eqn.\eqref{eqn:Wilson Coefficients} describing the long-distance contributions to $C_{9}^{eff}(\mu)$, associated with intermediate light vector mesons, i.e., $\rho$, $\omega$ and $\phi$ mesons, and vector charmonium mesons, i.e., $J/\psi\text{ and }\psi(2S)$ is expressed as \cite{Nayek:2018rcq,Jin:2020jtu}
\begin{equation}
\begin{split}
Y_{res}(q^{2},\mu)=&-\frac{3\pi}{\alpha_{e}^{2}}\biggl\{C(\mu)\sum_{V_{i}=J/\psi,\psi(2S)}\frac{m_{V_{i}}\mathcal{B}\left(V_{i}\rightarrow \ell^{+}\ell^{-}\right)\Gamma_{V_{i}}}{q^{2}-m_{V_{i}}^{2}+im_{V_{i}}\Gamma_{V_{i}}}\\
&-\lambda_{u}g(0,\hat{s})(3C_{1}(\mu)+C_{2}(\mu))\sum_{V_{i}=\rho,\phi,\omega}\frac{m_{V_{i}}\mathcal{B}(V_{i}\rightarrow \ell^{+}\ell^{-})\Gamma_{V_{i}}}{q^{2}-m_{V_{i}}^{2}+im_{V_{i}}\Gamma_{V_{i}}}\biggr\},
\end{split}
\end{equation}

where $m_{V_{i}}$ and $\Gamma_{V_{i}}$ are the mass and the total decay width of the vector meson, $V_{i}$ respectively, and $\mathcal{B}(V_{i}\rightarrow \ell^{+}\ell^{-})$ is the branching fraction of each of the dilepton decay mode. The numerical values of these parameters that we have taken as input in this work have been supplied by PDG \cite{ParticleDataGroup:2022pth} and are compiled in Table \ref{table:tabresinp}. Although $\psi(3770)$ and $\psi(4040)$ lie inside the physical $q^{2}$ region, their contribution to $Y_{res}$ is negligible, and their effects do not appear in the plots of any of the angular coefficients. This is primarily because their leptonic decay widths are highly suppressed compared to that of $J/\psi$ and $\psi(2S)$.

\begin{table}[htb!]
	\renewcommand{\arraystretch}{1.3}
\centering
\begin{tabular}{|c|c|c|c|c|}
\hline
Intermediate Meson& $m_{V_{i}}$  & $\Gamma_{V_{i}}$ & \multicolumn{2}{c|} {$\mathcal{B}(V_{i}\rightarrow \ell^{+}\ell^{-})$ with}\\
\cline{4-5}
$V_{i}$&in MeV& in MeV& $l=\mu$&$l=\tau$\\
\hline
$\rho$&775.26(23)&147.4(8)&4.55(28)$\times$ $10^{-5}$&-\\
$\omega$&782.66(13)&8.68(13)&7.4(1.8)$\times$ $10^{-5}$&-\\
$\phi$&1019.461(16)&4.249(13)&2.85(19)$\times$ $10^{-4}$&-\\
$J/\psi$&3096.900(6)&0.0926(17)&5.961(33)$\times$ $10^{-2}$&-\\
$\psi(2S)$&3686.10(6)&0.294(8)&8.0(6)$\times$ $10^{-3}$&3.1(4)$\times$ $10^{-3}$\\
$\psi(3770)$&3773.7(7)&27.2(1.0)&9.6(7)$\times$ $10^{-6}$&-\\
$\psi(4040)$&4040.0(4.0)&84.0(12.0)&1.02(17)$\times$ $10^{-5}$&-\\
\hline
\end{tabular}
\caption{Masses, total decay widths and dilepton branching fractions of the intermediate vector mesons.}
\label{table:tabresinp}
\end{table} 
With the angular coefficients $I_{i}$ discussed, we separate the CP-conserving and CP-violating effects by defining a set of 22 observables, 11 of them defining the CP conserving effects and the rest 11 defining CP-violating effects. They are expressed as \cite{Altmannshofer:2008dz,Li:2023mrj,Biswas:2022lhu}

\begin{equation}
S_{i}=\frac{I_{i}+\bar{I}_{i}}{d(\Gamma+\bar{\Gamma})/dq^{2}}, \qquad A_{i}=\frac{I_{i}-\bar{I}_{i}}{d(\Gamma+\bar{\Gamma})/dq^{2}},
\end{equation} 
where the observables have been normalized by the CP-averaged differential decay width in order to reduce the uncertainties. In this work, however, we will primarily focus our analysis on CP-conserving observables, as CP-violating observables are suppressed in the SM. Moreover, their study would require the inclusion of weak annihilation corrections to the transversity amplitudes \cite{Altmannshofer:2008dz}, for which an in-depth study is needed regarding the exact nature of the corrections for the decay channel analysed in this work.

With the CP-conserving observables mentioned, several physical observables can be derived from them, which we will study in this work within the SM framework.
\begin{itemize}
	\item Integrating the expressions for the differential decay widths mentioned in Eqns.\eqref{eqn:decay width angular} and \eqref{eqn:decay width angular CP} over the angles $\theta_{l}\in[0,\pi]$, $\theta_{D^{*}}\in[0,\pi]$ and $\phi\in[0,2\pi]$ the CP-averaged differential decay width for $B_{c}^{-}\rightarrow D^{*-}(\rightarrow D^{0}\pi^{-}) \ell^{+}\ell^{-}$ can be expressed as
	\begin{equation}
	\begin{split}
	\frac{d\Gamma_{CPavg}}{dq^{2}}=\frac{1}{2}\left(\frac{d\Gamma}{dq^{2}}+\frac{d\bar{\Gamma}}{dq^{2}}\right)
								  =\frac{1}{4}\left(3I_{1c}+6I_{1s}-I_{2c}-2I_{2s}\right).
	\end{split}
	\label{eqn:CP averaged branching ratio}
	\end{equation}
	\item The CP-averaged lepton forward-backward asymmetry can be expressed as
	\begin{equation}
	A_{FB}=\frac{3}{4}S_{6s}.
	\label{eqn:forward backward asymmetry}
	\end{equation}
	\item The longitudinal and transverse polarization fractions of the $D^{*}$ meson can be expressed as
	\begin{equation}
	\begin{split}
	F_{L}=&\frac{1}{4}(3S_{1c}-S_{2c}),\\
	F_{T}=&\frac{1}{2}(3S_{1s}-S_{2s}),
	\end{split}
	\label{eqn:lepton polarization}
	\end{equation}
	respectively.
	\item In addition, there are a number of clean angular observables $P_{1,2,3}$ and $P^{'}_{4,5,6,8}$ associated with the CP-conserved angular coefficients expressed as \cite{Matias:2012xw,Descotes-Genon:2013vna}
	\begin{eqnarray}
		\begin{split}
	P_{1}&=\frac{S_{3}}{2S_{2s}},\qquad &P_{4}^{'}&=\frac{S_{4}}{\sqrt{S_{1c}S_{2s}}},\\
	P_{2}&=\frac{\beta_{l}S_{6s}}{8S_{2s}},\qquad &P_{5}^{'}&=\frac{\beta_{l}S_{5}}{2\sqrt{S_{1c}S_{2s}}},\\
	P_{3}&=-\frac{S_{9}}{4S_{2s}},\qquad &P_{6}^{'}&=-\frac{\beta_{l}S_{7}}{2\sqrt{S_{1c}S_{2s}}},\\
							&	  \qquad	&P_{8}^{'}&=-\frac{S_{8}}{\sqrt{S_{1c}S_{2s}}}.
	\end{split}
	\label{eqn:Clean observables}
	\end{eqnarray}
\end{itemize}
\subsection{Form Factors}
\label{subsection:form factors}

Having discussed the physical observables, we now proceed to the discussion of the relevant form factors, whose information will play a pivotal role in our entire analysis. Feynman diagrams of the relevant decay channels are already shown in Figs.\ref{fig:charged current feynman diagram BcDStar} and \ref{fig:FCNC feynman diagram BcDStar}. Depending on whether the final state meson is pseudoscalar or vector, the transition matrix elements of each decay channel can be parametrized in terms of certain form factors. But before that, it would be more convenient to first present the decay kinematics along with all the momenta and other relevant terms.  

Considering the frame of reference to be the one in which the initial state meson, i.e., the $B$ or $B_{c}$ meson to be at rest, the momenta of the initial state and final state mesons can be defined in the light cone coordinate system as
\begin{equation}
P_{1}=\frac{M}{\sqrt{2}}(1,1,0_{\perp}),\qquad P_{2}=\frac{M}{\sqrt{2}}(r\eta^{+},r\eta^{-},0_{\perp}),
\label{eqn: kinematics momentum}
\end{equation}
respectively with $r=m/M$ and $\eta^{\pm}=\eta\pm\sqrt{\eta^{2}-1}$, $M$ and $m$ representing the masses of the initial and final state mesons, respectively. The term $\eta$ is expressed as
\begin{equation}
\eta=\frac{1+r^{2}}{2r}-\frac{q^{2}}{2rM^{2}}.
\end{equation}  
Momenta of the spectator quarks in the initial and final state mesons can be expressed as
\begin{equation}
k_{1}=\left(0,x_{1}\frac{M}{\sqrt{2}},k_{1\perp}\right),\qquad k_{2}=\left(x_{2}\frac{M}{\sqrt{2}}r\eta^{+},x_{2}\frac{M}{\sqrt{2}}r\eta^{-},k_{2\perp}\right),
\end{equation}
with $x_{1}$ and $x_{2}$ being the fraction of the total momentum carried by the respective quarks.

With the momenta of the mesons defined, we can now return to the discussion on form factors. For $B\rightarrow D$ and $B_{c}\rightarrow D$ transitions the matrix element constructed by the vector current can be parametrized in terms of the form factors $F_{+}(q^{2})$ and $F_{0}(q^{2})$ as \cite{Hu:2019bdf,Kurimoto:2001zj,PhysRevD.90.094018}
\begin{equation}
\begin{split}
\langle D(P_{2})|\bar{q}(0)\gamma_{\mu}b(0)|B_{(c)}(P_{1})\rangle=&\left[(P_{1}+P_{2})_{\mu}-\frac{M^{2}-m^{2}}{q^{2}}q_{\mu}\right]F_{+}(q^{2})\\&+\left[\frac{M^{2}-m^{2}}{q^{2}}q_{\mu}\right]F_{0}(q^{2}),
\end{split}
\label{eqn:matrix element BD}
\end{equation}
where $q_{\mu}=P_{1\mu}-P_{2\mu}$ is the momentum carried away by the lepton pair. These form factors are not independent but are connected by some constraints that specifically arise at $q^{2}=0$ in order to cancel the poles that we can see manifesting at maximum recoil. For form factors in Eqn.\eqref{eqn:matrix element BD}, the constraint equation takes the form
\begin{equation}
F_{+}(0)=F_{+}(0).
\end{equation} 
However, for calculations in pQCD, it is much more convenient to express the form factors in terms of two auxiliary form factors $f_{1}(q^{2})$ and $f_{2}(q^{2})$ defined as
\begin{equation}
\left\langle D(P_{2})|\bar{q}(0)\gamma_{\mu}b(0)|B_{(c)}(P_{1})\right\rangle=f_{1}(q^{2})P_{1\mu}+f_{2}(q^{2})P_{2\mu},
\end{equation}
and are related to $F_{+}(q^{2})$ and $F_{0}(q^{2})$ as
\begin{equation}
\begin{split}
F_{+}(q^{2})=&\frac{1}{2}[f_{1}(q^{2})+f_{2}(q^{2})],\\
F_{0}(q^{2})=&\frac{1}{2}f_{1}(q^{2})\left[1+\frac{q^{2}}{M^{2}-m^{2}}\right]+\frac{1}{2}f_{2}(q^{2})\left[1-\frac{q^{2}}{M^{2}-m^{2}}\right].
\end{split}
\end{equation}

In addition to $F_{+}(q^{2})$ and $F_{0}(q^{2})$ for $B_{c}\rightarrow D$ transition, the matrix element constructed by the tensor current induces the tensor form factor $F_{T}(q^{2})$ defined as \cite{PhysRevD.90.094018,Colangelo:1995jv}
\begin{equation}
\left\langle D(P_{2})|\bar{q}(0)\sigma_{\mu\nu}b(0)|B_{c}(P_{1})\right\rangle=i[P_{2\mu}q_{\nu}-q_{\mu}P_{2\nu}]\frac{2F_{T}(q^{2})}{M+m}.
\end{equation} 

As for the $B\rightarrow D^{*}$ and $B_{c}\rightarrow D^{*}$ transitions the matrix elements constructed by the vector and axial vector currents can be parametrized in terms of four form factors $A_{0,1,2}(q^{2})$ and $V(q^{2})$ as in \cite{Hu:2019bdf,Kurimoto:2001zj,PhysRevD.90.094018}
\begin{equation}
\langle D^{*}(P_{2})|\bar{q}(0)\gamma_{\mu}b(0)|B_{(c)}(P_{1})\rangle=\epsilon_{\mu\nu\alpha\beta}\epsilon^{\nu*}P_{1}^{\alpha}P_{2}^{\beta}\frac{2\cdot V(q^{2})}{M+m},
\end{equation}
and
\begin{equation}
\begin{split}
\langle D^{*}(P_{2})|\bar{q}(0)\gamma_{\mu}\gamma_{5}b(0)|B_{(c)}(P_{1})\rangle=&i\left[\epsilon^{*}_{\mu}-\frac{\epsilon^{*}\cdot q}{q^{2}}q_{\mu}\right](M+m)A_{1}(q^{2})\\
&-i\left[(P_{1}+P_{2})_{\mu}-\frac{M^{2}-m^{2}}{q^{2}}q_{\mu}\right](\epsilon^{*}\cdot q)\frac{A_{2}(q^{2})}{M+m}\\
&+i\left[\frac{2m(\epsilon^{*}\cdot q)}{q^{2}}q_{\mu}\right]A_{0}(q^{2}),
\end{split}
\end{equation}
with $\epsilon^{\mu}$ representing the polarization vector of the $D^{*}$ meson. In addition to the above for $B_{c}\rightarrow D^{*}$ transition, the tensor current induces three additional tensor form factors $T_{1,2,3}(q^{2})$ defined by \cite{PhysRevD.90.094018,Colangelo:1995jv}
\begin{equation}
\begin{split}
\langle D^{*}(P_{2})|\bar{q}(0)\sigma_{\mu\nu}q^{\nu}(1+\gamma_{5})&b(0)|B_{c}(P_{1})\rangle=i\epsilon_{\mu\nu\alpha\beta}\epsilon^{*\nu}P_{1}^{\alpha}P_{2}^{\beta}2T_{1}(q^{2})\\
&+\left[\epsilon_{\mu}^{*}(M^{2}-m^{2})-(\epsilon^{*}\cdot q)(P_{1}+P_{2})_{\mu}\right]T_{2}(q^{2})\\
&+(\epsilon^{*}\cdot q)\left[q_{\mu}-\frac{q^{2}}{M^{2}-m^{2}}(P_{1}+P_{2})_{\mu}\right]T_{3}(q^{2}).
\end{split}
\end{equation} 
Similar to the form factors $F_{+}(q^{2})$ and $F_{0}(q^{2})$, the form factors $A_{0,1,2}(q^{2})$ and $T_{1,2}(q^{2})$ satisfy constraint relations at $q^{2}=0$ and are of the form as
\begin{equation}
2rA_{0}(0)=(1+r)A_{1}(0)-(1-r)A_{2}(0),\qquad T_{1}(0)=T_{2}(0),
\label{eqn:QCD constraints}
\end{equation} 
respectively. Now, of all the theoretical approaches available in existing literature to calculate the form factors, we do so in the perturbative QCD framework, in which the effective Hamiltonian, or the decay amplitudes, or rather the form factors, can be expressed as the convolution of the initial and the final state meson distribution amplitudes and a hard amplitude. The distribution amplitudes absorb the non-perturbative dynamics of the process and are effectively process-independent, while the hard amplitude encodes contributions from all the hard sub-processes occurring, such as the exchange of hard gluons between the decaying quark and the spectator quark, and is perturbatively calculable and process-dependent.

As has been explained in \cite{Li:1994iu} for $B\rightarrow \pi$ form factor and in \cite{Kurimoto:2002sb} for $B\rightarrow D^{(*)}$ form factors, radiative corrections to the meson wave functions and hard amplitudes, with the processes having kinematics that have already been shown in Eqn \eqref{eqn: kinematics momentum}, generate double logarithms from the overlap of collinear and soft radiative corrections of the form $\alpha_{s}\ln^{2}(x)$ and $\alpha_{s}\ln^{2}(k_{T})$, $k_{T}$ representing the transverse momenta of the quarks. These terms generate divergences at the endpoint regions of $x$ and $k_{T}$. To remove the divergences, the double logarithm terms are resummed into what are termed the Sudakov factors. The first kind of double logarithms are organised into a jet function $S_{t}(x)$ as a consequence of threshold resummation \cite{Li:2001ay}. The jet function is expressed as
\begin{equation}
\label{eqn:jet function}
S_{t}(x)=\frac{2^{1+2c}\Gamma(\frac{3}{2}+c)}{\sqrt{\pi}\Gamma(1+c)}[x(1-x)]^{c},
\end{equation}
with c=0.3. The second kind of double logarithms are also organised into all orders into a Sudakov factor as a consequence of $k_{T}$ resummation \cite{Nagashima:2002ia}. The Sudakov factor thus obtained fixes the infrared divergences in $k_{T}$ space. The factorization formulae thus obtained in $k_{T}$ space, upon Fourier transform, are translated to the impact parameter space. Details of the derivation for form factors has been shown in \cite{Li:1994iu}. Following the same procedure, authors of \cite{Rui:2016opu} have derived the form factors for $B_{c}\rightarrow P,V$ channels, which we have adopted in this work for the analysis of semileptonic and rare decays of $B_{c}$ meson. However, we have introduced a modification to the Sudakov factor arising from $k_{T}$ resummation. \\
\indent Analysis of $B_{c}$ meson, being a heavy-heavy system, involves multiple scales and may be studied in the formalism for heavy quarkonium decays. Resummation of such systems is much more complicated than for B meson decays. However, taking the limit $m_{b}\rightarrow \infty$ with $m_{c}$ remaining finite, the $B_{c}$ meson can be treated as a heavy-light system and analysis of the decays can be carried out in the conventional pQCD approach for B meson decays \cite{Kurimoto:2002sb}. We have modified the Sudakov factors by considering the effects a finite charm quark mass scale will have on $k_{T}$ resummation formalism. Details of the formalism have been shown in \cite{PhysRevD.97.113001}. The Sudakov exponent thus derived, taking the charm quark mass effect in the impact parameter space, has been derived to have the form
\begin{equation}
\begin{split}
s_{c}(Q,b)=&s(Q,b)-s(m_{c},b),\\
=&\int_{m_{c}}^{Q}\frac{d\mu}{\mu}\left[\int_{1/b}^{\mu}\frac{d\bar{\mu}}{\bar{\mu}}A(\alpha_{s}(\bar{\mu}))+B(\alpha_{s}(\bar{\mu}))\right],
\end{split}
\end{equation}
where the expressions for $s(Q,b)$ representing the Sudakov exponent obtained by $k_{T}$ resummation of an energetic light quark, $A(\alpha_{s}(\bar{\mu}))$ and $B(\alpha_{s}(\bar{\mu}))$ has been taken from \cite{Li:1999kna}.\\
\indent However, it is to be noted that in the case of $B$ mesons, this modified pQCD framework is not necessary. This is because, unlike $B_{c}$ mesons, the $B$ meson is formed by the $b$ and $u$ quarks, the former being a heavy quark and the latter a light quark. Thus, the analysis of $B$ mesons can be carried out using the conventional pQCD approach, and the introduction of a finite charm quark mass scale is not required in this case.

\subsection{Light Cone Distribution amplitudes}\label{subsection:LCDAs}
 
With the form factors for the relevant decay channels discussed and their analytical expressions presented in the appendix, in this subsection, we shift our focus to expressions of the LCDAs of all the participating mesons. We are going to discuss LCDAs of $B$, $B_{c}$ and $D^{(*)}$ mesons. The reason for considering $B$ mesons, along with the other mesons, even though we are not studying $B$ meson decays in this work, will be discussed in the next section. 
 
 \begin{itemize}
 	\item As for the B meson, the wavefunction has the form as
 	\begin{equation}
 	\Phi_{B}(p,x)=\frac{i}{\sqrt{2N_{c}}}\left(\slashed{p}_{B}+m_{B}\right)\gamma_{5}\phi_{B}(x,b),
 	\end{equation}
 	where $\phi_{B}(x,b)$ represents the B meson LCDA, which in this work is considered to have an approximate Gaussian form as \cite{Kurimoto:2002sb,Xiao:2011tx} 
 	\begin{equation}
 	\phi_{B}(x,b)=\frac{f_{B}}{2\sqrt{2N_{c}}}N_{B}x^{2}(1-x)^{2}\exp\left[-\frac{x^{2}m_{B}^{2}}{2\omega_{B}^{2}}-\frac{1}{2}\omega_{B}^{2}b^{2}\right],
 	\end{equation}
 	with $\omega_{B}$ being the shape parameter defining the shape of the distribution amplitude, and $N_{B}$ the normalization constant fixed by the relation
 	\begin{equation}
 	\int_{0}^{1}\phi_{B}(x,b=0)dx=\frac{f_{B}}{2\sqrt{2N_{c}}},
 	\end{equation}
 	where $f_{B}$ is the decay constant of $B$ meson.
 	\item The form of the $B_{c}$ meson, the wavefunction has the form as
 	\begin{equation}
 	\Phi_{B_{c}}(p,x)=\frac{i}{\sqrt{2N_{c}}}\left(\slashed{p}_{B_{c}}+m_{B_{c}}\right)\gamma_{5}\phi_{B_{c}}(x,b),
 	\end{equation}
 	where $\phi_{B_{c}}(x,b)$ represents the $B_{c}$ meson LCDA, which in this work is also considered to have an approximate Gaussian form as \cite{PhysRevD.97.113001}
 	\begin{equation}
 	\begin{split}
 	\phi_{B_{c}}(x,b)=\frac{f_{B_{c}}}{2\sqrt{2N_{c}}}N_{B_{c}} x (1-x) exp\left[ -\frac{(1-x)m_{c}^{2}+xm_{b}^{2}}{8\omega_{B_{c}}^{2}x(1-x)}\right]\exp[-2\hspace{0.05cm}\omega_{B_{c}}^{2}x(1-x)b^{2}],
 	\end{split}
 	\end{equation}
 	The normalization constant $N_{B_{c}}$ is fixed by the relation
 	\begin{equation}
 	\int_{0}^{1}\phi_{B_{c}}(x,b=0)dx=\frac{f_{B_{c}}}{2\sqrt{2N_{c}}},
 	\end{equation}
 	and the parameter $b$ being the impact parameter, which is in fact the Fourier conjugate to the transverse momentum $k_{T}$, $\omega_{B_{c}}$ being the shape parameter of the $B_{c}$ meson distribution amplitude and $f_{B_{c}}$ the decay constant of $B_{c}$ meson.
 	\item As for the $D$ and $D^{*}$ mesons, the wavefunction has the same form as in 
 	\begin{equation}
 		\footnotesize
 	\begin{split}
 	\Phi_{D}(p,x)=&\frac{i}{\sqrt{2N_{c}}}\gamma_{5}\left(\slashed{p}_{D}+m_{D}\right)\phi_{D}(x,b),\\
 	\Phi_{D^{*}}(p,x)=&-\frac{i}{\sqrt{2N_{c}}}\left[\slashed{\epsilon}_{L}\left(\slashed{p}_{D^{*}}+m_{D^{*}}\right)\phi_{D^{*}}^{L}(x,b)+\slashed{\epsilon}_{T}\left(\slashed{p}_{D^{*}}+m_{D^{*}}\right)\phi_{D^{*}}^{T}(x,b)\right],
 	\end{split}
 	\end{equation}
 	where $\phi_{D}(x,b)$, $\phi_{D^{*}}^{L}(x,b)$ and $\phi_{D^{*}}^{T}(x,b)$ the LCDAs of the respective mesons, which in this work we consider to have a simple polynomial form as \cite{Li:2008ts}
 	\begin{equation}
 	\label{eqn:wf1}
 	\phi_{D^{(*)}}(x,b)=\frac{f_{D^{(*)}}}{2\sqrt{2N_{c}}}6x(1-x)\left[1+C_{D^{(*)}}(1-2x)\right]\cdot \exp\left[-\frac{\omega_{D^{(*)}}^{2}b^{2}}{2}\right],
 	\end{equation}
 	with $C_{D^{(*)}}$ and $\omega_{D^{(*)}}$ are parameters representing the shape of the corresponding distribution amplitude, and $\phi_{D^{(*)}}$ satisfying the normalization condition
 	\begin{equation}
 	\int_{0}^{1}\phi_{D^{(*)}}(x,0)dx=\frac{f_{D^{(*)}}}{2\sqrt{2N_{c}}},
 	\end{equation}
 	with $f_{D^{(*)}}$ being the decay constant of the respective meson.
 \end{itemize}

Now, in all existing works available in literature that explores the decays of $B$ and $B_{c}$ mesons in the pQCD framework \cite{PhysRevD.90.094018,PhysRevD.97.113001,Hu:2019bdf}, the authors, in their analysis, have considered the shape parameters $\omega_{B}$, $\omega_{B_{c}}$, $C_{D^{(*)}}$ and $\omega_{D^{(*)}}$ as inputs having a fixed numerical value. The lattice QCD information was incorporated only at the kinematic endpoint $q^{2} = q^{2}_{\text{max}}$. In contrast, our work provides a fully data-driven extraction of LCDA shape parameters $\omega_{B_c}$, $C_{D^{(*)}}$, and $\omega_{D^{(*)}}$. Unlike previous studies that fixed these parameters to model-dependent values, we treat them as free parameters and determine them through a global $\chi^2$ minimization. Performing such a joint fit using inputs from both $B_c \to D$ and $B \to D^{(*)}$ form factors have enabled us to achieve a unified determination of LCDA shape parameters, which ultimately allows us to extract the full correlation matrix among parameters, such as between $\omega_{B_c}$ and $C_{D^*}$, which would not be possible if the channels were analyzed separately. These correlations are essential for providing reliable uncertainty estimates for the predicted $B_c \to D^*$ pQCD form factors.

	\begin{table}[htb!]
		\renewcommand{\arraystretch}{1.3}
		\centering
		\begin{tabular}{|c|lll|}
			\hline 
			\textbf{Mass (GeV)} & $m_{B_{c}}=6.277$ & $m_{B}=5.28$ &  \\ 
			
			& $m_{D}=1.865$ & $m_{D^{*}}=2.007$ &  \\ 
			
			& $m_{b}=4.18$ & $m_{c}=1.275$&\\
			& $m_{e}=0.511\times 10^{-3}$&$m_{\nu}=0.105$&$m_{\tau}=1.776$\\
			\hline 
			\textbf{Decay}&$f_{B_{c}}=0.427(6)$\cite{McNeile:2012qf}&$f_{B}=0.193(6)$\cite{Lubicz:2017asp}&\\
			\textbf{Constants (GeV)}&$f_{D}=0.2074(38)$\cite{Lubicz:2017asp}&$f_{D^{*}}$=\textcolor{blue}{0.2235(87)}\cite{Lubicz:2017asp}&\\
			\hline
			\textbf{CKM}&$V_{cb}=0.0411(14)$&&\\
			\textbf{matrix elements}&$V_{ud}=0.97367(32)$&$V_{ub}=3.82(20)\times 10^{-3}$&\\
			&$V_{td}=8.6(2)\times 10^{-3}$&$V_{tb}=1.010(27)$&\\
			\hline
			\textbf{Lifetime (psec)}&$\tau_{B_{c}}=0.510(9)$\cite{ParticleDataGroup:2022pth}&&\\
			\hline
		\end{tabular}
	\caption{Values of input parameters used in this work.}
	\label{table:input parameters}
	\end{table}
	
 \section{Extraction of LCDA shape parameters and form factors at $q^{2}=0$}
\label{section:Extraction of LCDA parameters}

With all the relevant theoretical prerequisites and the various input parameters discussed, we now move on to the extraction of the shape parameters of LCDAs of the participating mesons that were introduced in subsection \ref{subsection:LCDAs}. We do this by constructing a chi-square function, maintaining the shape parameters as free parameters, and then optimising it to extract the required parameters. The chi-square function to be constructed has the form
\begin{equation}
\label{eqn:chi square function}
\chi^{2}=\sum_{i,j}(\mathcal{O}_{i}^{th}-\mathcal{O}_{i}^{data})V_{ij}^{-1}(\mathcal{O}_{j}^{th}-\mathcal{O}_{j}^{data})^{T}+\chi^{2}_{nuis},
\end{equation}

with $\mathcal{O}_{i}^{th}$ representing the pQCD expressions for form factors at $q^{2}=0$, $\mathcal{O}_{i}^{data}$ representing the information on the corresponding form factors at $q^{2}=0$, $\chi^{2}_{nuis}$ representing the chi-square function for the relevant nuisance parameters and $V_{ij}$ representing the covariance matrix between the inputs. In this step, our primary aim is to constrain $\omega_{B_{c}}$, the shape parameter of $B_{c}$ meson LCDA, and $C_{D^{(*)}}$, the shape parameters of $D^{(*)}$ LCDAs. For this we are going to consider the form factors of $B_{c}\rightarrow D$ semileptonic channel at $q^{2}=0$, calculated using the pQCD approach and discussed in subsection \ref{subsection:form factors} and in Appendix \ref{section:appendix PQCD form factor expressions}. The data inputs are taken from HPQCD \cite{Cooper:2021bkt}, who has recently supplied information of the relevant form factors using Bourrely-Caprini-Lellouch (BCL) \cite{Bourrely:2008za} parametrization. Using this information, we extract our values of the form factors at $q^{2}=0$. To better constrain $C_{D^{(*)}}$ and $\omega_{D^{(*)}}$, we take form factors of $B\rightarrow D^{(*)}$ semileptonic channels at $q^{2}=0$ as additional inputs, pQCD expressions for which have been presented in Appendix  \ref{section:appendix PQCD form factor expressions}. As for the inputs, HPQCD \cite{Na:2015kha} and MILC \cite{FermilabLattice:2021cdg} collaborations have already presented their fit results for the z-expansion coefficients for $B\rightarrow D$ and $B\rightarrow D^{*}$ form factors, respectively. The $B\rightarrow D$ and $B\rightarrow D^{*}$ form factors have been defined using the BCL and Boyd-Grinstein-Lebed (BGL) \cite{Boyd:1995cf,Boyd:1995sq,Boyd:1997kz} parametrizations for the z-expansion, respectively. We use them to obtain synthetic data points at $q^{2}$. In addition to the lattice inputs, we have also used inputs of the respective form factors presented in the LCSR approach \cite{Bordone:2019guc}. The numerical values of the form factor data for the first chi-square minimization are shown in Table \ref{table:lattice inputs for BDStar form factors}.

\begin{table}[htb!]
	\centering
	\begin{tabular}{|c|c|c|c|c|}
		\hline
		\textbf{Decay}&\textbf{Form}&\multicolumn{3}{c|}{\textbf{Values at }$\boldsymbol{q^{2}=0}$}\\
		\cline{3-5}
		\textbf{Channel}&\textbf{Factors}&\multicolumn{2}{c|}{\textbf{Lattice}}&\textbf{LCSR}\\
		\cline{3-4}
		&&\textbf{Group}&\textbf{Value}&\\
		\hline
		$B\rightarrow D l\nu_{l}$&$F_{+}(0)$&HPQCD&0.663(34)&0.649(79)\\
		\hline
		&$A_{0}(0)$&&0.539(44)&0.678(115)\\
		$B\rightarrow D^{*} l\nu_{l}$&$A_{1}(0)$&MILC&0.597(34)&0.599(92)\\
		&$V(0)$&&0.656(83)&0.691(129)\\
		\hline
		$B_{c}\rightarrow D$&$F_{+}(0)$&HPQCD&0.168(29)&-\\
		\hline
	\end{tabular}
\caption{Form factor data for $B\rightarrow D^{(*)}$ and $B_{c}\rightarrow D$ semileptonic channels at $q^{2}=0$.}
\label{table:lattice inputs for BDStar form factors}
\end{table}

Additionally, we take the charm and bottom quark masses, $m_{c}$ and $m_{b}$, as the arithmetic averages of their values in the pole, $ \overline{\text{MS}}$, and kinetic schemes. To ensure a scheme-independent and inclusive treatment of mass uncertainties, we assign relative errors of 25\% for $m_{c}$ and 10\% for $m_{b}$, chosen to encompass the full range of variation across these schemes. The quark masses are presented in Table \ref{table:scheme dependent masses}.

\begin{table}[htb!]
	\centering
	\begin{tabular}{|c|cc|}
		\hline
		\textbf{Scheme}&$\boldsymbol{m_{b}}$ (GeV)&$\boldsymbol{m_{c}}$ (GeV)\\
		\hline
		Pole mass&4.78&1.67\\
		$\overline{MS}$&4.18&1.273\\
		Kinetic&4.56&1.091\\
		\hline
		Average&4.506(451)&1.345(336)\\
		\hline
	\end{tabular}
	\caption{Values of $m_{b}$ and $m_{c}$ in three different schemes and their average value \cite{ParticleDataGroup:2022pth}.}
	\label{table:scheme dependent masses}
\end{table}

With these inputs established, we proceed to construct the chi-square function and perform its minimization to extract the shape parameters of the $B$, $B_{c}$, and $D^{(*)}$ mesons. It is important to clarify that our current analysis already incorporates radiative corrections up to $\mathcal{O}(\alpha_s)$ and $\mathcal{O}(\alpha_s^2)$, following the calculations in Ref.~\cite{Keum:2000wi} for $B$ mesons and in \cite{Liu:2020upy} for $B_c$ meson wave functions, which are directly relevant for the $B \to P(V)$ and $B_c \to P(V)$ form factors within the PQCD framework.
Furthermore, it is important to emphasize that the PQCD form factors employed in this analysis incorporate only the leading-order (LO) contributions in the hard kernel, and the meson wave functions are defined within the leading-twist approximation.

The LCDAs describe the momentum fraction distribution of partons inside a hadron projected onto the light-cone. They are essential in QCD factorization for exclusive processes. We are using a QCD-inspired DA refers to a light-cone wave function or amplitude that is derived or constrained using principles of Quantum Chromodynamics (QCD), rather than being purely phenomenological like a Gaussian ansatz. The Gaussian ansatz is purely phenomenological and often suppresses endpoint contributions too strongly. The QCD-inspired DAs differ from a simple Gaussian ansatz in several ways. For example, they incorporate constraints from QCD sum rules, lattice QCD, and perturbative QCD asymptotics and exhibit correct endpoint behavior, avoiding excessive suppression near $x \to 0$ or $x \to 1$.

We have mentioned that in our analysis, we define the full light-cone wave function as a product of the longitudinal distribution amplitude and a transverse profile:
\begin{equation}
	\Psi(x,\mathbf{k}_\perp,\mu)
	= \phi(x,\mu)\,\psi(\mathbf{k}_\perp),
\end{equation}
A commonly used (scale-independent) Gaussian ansatz for the transverse profile is
\begin{equation}
	\psi(\mathbf{k}_\perp)
	\propto \exp\!\left[-\frac{\mathbf{k}_\perp^2}{2\,\beta^2}\right],
\end{equation}
with \(\beta\) setting the typical transverse-momentum width. QCD-inspired DAs keep the simplicity of a Gaussian-like transverse profile but improve the longitudinal part and include QCD-based constraints.

In the light-cone formalism, the twist of an operator is defined as its dimension minus spin. The leading-twist (twist-2) contribution corresponds to the simplest quark--antiquark configuration and dominates at large momentum transfer. In our analysis, we restrict ourselves to the leading-twist contribution in all light-cone distribution amplitudes (LCDA). Higher-twist effects (twist-3, twist-4, etc.) originate from several sources, such as:
\begin{itemize}
	\item Quark transverse momentum effects,
	\item Quark--gluon interactions,
	\item Multi-parton correlations.
\end{itemize}
In general, these contributions are suppressed by powers of $\Lambda_{\text{QCD}}/m_Q$. Nevertheless, the higher-twist effects can become relevant, particularly at low and intermediate $q^2$. Twist-3 distribution amplitudes typically involve pseudoscalar and tensor components, while twist-4 contributions include quark-gluon operators. Incorporating these terms introduces additional nonperturbative parameters and can improve the precision of form factor predictions.

 However, in the present work, we lack sufficient input to reliably constrain all higher-twist contributions. To account for the potential impact of these missing pieces, we have included an additional uncertainty in our estimates of the LCDA shape parameters and the resulting form factors.	Previous studies of $B \to \pi$ and $B \to \rho$ PQCD form factors (see Refs. \cite{Wang:2012ab,Cheng:2014fwa,Mahajan:2004dx}) indicate that next-to-leading-order (NLO) corrections to the hard kernel, as well as higher-twist effects, can introduce an additional error in the form factors of approximately 20–30\% relative to the LO predictions. To account for these theoretical uncertainties arising from neglected higher-order loop corrections and higher-twist LCDA contributions, we conservatively assign a 30\% uncertainty to the LO form factor predictions. This is implemented by introducing a multiplicative nuisance parameter, $\delta_{f_{i}}$, constrained within this range. By treating this correction as a nuisance parameter during the minimization procedure, we ensure that the associated uncertainties are rigorously propagated to the final error estimates of the extracted shape parameters\footnote{When constructing the $\chi^{2}$ function, we have not taken the decay constants of the participating mesons into $\chi^{2}_{nuis}$, since their uncertainties are small, typically of the order of 2-3\%, as can be seen in Table \ref{table:input parameters}. Thus their variation has negligible impact on the fit results, compared to the larger theoretical uncertainties associated with the form factors.}. The extracted parameters and the corresponding correlation matrix are shown in Tables \ref{table:extracted value of Bc and D shape parameters} and \ref{table: correlation LCDA BcD} respectively\footnote{This approach ensures that our predictions remain robust while acknowledging the limitations imposed by neglecting higher-twist effects. Nevertheless, we acknowledge that higher-order corrections to the hard kernel and higher-twist contributions to the wave functions remain unaccounted for and could impact the extraction of shape parameters. Dedicated calculations of these effects are essential for achieving ultimate precision.}.
\begin{table}[t!]
	\centering
	\begin{tabular}{|cl|cl|}
		\hline
		\multicolumn{2}{|c|}{\textbf{Free Parameters}}  &  \multicolumn{2}{c|}{\textbf{Nuisance Parameters}} \\
		\hline
		\textbf{Parameters}& \textbf{Fit Results} & \textbf{Parameters}& \textbf{Fit Results} \\
		\hline
		$\omega_{B_{c}}$ &1.064(108)~GeV & $m_{b}$ & 4.500(68)~GeV \\
		$\omega_{B}$ &0.499(140)~GeV & $m_{c}$ & 1.337(196)~GeV \\
		$C_{D}$ & 0.504(52) & $\delta_{f_{1}}^{B\to D}$ & -0.070(197) \\
		$C_{D^{*}}$ & 0.492(97) & $\delta_{f_{2}}^{B\to D}$ & 0.047(257) \\
		$\omega_{D}$& 0.105(43) & $\delta_{A_{0}}^{B\to D^{*}}$ & -0.055(304) \\
		$\omega_{D^{*}}$&0.099(16)&$\delta_{A_{1}}^{B\to D^{*}}$ & -0.010(137)\\
		&&$\delta_{V}^{B\to D^{*}}$ & 0.007(57)\\
		&&$\delta_{f_{1}}^{B_{c}\to D}$ & -0.131(58)\\
		&&$\delta_{f_{2}}^{B_{c}\to D}$ & -0.086(176)\\
		\hline
		\textbf{DOF}&\multicolumn{3}{|c|}{3}\\
		\hline
		\textbf{p-Value}&\multicolumn{3}{|c|}{18.97\%}\\
		\hline
	\end{tabular}
	\caption{Extracted values of LCDA parameters obtained by fitting pQCD form factors of $B_{c}\rightarrow D$ and $B\rightarrow D^{(*)}$ transitions with corresponding lattice and LCSR inputs at $q^{2}=0$.}
	\label{table:extracted value of Bc and D shape parameters}
\end{table}

From Table \ref{table:extracted value of Bc and D shape parameters} we observe that the shape parameter of $B_{c}$ meson LCDA, $\omega_{B_{c}}$ is well constrained and it agrees well with our previous estimate \cite{Dey:2025xdx}, thereby justifying its value through two independent sets of inputs. The shape parameters of $B$ and $D^{(*)}$ meson LCDAs are also well constrained and agree well with existing literature. The nuisance parameters involve mass of bottom and charm quarks. 

With these extracted parameters, we present our predictions for the $B_c \to D^{(*)}$ form factors at $q^2=0$, along with a comparison to existing literature predictions; these results are showcased in Table \ref{table:form factor predictions BcDStar}. In these predictions, we have explicitly propagated a $30\%$ correction factor to account for uncertainties arising from loop-level corrections and contributions from next-to-leading twist LCDAs. Furthermore, the errors associated with the decay constants have also been propagated independently to the final results.
\begin{table}[htb!]
	\renewcommand{\arraystretch}{1.2}
		\centering
		\begin{tabular}{|c|cccc|}
		\hline
		\textbf{Form Factors}& \textbf{This work}&\textbf{Previous PQCD}\cite{PhysRevD.90.094018}&\textbf{CLFQM}\cite{Wang:2008xt}&  \\
		\hline
		$F_{+}^{B_{c}\rightarrow D}(0)$& 0.179(32) & 0.19(3) &$0.16^{+0.02+0.02}_{-0.02-0.01}$&\\
		$F_{T}^{B_{c}\rightarrow D}(0)$& 0.160(61) & 0.20(3) &- &\\
		\hline
		$A_{0}^{B_{c}\rightarrow D^{*}}(0)$& 0.201(70) & 0.17(4) &$0.09^{+0.01+0.01}_{-0.01-0.01}$ &\\
		$A_{1}^{B_{c}\rightarrow D^{*}}(0)$& 0.188(66) & 0.18(4) &$0.08^{+0.01+0.01}_{-0.01-0.01}$ &\\
		$A_{2}^{B_{c}\rightarrow D^{*}}(0)$& 0.171(60) & 0.20(4) & $0.07^{+0.01+0.01}_{-0.01-0.01}$&\\
		$V^{B_{c}\rightarrow D^{*}}(0)$& 0.185(66) & 0.25(5)& $0.13^{+0.01+0.02}_{-0.02-0.02}$ &\\
		$T_{1}^{B_{c}\rightarrow D^{*}}(0)$& 0.216(77) & 0.22(4) &- &\\
		$T_{3}^{B_{c}\rightarrow D^{*}}(0)$& 0.149(56) & 0.20(4) &- &\\
		\hline
	\end{tabular}
	\caption{Prediction of form factors of $B_{c}\rightarrow D^{(*)}$ transition at $q^{2}=0$ along-with comparison with other predictions.}
		\label{table:form factor predictions BcDStar}
\end{table}

Revisiting the form factor expressions in Appendix \ref{section:appendix PQCD form factor expressions}, we can see that the integrations over $b_{1}$ and $b_{2}$ have been done upto a cut-off $b_{cut}$. In this work we have set it at around 90\% of $1/\Lambda_{QCD}$. This has been done to keep our calculations well within the perturbative QCD region and to avoid it from taking up any non-perturbative contributions. Regarding the error estimates of each form factor in Table \ref{table:form factor predictions BcDStar}, we can observe that our prediction of $F_{+}$ in the first row is comparable to the lattice QCD and previous pQCD predictions. The slight 1-2\% larger error is due to the decay constants, which are propagated independently into the form factors. As for the rest of the rows, they have a much larger error estimate than previous pQCD predictions. This is mainly due to the 30\% uncertainty that we have propagated as systematic error. In addition, we have the parameters $\omega_{B_{c}}$, $C_{D}^{(*)}$, $\omega_{D^{(*)}}$, $m_{b}$ and $m_{c}$ that also contribute to the total error. On the contrary, the predictions in ref. \cite{PhysRevD.90.094018} have used model-dependent values of all these parameters, and no errors are considered in $C_{D}^{(*)}$ and $\omega_{D^{(*)}}$, only a 10\% error has been introduced in $\omega_{B_{c}}$.

We also calculate correlation between each of the predicted form factors, and present them in Table \ref{table:correlation form factors BcDStar} in Appendix \ref{section:appendix correlation matrices}. The high correlation between the $B_{c}\rightarrow D^{*}$ form factors obtained in the table is particularly due to their shared sensitivity to a majority of the input parameters. Similarly, despite minor diferences between the LCDA input parameters between the $B_{c}\rightarrow D$ and $B_{c}\rightarrow D^{*}$ form factors, the error propagation is largely controlled by the shared inputs leading to a high correlation between them.

\section{Obtaining form factor information over full physical $q^{2}$ range}
\label{section:Extrapolation to full physical range}

In the previous section, we calculated the $B_{c}\rightarrow D^{(*)}$ form factors at $q^{2}=0$. However, this is not enough to obtain predictions for the relevant physical observables discussed before. As evident from their expressions, we would require information about the form factors over the complete semileptonic region. The pQCD framework alone cannot do the job, as its calculations are reliable only at low $q^{2}$ values. To overcome this limitation, we use certain symmetry relations that exist between the form factors, to obtain the information of the unknown form factor. Then we adopt a suitable parametrization to obtain their information over the full physical $q^{2}$ range. This section is organised into subsections, each detailing a specific step of the procedure and presenting the corresponding results.

\subsection{Connecting $B_{c}\rightarrow D^{*}$ with $B_{c}\rightarrow D$ form factors}
\label{subsection: connecting the form factors}

The plan in this subsection is to use the information of $B_{c}\rightarrow D$ form factors supplied by HPQCD \cite{Cooper:2021bkt} to obtain the $B_{c}\rightarrow D^{*}$ form factors. We do so by constructing a unified description of both the channels, using the Heavy Quark Effective Theory (HQET), utilizing the trace formalism, and retaining both leading and subleading contributions in the heavy quark expansion. This would allow us to express all 10 form factors, 3 for the $B_{c}\rightarrow D$ channel and 7 for the $B_{c}\rightarrow D^{*}$ channel, in terms of just two universal soft functions.

Following the trace formalism developed in \cite{Jenkins:1992nb,Colangelo:2021dnv}, the weak  matrix element in HQET for transitions with a generic Dirac bilinear $\Gamma$ can be expressed as
\begin{equation}
	\langle D^{(*)}(v,k)|\bar{q}\Gamma Q|B_{c}(v)\rangle=-\sqrt{M m}~\text{Tr}[\bar{H}^{(\bar{c})}\Sigma(v,a_{0}k)\Gamma H^{(c\bar{b})}],
\end{equation} 
where the initial state meson and the final state mesons carry four momenta $P_{1}=Mv$ and $P_{2}=mv+k$, respectively, with $k$ being a small residual momentum. The $B_{c}$ and $B_{c}^{*}$ doublet comprising of two heavy quarks $\bar{b}$ and $c$ is represented by the effective field
\begin{equation}
	H^{c\bar{b}}=\frac{1+\slashed{v}}{2}[B_{c}^{*\mu}\gamma_{\mu}-B_{c}\gamma_{5}]\frac{1-\slashed{v}}{2},
\end{equation}
and the $D$ and $D^{*}$ doublet with a single heavy quark $c$ is represented by the effective fields
\begin{equation}
	H^{c}=[D^{*\mu}\gamma_{\mu}-D\gamma_{5}]\frac{1-\slashed{v}}{2}.
\end{equation}

The term $\Sigma(v,a_{0}k)$ can be generalized in terms of two dimensionless functions, as
\begin{equation}
	\Sigma(v,a_{0}k)=\Sigma_{1}+\slashed{k}a_{0}\Sigma_{2},
\end{equation}
with $\Sigma_{1}$ and $\Sigma_{2}$ being the universal form factors that encapsulate the non-perturbative QCD dynamics of the light degrees of freedom. The function $\Sigma_{1}$ arises at leading order contribution to the HQET expansion of the matrix element and reflects the structure dictated by Heavy Quark Spin Symmetry (HQSS), which becomes exact in the infinite mass limit. The second function $\Sigma_{2}$, however, captures the leading corrections to this symmetry. These corrections are associated with the residual momentum $k$ of $D^{(*)}$ meson, and introduce the $\mathcal{O}(1/m_{Q})$ symmetry-breaking effects. $a_{0}$ has the dimension of length and is typically the Bohr radius of the $B_{c}$ meson. The $a_{0}$ factor suppresses $\Sigma_{2}$ compared to $\Sigma_{1}$, thus making the symmetry-breaking effects small. The weak matrix element for $B_{c}\rightarrow D$ transition is then obtained as
\begin{equation}
	\langle D(v,k)|\bar{q}\gamma_{\mu}Q|B_{c}(v)\rangle=2\sqrt{Mm}[\Sigma_{1}(w)v_{\mu}+a_{0}\Sigma_{2}(w)k_{\mu}],
	\label{eqn:matrix element HQET}
\end{equation}
where the recoil parameter $w$ is related is $q^{2}$ as
	\begin{equation}
	w=\frac{M^{2}+m^{2}-q^{2}}{2Mm}.
\end{equation}
Following the same method, the $B_{c}\rightarrow D$ matrix element induced by the tensor current can be expressed as
	\begin{equation}
	\langle D(v)|\bar{q}\sigma_{\mu\nu}Q|B_{c}(v)\rangle=-2i\sqrt{Mm}a_{0}\Sigma_{2}(w)(v_{\mu}k_{\nu}-v_{\nu}k_{\mu}),
	\label{eqn:matrix element HQET FT}
\end{equation}
and the $B_{c}\rightarrow D^{*}$ matrix elements induced by vector, axial-vector and tensor currents can be expressed as
	\begin{equation}
	\begin{split}
		\langle D^{*}(v,k,\epsilon)|\bar{q}\gamma_{\mu}Q|B_{c}(v)\rangle=2\sqrt{Mm}~\epsilon_{\mu\nu\alpha\beta}\epsilon^{*\nu}v^{\alpha}k^{\beta}a_{0}\Sigma_{2}(w),
	\end{split}
	\label{eqn:matrix element HQET 1}
\end{equation}
\begin{equation}
	\begin{split}
	\langle D^{*}(v,\epsilon)|\bar{q}\gamma_{\mu}\gamma_{5}Q|B_{c}(v)\rangle=2i\sqrt{Mm}&\bigg[\epsilon^{*}_{\mu}\left(\Sigma_{1}(w)+v\cdot k a_{0}\Sigma_{2}(w)\right)\\&-\left(v_{\mu}-\frac{k_{\mu}}{m}\right)\epsilon^{*}\cdot k a_{0}\Sigma_{2}(w)\bigg],
	\end{split}
		\label{eqn:matrix element HQET 2}
\end{equation}
\begin{equation}
	\normalsize
	\begin{split}
		\langle D^{*}(v,\epsilon)|\bar{q}\sigma_{\mu\nu}q^{\nu}Q|B_{c}(v)\rangle=2i\sqrt{Mm}~\epsilon_{\mu\nu\alpha\beta}\epsilon^{*\alpha}\left[v^{\beta}\Sigma_{1}(w)+k^{\beta}a_{0}\Sigma_{2}(w)\right]\left[(M-m)v^{\nu}-k^{\nu}\right],
	\end{split}
	\label{eqn:matrix element HQET 3}
\end{equation}
and
\begin{equation}
	\begin{split}
	\langle D^{*}(v,\epsilon)|\bar{q}\sigma_{\mu\nu}q^{\nu}\gamma_{5}Q|B_{c}(v)\rangle=&2\sqrt{Mm}\bigg[ \epsilon^{*}_{\mu}(v_{\nu}\Sigma_{1}(w)+k_{\nu}a_{0}\Sigma_{2}(w))\\&-\epsilon^{*}_{\nu}(v_{\mu}\Sigma_{1}(w)+k_{\mu}a_{0}\Sigma_{2}(w))\bigg]\left[(M-m)v^{\nu}-k^{\nu}\right],
	\end{split}
		\label{eqn:matrix element HQET 4}
\end{equation}
respectively. Matching these HQET matrix elements with the ones previously discussed in subsection \ref{subsection:form factors}, we can express all the full QCD form factors in terms of $\Sigma_{1}(w)$ and $\Sigma_{2}(w)$. For $B_{c}\rightarrow D$ form factors, we get the relations as
	\begin{equation}
		\begin{split}
			F_{+}(w)&=\sqrt{\frac{m}{M}}\left(\Sigma_{1}(w)+(M-m)a_{0}\Sigma_{2}(w)\right),\\
			F_{0}(w)&=\frac{2\sqrt{Mm}}{M^{2}-m^{2}}\left[(M-mw)\Sigma_{1}(w)+m(M+m)(w-1)a_{0}\Sigma_{2}(w)\right],\\
			F_{T}(w)&=\sqrt{\frac{m}{M}}(M+m)a_{0}\Sigma_{2}(w),
		\end{split}
		\label{eqn:Full QCD to HQET Bc to D form factors}
	\end{equation}
	and for the $B_{c}\rightarrow D^{*}$ form factors, we get the relations as
	\begin{equation}
		\begin{split}
		V(w)&=\sqrt{\frac{m}{M}}(M+m)a_{0}\Sigma_{2}(w),\\
		A_{0}(w)&=\frac{2M\Sigma_{1}(w)+((M-2m)(2mw-M)+(2m^{2}+M^{2}-2Mmw))a_{0}\Sigma_{2}(w)}{2\sqrt{Mm}},\\
		A_{1}(w)&=\frac{2\sqrt{Mm}}{M+m}\left(\Sigma_{1}(w)+m(w-1)a_{0}\Sigma_{2}(w)\right),\\
		A_{2}(w)&=\frac{\sqrt{Mm}(M-2m)(M+m)a_{0}\Sigma_{2}(w)}{M^{2}},
		\end{split}
		\label{eqn:Full QCD to HQET Bc to D* form factors 1}
	\end{equation}
	for the vector and axial-vector form factors, and
	\begin{equation}
		\begin{split}
			T_{1}(w)&=\sqrt{\frac{m}{M}}\left[\Sigma_{1}(w)+(M-m)a_{0}\Sigma_{2}(w)\right],\\
			T_{2}(w)&=\frac{2\sqrt{Mm}}{M^{2}-m^{2}}\left[\Sigma_{1}(w)\left(M-mw\right)+\left(m(M+m)(w-1)\right)a_{0}\Sigma_{2}(w)\right],\\
			T_{3}(w)&=\sqrt{\frac{m}{M}}\left[-\Sigma_{1}(w)+(M+m)a_{0}\Sigma_{2}(w)\right],
		\end{split}
		\label{eqn:Full QCD to HQET Bc to D* form factors 2}
	\end{equation}
	for the tensor form factors. In all the calculations, since we have not neglected the symmetry-breaking corrections, we have not neglected $v\cdot k$, and also have considered the contributions coming from $k_{\mu}/m$.

		\begin{table}[t]
	\renewcommand{\arraystretch}{1.5}
	\footnotesize
	\centering
	\begin{tabular}{|c|c|cccccc|}
		\hline
		\textbf{Soft}&\textbf{Our}&&&\textbf{Correlation}&&&\\
		\cline{3-8}
		\textbf{functions}&\textbf{estimates}&$\Sigma_{1}(1.00)$&$\Sigma_{1}(1.15)$&$\Sigma_{1}(1.3)$&$a_{0}\Sigma_{2}(1.00)$&$a_{0}\Sigma_{2}(1.15)$&$a_{0}\Sigma_{2}(1.3)$\\
		\hline
		$\Sigma_{1}(1.00)$&0.729(40)&1.0&0.639&0.425&-0.086&-0.077&-0.057\\
		$\Sigma_{1}(1.15)$&0.383(44)&&1.0&0.945&-0.672&-0.746&-0.727\\
		$\Sigma_{1}(1.3)$&0.217(54)&&&1.0&-0.708&-0.847&-0.864\\
		$a_{0}\Sigma_{2}(1.00)~(GeV^{-1})$&0.439(106)&&&&1.0&0.925&0.792\\
		$a_{0}\Sigma_{2}(1.15)~(GeV^{-1})$&0.267(57)&&&&&1.0&0.964\\
		$a_{0}\Sigma_{2}(1.3)~(GeV^{-1})$&0.179(39)&&&&&&1.0\\
		\hline
	\end{tabular}
	\caption{Extracted estimates of soft functions $\Sigma_{1}$ and $a_{0}\Sigma_{2}$ at $w=1.0,~1.15$ and $1.3$.}
	\label{table:Soft functions at w values}
\end{table}
	
\subsection{Extracting the universal functions $\Sigma_{1}$ and $\Sigma_{2}$}\label{subsection:extracting the universal functions}
	
In the previous subsection, we were able to express the $B_{c} \rightarrow D$ and $B_{c} \rightarrow D^{*}$ form factors in terms of just two universal functions, $\Sigma_1$ and $\Sigma_2$. Our next course of action will be to extract the $w$ distribution of these two functions. The universal functions, as we know, are well defined near the zero recoil point, that is $q^{2}=q_{max}^{2}$ or $w=1$, where the initial and the final mesons have the same velocity. Thus, we can expand each universal function in a Taylor series around $w=1$, allowing us to express them in a systematic parametric form as
	\begin{equation}
		\begin{split}
			\Sigma_{1}(w)&=\Sigma_{1}(1)+\Sigma_{1}^{\prime}(w-1)+\frac{1}{2}\Sigma_{1}^{\prime\prime}(w-1)^{2},\\
			a_{0}\Sigma_{2}(w)&=a_{0}\Sigma_{2}(1)+a_{0}\Sigma_{2}^{\prime}(w-1)+\frac{1}{2}a_{0}\Sigma_{2}^{\prime\prime}(w-1)^{2},\\
		\end{split}
		\label{eqn:sigma functions}
	\end{equation}
where $\Sigma_{1}(1)$, $\Sigma_{1}^{\prime}$, $\Sigma_{1}^{\prime\prime}$, $a_{0}\Sigma_{2}(1)$, $a_{0}\Sigma_{2}^{\prime}$ and $a_{0}\Sigma^{\prime\prime}_{2}$ are the parameters that we intend to extract. To extract these parameters, we follow a two-step process. The first step involves extracting information of $\Sigma_{1}$ and $\Sigma_{2}$ at certain $w$ values. They can be obtained from the expressions of $F_{+}$ and $F_{0}$ in Eq. \eqref{eqn:Full QCD to HQET Bc to D form factors} as
	\begin{equation}
		\begin{split}
		\Sigma_{1}(w) &=\frac{M+m}{2q^{2}\sqrt{Mm}}\left[(M-m)^{2}(F_{0}(w)-F_{+}(w))+q^{2}F_{+}(w)\right],\\
		a_{0}\Sigma_{2}(w)&=\frac{1}{2q^{2}\sqrt{Mm}}\left[(M^{2}-m^{2})(F_{+}(w)-F_{0}(w))+q^{2}F_{+}(w)\right],
		\end{split}
		\label{eqn:sigma1 and sigma2 in terms of form factors}
	\end{equation}
with $q^{2}$ being appropriately replaced with $w$. Using the available HPQCD information of $B_{c}\rightarrow D$ form factors at three $w$ points, shown in Table \ref{table:input BcD} in the Appendix, we construct a set of six equations, which, on solving simultaneously, gives numerical estimates of the soft functions at the same $w$ values and is shown in Table \ref{table:Soft functions at w values}.
	
\begin{figure}[htb!]
		\centering
		\subfloat[\centering]{\includegraphics[width=7.0cm]{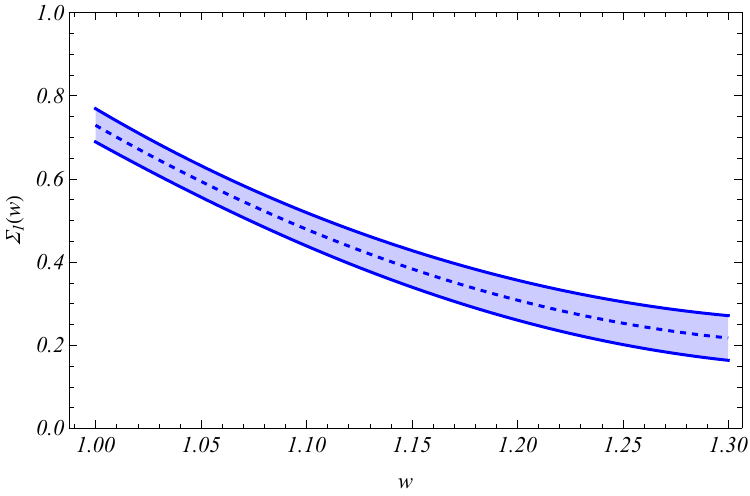}}
		\qquad
		\subfloat[\centering]{\includegraphics[width=7.0cm]{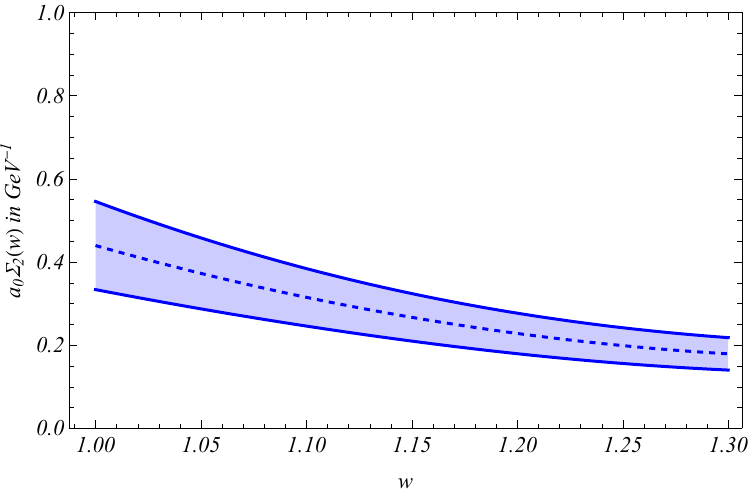}}
		\caption{Plots showing the $w$ distribution of $\Sigma_{1}(w)$ and $a_{0}\Sigma_{2}(w)$.}
		\label{fig:universal functions HQSS}
	\end{figure}

\begin{figure}[htb!]
		\centering
		\subfloat[\centering]{\includegraphics[width=7.0cm]{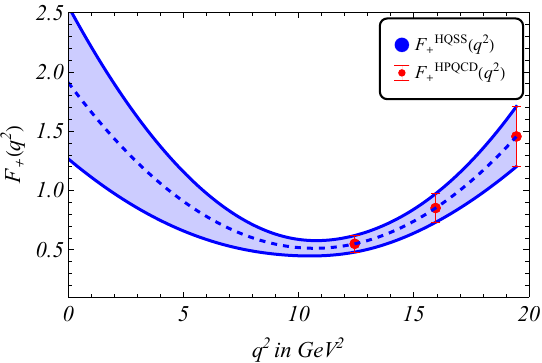}}
		\qquad
		\subfloat[\centering]{\includegraphics[width=7.0cm]{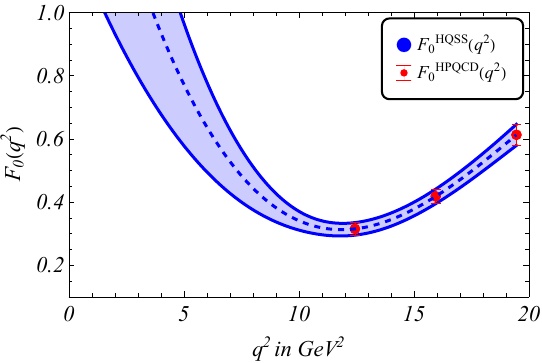}}
		\caption{Plots showing the $q^{2}$ distribution of $B_{c}\rightarrow D$ form factors $F_{+}$ and $F_{0}$. The red markers denote the synthetic data points generated using the BCL parameters supplied by HPQCD.}
		\label{fig:Form factors F+F0}
\end{figure}

From Table \ref{table:Soft functions at w values} we observe that our estimates of $\Sigma_{1}$ are more precise compared to $\Sigma_{2}$, with an error estimate of around 5\% for the former compared to around 24\% for the latter. This behaviour is an artefact of the errors of $F_{+}$ and $F_{0}$ propagating through the expressions in Eqs.\eqref{eqn:sigma1 and sigma2 in terms of form factors}. We next use the values of the soft function, as extracted, and their series-expanded form previously shown in Eqn.\eqref{eqn:sigma functions}, to construct a set of six equations, which we solve to extract the series parameters. Our estimates of the thus extracted parameters are shown in Table \ref{table:Soft function parameters}. Furthermore, we observe that the leading order parameters in the Taylor expansion, $\Sigma_{1}(1)$ and $\Sigma_{2}(1)$ are well constrained, but the uncertainties increase progressively for higher order parameters, signifying the diminishing sensitivity of the slope on the parameter as we move to higher order. Having estimates of these parameters enables us to get the $w$ distribution of both $\Sigma_{1}$ and $\Sigma_{2}$,  which we present in Fig.\ref{fig:universal functions HQSS}. Since this method is more reliable near $w=1$, we confine our plots in the region $w\in (1,1.3)$. This is also visible in the $q^2$ shapes of $F_{+,0}(q^2)$ in Fig.\ref{fig:Form factors F+F0}, which we have obtained using our results of $\Sigma_{1}(w)$ and $\Sigma_{2}(w)$ and Eqn.~\eqref{eqn:Full QCD to HQET Bc to D form factors} in the full kinematically allowed region of $q^2$ or $w$. Note that below $q^2< 10$ GeV$^2$, which corresponds to $w>1.3$, the shapes of the form factors are not well behaved. Hence, we restrict the validity of the above results upto $w = 1.3$.  


		\begin{table}[t]
			\renewcommand{\arraystretch}{1.3}
		\centering
		\begin{tabular}{|c|c|cccccc|}
			\hline
			\textbf{Parameters}&\textbf{Our}&&&\textbf{Correlation}&&&\\
			\cline{3-8}
			&\textbf{estimates}&$\Sigma_{1}(1)$&$a_{0}\Sigma_{2}(1)$&$\Sigma_{1}^{\prime}$&$a_{0}\Sigma_{2}^{\prime}$&$\Sigma_{1}^{\prime\prime}$&$a_{0}\Sigma_{2}^{\prime\prime}$\\
			\hline
			$\Sigma_{1}(1)$&0.729(40)&1.0&-0.086&-0.319&0.079&0.261&-0.068\\
			$a_{0}\Sigma_{2}(1)~(GeV^{-1})$&0.439(106)&&1.0&-0.744&-0.908&0.714&0.842\\
			$\Sigma_{1}^{\prime}$&-2.898(313)&&&1.0&0.537&-0.927&-0.432\\
			$a_{0}\Sigma_{2}^{\prime}~(GeV^{-1})$&-1.435(505)&&&&1.0&-0.617&-0.989\\
			$\Sigma_{1}^{\prime\prime}$&7.953(1.099)&&&&&1.0&0.566\\
			$a_{0}\Sigma_{2}^{\prime\prime}~(GeV^{-1})$&3.776(1.648)&&&&&&1.0\\
			\hline
		\end{tabular}
		\caption{Extracted estimates of parameters of $\Sigma_{1}$ and $a_{0}\Sigma_{2}$.}
		\label{table:Soft function parameters}
	\end{table}

	\subsection{Obtaining $q^{2}$ distribution of rest of the $B_{c}\rightarrow D^{(*)}$ form factors}
	\label{subsection:Bc to D* form factors}
	
	Now that we have the required parameters of the soft functions and their information around $w=1$, we can use them to get values of $B_{c}\rightarrow D$ form factor $F_{T}$ and $B_{c}\rightarrow D^{*}$ form factors $A_{0,1,2}$, $V$ and $T_{1,2,3}$ using the relations previously derived in Eqns.\eqref{eqn:Full QCD to HQET Bc to D form factors}, \eqref{eqn:Full QCD to HQET Bc to D* form factors 1} and \eqref{eqn:Full QCD to HQET Bc to D* form factors 2}. We calculate the form factors at three $w$ points, $w=1.0,~1.15$ and $1.3$, whose numerical values are presented in Tables \ref{table: inputs BcD FT}, \ref{table: inputs BcDStar A012V} and \ref{table: inputs BcDStar T123} in Appendix \ref{section:appendix synthetic data of form factors}. However, there is a limitation to this approach. As we discussed before, the soft functions were expanded around $w=1$, making them unreliable at $w\gg 1$ or towards the low $q^{2}$ region. To obtain reliable information on the form factors over the full $q^{2}$ region, we need to parametrize them. In this work, we adopt the BCL parametrization, due to its rapid convergence over the full $q^{2}$ region, making it reliable with only 3 to 4 parameters. It also respects analyticity of the form factors and avoids unphysical behaviour due to intermediate poles through a pole factor. The form factors defined in BCL parametrization have the form as
	\begin{equation}
		f_{i}(q^{2})=\frac{1}{P(q^{2})}\sum_{n=0}^{2}a_{n}^{i}z(q^{2})^{n},
	\end{equation}
	where $a_{n}^{i}$ are the parameters we intend to extract. $z(q^{2})$, the conformal variable is defined as
	\begin{equation}
		z(q^{2})=\frac{\sqrt{t_{+}-q^{2}}-\sqrt{t_{+}-t_{0}}}{\sqrt{t_{+}-q^{2}}+\sqrt{t_{+}-t_{0}}},
	\end{equation}
	where $t_{+}=(m_{B}+m_{\pi/\rho})^{2}$ and $t_{0}$, an arbitrary reference point is chosen as \cite{Biswas:2022lhu}
		\begin{equation}
			t_{0}=t_{opt}=t_{+}\left(1-\sqrt{1-\frac{t_{-}}{t_{+}}}\right),
		\end{equation}
		to centre the z-expansion around mid $q^{2}$ region, so as to facilitate a faster convergence\footnote{Note that in this work we adopt a different choice of $t_{0}$ compared to HPQCD \cite{Cooper:2021bkt}. This choice of $t_{0}$ maps the physical semileptonic region into a symmetric interval in $z$, i.e., $|z(q^{2}=0)|=|z(q^{2}=t_{-})|$, so that the physical $q^{2}$ region corresponds to $-z_{max}\leq z\leq z_{max}$. As a result, the maximum $|z|$ is minimized, compared to other choices of $t_{0}$, which means the truncated z-series converges much faster compared to either $t_{0}=0$ or $t_{0}=t_{-}$.}. The function $P(q^{2})$ represents the Blaschke factor and prevents divergence in the BCL expansion by factoring out known resonance poles from the form factors, thus ensuring that the remaining piece is analytic and can be safely expanded in a power series in $z(q^{2})$. It has the form as
	\begin{equation}
		P(q^{2})=1-\frac{q^{2}}{M_{res}^{2}}.
	\end{equation}
	$M_{res}$  represents the resonance pole masses, with their values used in this work are shown in Table \ref{table:resonance masses} and has been taken from \cite{Ball:2004rg}.
	\begin{table}[htb!]
		\centering
		\begin{tabular}{|c|cccccccc|}
			\hline
			Form Factor& $F_{T}$& $A_{0}$ & $A_{1}$ & $A_{2}$ & $V$&$T_{1}$&$T_{2}$&$T_{3}$\\
			\hline
			$M_{res}$ in GeV&5.325&5.28&5.68&5.68&5.325&5.325&5.68&5.68\\
			\hline
		\end{tabular}	
		\captionof{table}{Masses of the low lying $B^{*}$ resonances.}
		\label{table:resonance masses}
	\end{table}
	
	To extract the BCL parameters we follow the frequentist approach based on the minimization of a chi-square function. We construct two chi-square functions, one for $B_{c}\rightarrow D$ form factors, and the other for all the $B_{c}\rightarrow D^{*}$ form factors:
	\begin{itemize}
		\item
	To extract the $B_{c}\rightarrow D$ form factor information, especially the tensor form factor $F_{T}$, whose distribution, as of now, has not been supplied by any lattice groups, we plan to extract the BCL parameters of the form factor using the values obtained using Eqn.\eqref{eqn:Full QCD to HQET Bc to D form factors} and shown in Table \ref{table: inputs BcD FT}. But extracting just $F_{T}$ is not going to be helpful, since we plan to predict observables that depend on all three form factors, $F_{+}$, $F_{0}$ and $F_{T}$, and hence the correlation between all the BCL parameters. becomes necessary Therefore, we take HPQCD supplied $F_{+}$ and $F_{0}$ values, presented in Table \ref{table:input BcD}, and club them with our $F_{T}$ values presented in Table \ref{table: inputs BcD FT}. This constrains the form factor slopes at high $q^{2}$ region. To constrain them at the low $q^{2}$ region, we add our pQCD estimates of the form factors at $q^{2}=0$, previously presented in Table \ref{table:form factor predictions BcDStar}. With all the inputs organised, we then construct the chi-square function and minimize it to extract the required BCL parameters. We present our estimates of the extracted parameters in Table \ref{table:BCL Bc to D}, along with the corresponding correlation matrix in Table \ref{table:correlation BCL parameters BcD}.
	
\begin{table}[htb!]
	\centering
	\begin{tabular}{|c|ccc|}
		\hline
		\textbf{Form factors}&$\boldsymbol{a_{0}}$&$\boldsymbol{a_{1}}$&$\boldsymbol{a_{2}}$\\
		\hline
		$F_{+}$&0.309(19)&-1.105(136)&0.980(522)\\
		$F_{0}$&-&-0.217(43)&0.777(273)\\
		$F_{T}$&0.351(33)&-1.173(213)&0.0435(1.240)\\
		\hline
		\textbf{DOF}&&3&\\
		\hline
		\textbf{p-Value}&&54.35\%&\\
		\hline
	\end{tabular}
	\caption{Extracted BCL parameters for $B_{c}\rightarrow D$ form factors.}
	\label{table:BCL Bc to D}
\end{table}

From Table \ref{table:BCL Bc to D} we note that our estimates of BCL parameters of $F_{+}$ and $F_{0}$ do not match with the ones extracted by HPQCD in \cite{Cooper:2021bkt}. This is entirely due to the difference in the choice of $t_{0}$. Further observation highlights an increase in error estimates of BCL coefficients as the order increases, thereby signifying a reduced sensitivity of the form factors to the BCL coefficients as the order increases. The convergence of the z-series is verified since the extracted parameters follow the hierarchy $a_{2}z(q^{2})^{2} < a_{1}z(q^{2}) < a_{0}$. In this work, since we have truncated the BCL series up to quadratic order, i.e., up to the second order in $z$. We introduce an additional error to account for the error associated with missing higher-order terms. In particular, we estimate the possible effect of the next higher-order term in the expansion, neglecting the relatively small higher-order terms after that, as prescribed in \cite{Bourrely:2008za}. We then propagate this additional error as a systematic uncertainty into the form factors. The estimate of this systematic error is defined as follows:
\begin{equation}
	\delta f_{i}(q^{2})=\frac{a_{3}^{max}|z(q^{2})^{3}|}{P(q^{2})}.
	\label{eqn:truncation}
	\end{equation}
For the coefficient $a_{3}^{\text{max}}$, we adopt a conservative estimate by assuming that the cubic term contributes around 50\% of the quadratic term. Although ideally the cubic term contribution is expected to be much less than the quadratic one in order to ensure convergence of the z-series, i.e., $a_{3}z^{3}<<a_{2}z^{2}$, our prescription ensures that the systematic error is not underestimated. We deliberately take an enlarged estimate of the systematic error to avoid overly precise predictions, thereby maintaining a conservative error budget.

In Fig.\ref{fig:form factors Bc to D FT} we present the same for the form factor $F_{T}$.  With the thus extracted BCL parameters and truncation uncertainty, we can now obtain $q^{2}$ distribution of the $B_{c}\rightarrow D$ form factors. We do not present our BCL plots for $F_{+}$ and $F_{0}$, since HPQCD has already obtained them, and presenting them here will not add any new information.
		\begin{figure}[htb!]
	\centering
	\includegraphics[width=9.0cm]{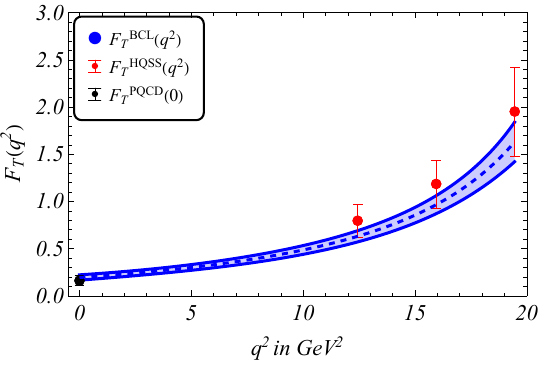}
	\caption{$q^{2}$ distribution of $B_{c}\rightarrow D$ tensor form factor $F_{T}$. The blue curve represents the distribution obtained using the extracted BCL parameters. The red and black markers denote the form factor values obtained using the extracted soft functions and pQCD, respectively. }
	\label{fig:form factors Bc to D FT}
\end{figure}

\item

Next, to extract the BCL parameters for $B_{c}\rightarrow D^{*}$ form factors, we construct a chi-square function similar to the previous case, replacing the $B_{c}\rightarrow D$ form factors with $B_{c}\rightarrow D^{*}$ ones. To constrain form factor slopes at high $q^{2}$ regions, we take the form factor values obtained using HQET expressions previously derived in Eqns.\eqref{eqn:Full QCD to HQET Bc to D* form factors 1} and \eqref{eqn:Full QCD to HQET Bc to D* form factors 2}, and presented in Tables \ref{table: inputs BcDStar A012V} and \ref{table: inputs BcDStar T123}. Additionally, to constrain the slope in the low $q^{2}$ region, we utilize previously obtained pQCD values of the form factors at $q^{2}=0$ as additional inputs. With these inputs, we construct our chi-square function and then minimize it to extract the required BCL parameters. We present our extracted estimates of the parameters in Table \ref{table:BCL Bc to D*} along with the corresponding correlation matrix in Table \ref{table:correlation BCL parameters BcDStar}.
\begin{table}[htb!]
	\centering
	\begin{tabular}{|c|ccc|}
		\hline
		\textbf{Form factors}&$\boldsymbol{a_{0}}$&$\boldsymbol{a_{1}}$&$\boldsymbol{a_{2}}$\\
		\hline
		$A_{0}$&0.430(14)&-3.619(218)&10.761(4.023)\\
		$A_{1}$&0.188(10)&-0.569(140)&4.996(2.337)\\
		$A_{2}$&-&-1.061(242)&6.716(4.954)\\
		$V$&0.531(39)&-2.995(389)&-11.944(5.826)\\
		$T_{1}$&0.345(17)&-1.795(193)&3.921(2.172)\\
		$T_{2}$&-&-0.296(110)&2.861(1.990)\\
		$T_{3}$&0.487(68)&-3.087(496)&-9.856(10.065)\\
		\hline
		\textbf{DOF}&&7&\\
		\hline
		\textbf{p-Value}&&99.98\%&\\
		\hline
	\end{tabular}
	\caption{Extracted BCL parameters for $B_{c}\rightarrow D^{*}$ form factors.}
	\label{table:BCL Bc to D*}
\end{table}

Checking Table \ref{table:BCL Bc to D*}, we observe that error estimates of the parameters for each form factor increase as we go up in order of z-expansion series, signifying a reduced sensitivity of the form factor slopes with increasing order of the parameters. Further, we also observe that the extracted parameters follow the hierarchy $a_{2}z(q^{2})^{2}<a_{1}z(q^{2})<a_{0}$, implying that our z-expansion is converging in nature. Similar to the previous case, we also truncate the z-series at quadratic order here and include the missing higher-order terms as a systematic error. We estimate this error by following the same steps as in Eqn.\eqref{eqn:truncation} and propagate it as a systematic uncertainty into the form factors.

 With the BCL parameters extracted and the truncation uncertainty included, we have all the information necessary to get the $q^{2}$ distribution of the form factors over the full semileptonic region. In Figs. \ref{fig:form factors Bc to D* 1} and \ref{fig:form factors Bc to D* 2} we showcase these $q^{2}$ distribution plots.

\begin{figure}[htb!]
	\centering
	\subfloat[\centering]{\includegraphics[width=7.0cm]{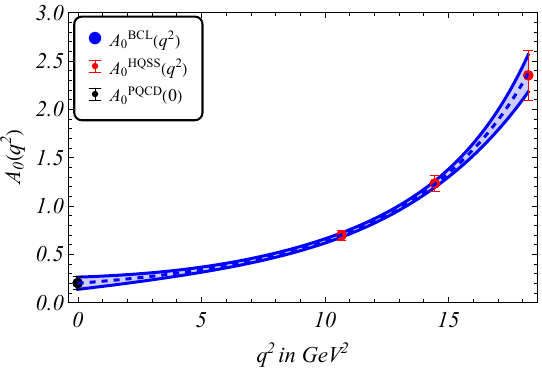}}
	\qquad
	\subfloat[\centering]{\includegraphics[width=7.0cm]{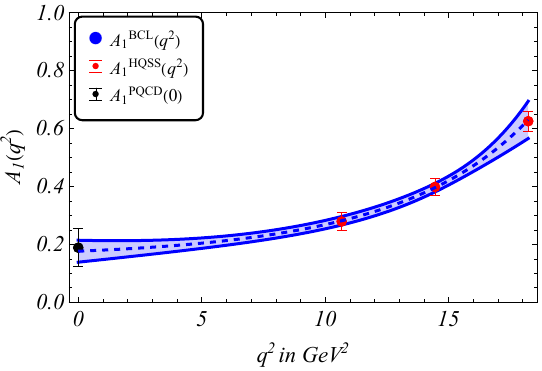}}
	\qquad
	\subfloat[\centering]{\includegraphics[width=7.0cm]{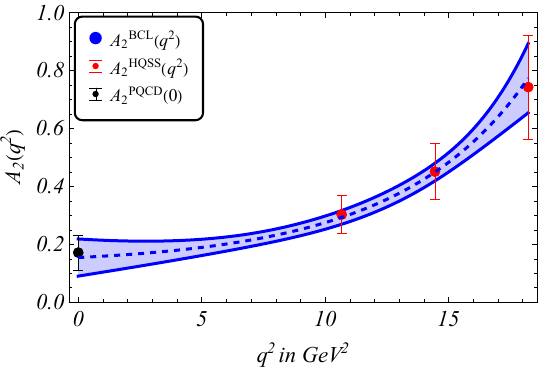}}
	\qquad
	\subfloat[\centering]{\includegraphics[width=7.0cm]{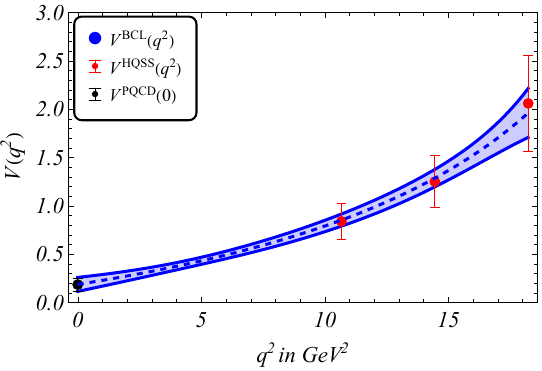}}
	\caption{$q^{2}$ distribution of $B_{c}\rightarrow D^{*}$ vector and axial-vector form factors. The blue curve represents the distribution obtained using the extracted BCL parameters. The red and black markers denote the form factor values obtained using the extracted soft functions and pQCD, respectively. }
	\label{fig:form factors Bc to D* 1}
\end{figure}

		\begin{figure}[htb!]
	\centering
	\subfloat[\centering]{\includegraphics[width=7.0cm]{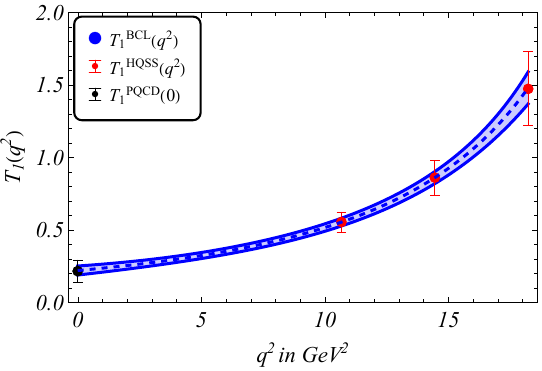}}
	\qquad
	\subfloat[\centering]{\includegraphics[width=7.0cm]{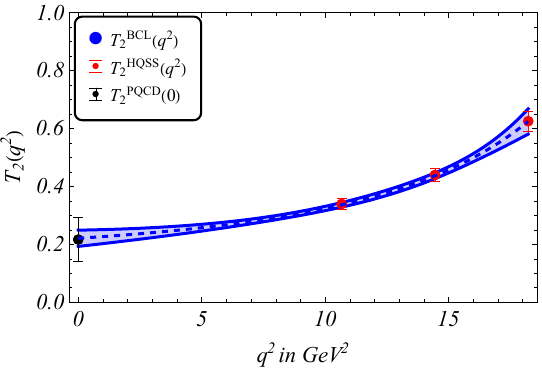}}
	\qquad
	\subfloat[\centering]{\includegraphics[width=7.0cm]{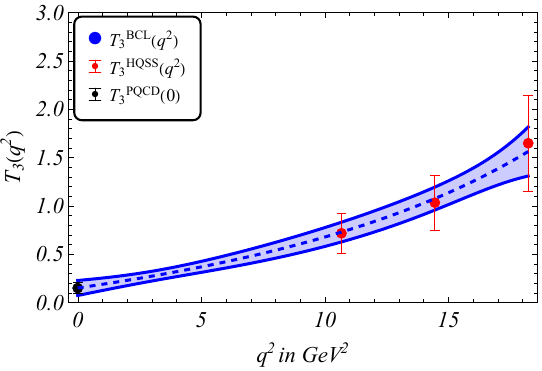}}
	\caption{$q^{2}$ distribution of $B_{c}\rightarrow D^{*}$ tensor form factors. The blue curve represents the distribution obtained using the extracted BCL parameters. The red and black markers denote the form factor values obtained using the extracted soft functions and pQCD, respectively. }
	\label{fig:form factors Bc to D* 2}
\end{figure}

\end{itemize}                            

\section{Prediction of physical observables}
\label{section:Prediction of physical observables}

With $q^{2}$ distribution of all the relevant form factors determined, we now proceed to the predictions of the physical observables previously discussed in subsection \ref{subsection:physical observables}. In subsection \ref{subsection:branchng ratios}, we present our predictions for the branching fractions and their ratios for all the semileptonic and rare decay channels considered in this work. In subsection \ref{subsection:angular observables}, we provide our predictions for the various angular observables associated with the cascade decay channel $B_{c}^{-}\rightarrow D^{*-}(\rightarrow D^{0}\pi^{-})\ell^{+}\ell^{-}$.

\subsection{Decay Width and branching fractions}
\label{subsection:branchng ratios}
\begin{itemize}

\item For the $B_{c}^{-}\rightarrow \bar{D}^{*0}\ell^{-}\bar{\nu}_{\ell}$ channels the underlying quark-level process is a $b\rightarrow u$ charged current one. We calculate the differential decay width as a function of $q^{2}$ using the previously mentioned Eqn.\eqref{eqn:differential decay width Bc_DStar}. In Fig. \ref{fig:decay width distribution Bc to DStar}
we present the $q^{2}$ distribution of its differential decay width over the full physical $q^{2}$ region. The decay width distributions rise uniformly as we move from $m_{\ell}^{2}$, i.e., minimum of the semileptonic region and attain a peak value at around $q^{2}=15.0~GeV^{2}$, after which they fall sharply at $q_{max}^{2}$, due to phase space suppression. Further, comparing the light lepton and heavy lepton plots, we observe that the separation between the two plots is more prominent at the low $q^{2}$ region, and it reduces as we move towards $q^{2}=15.0~GeV^{2}$. This behaviour arises primarily due to lepton mass suppression effects.

		\begin{figure}[htb!]
	\centering
	\includegraphics[width=8.0cm]{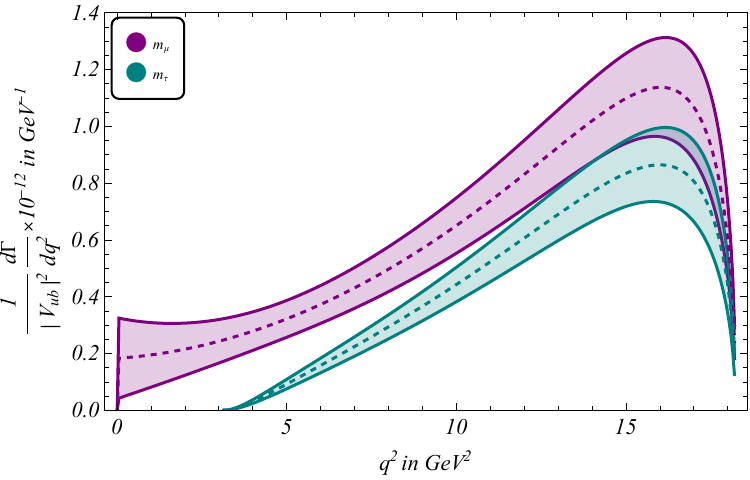}
	\caption{$q^{2}$ distribution of $|V_{ub}|^{-2}d\Gamma(B_{c}\rightarrow D^{*})/dq^{2}$ semileptonic channel. The violet and green plots denote the light and heavy lepton cases respectively.}
	\label{fig:decay width distribution Bc to DStar}
\end{figure}

Once the $q^2$ dependent decay width distributions are obtained, we compute the branching fractions of the relevant semileptonic modes by integrating over the full allowed semileptonic region i.e., $q^{2}\in (m_{\ell}^{2},(M-m)^{2})$. We present our predictions of branching fractions along with a comparison with previous pQCD predictions in Table \ref{table:predictions branching fractions semileptonic BcDStar}. The uncertainties in the calculation of branching fractions are estimated by propagating errors from the semileptonic form factors and CKM matrix elements. The total error is estimated by propagating the correlation matrix from Table \ref{table:correlation form factors BcDStar} through the calculation.
	\begin{table}[htb!]
		\renewcommand{\arraystretch}{1.3}
	\centering
	\begin{tabular}{|c|cc|}
		\hline
		\textbf{Decay modes}&\textbf{This work}&\textbf{Previous PQCD}\cite{PhysRevD.90.094018}\\
		\hline
		$\mathcal{B}(B_{c}^{-}\rightarrow \bar{D}^{*0}\ell^{-}\bar{\nu_{l}})$&1.254(146)$\times 10^{-4}$&1.09$^{+0.51}_{-0.43}\times 10^{-4}$\\
		$\mathcal{B}(B_{c}^{-}\rightarrow \bar{D}^{*0}\tau^{-}\bar{\nu_{\tau}})$&0.789(71)$\times 10^{-4}$&0.64$^{+0.30}_{-0.25}\times 10^{-4}$\\
		\hline
	\end{tabular}
	\caption{Predictions for branching fractions of the $B_{c}^{-}\rightarrow D^{*}$ charged current mediated semileptonic transitions, with $(\ell=e,\mu)$ along with a comparison to previous pQCD predictions.}
	\label{table:predictions branching fractions semileptonic BcDStar}
	\end{table}
	
	As seen in Table \ref{table:predictions branching fractions semileptonic BcDStar}, the branching ratio for the $\tau$ mode is noticeably smaller than the light lepton mode. This suppression arises primarily due to the difference in the phase space available, a consequence of the larger $\tau$ lepton mass when compared to the light lepton masses. Further comparing our predictions with previous pQCD predictions, we find that the central values exhibit good agreement, although there is a significant improvement in the error estimates. Hence, our predictions are far more precise compared to previous predictions.

In addition to branching fractions, we also define the ratio between the heavy lepton and light lepton branching fractions for the charged current processes, defining lepton flavor universality violation, and is expressed as
\begin{equation}
	R_{D^{*}}^{B_{c}^{-}\rightarrow \bar{D}^{*0}}=\frac{\mathcal{B}(B_{c}^{-}\rightarrow \bar{D}^{*0}\tau^{-}\bar{\nu}_{\tau})}{\mathcal{B}(B_{c}^{-}\rightarrow \bar{D}^{*0}\ell^{-}\bar{\nu}_{\ell})}.
\end{equation}
This observable is a far more cleaner due to the cancellation of uncertainties arising from the form factors. In Table \ref{table:predictions RD RDStar BcDStar} we present our predictions of $R_{D^{*}}^{B_{c}^{-}\rightarrow \bar{D}^{*0}}$ along-with a comparison with previous pQCD prediction.
	\begin{table}[htb!]
		\renewcommand{\arraystretch}{1.3}
	\centering
	\begin{tabular}{|c|ccc|}
		\hline
		\textbf{Ratios}&\textbf{This work}&\textbf{Previous PQCD}\cite{PhysRevD.90.094018}&\textbf{Lattice}\\
		\hline		
		$R_{D^{*}}^{B_{c}^{-}\rightarrow \bar{D}^{*0}}$&0.632(22)&0.59$^{+0.00+0.00}_{-0.01-0.01}$&-\\
		\hline
		\end{tabular}
		\caption{Prediction for $R_{D^{*}}^{B_{c}^{-}\rightarrow \bar{D}^{*0}}$ and comparison with other predictions.}
	\label{table:predictions RD RDStar BcDStar}
	\end{table}
	
	In this work we have not calculated branching fractions for $B_{c}^{-}\rightarrow \bar{D}^{0}\ell^{-}\bar{\nu}_{\ell}$ channel since the expression for the branching ratio is a function of $F_{+}$ and $F_{0}$, information of which has already been supplied by HPQCD in their work \cite{Cooper:2021bkt}, and our analysis is based on their form factor information as inputs. Calculating the branching fractions again in this work will not yield any new information.

\item For the $b\rightarrow d\text{ }\ell^{+}\ell^{-}(\nu\bar{\nu})$ FCNC processes, with $\ell=e,\mu$, the branching fractions are presented in Tables \ref{table:predictions branching fractions rare BcDStar 1} and \ref{table:predictions branching fractions rare BcDStar 2}. The calculation is done by integrating the differential decay width expressions previously shown in subsection \ref{subsection:physical observables} over the full physical $q^{2}$ region, i.e., $q^{2}\in (4m_{\ell}^{2},(m_{B_{c}}-m_{D^{*}})^{2})$.
	\begin{table}[htb!]
		\renewcommand{\arraystretch}{1.3}
	\centering
	\begin{tabular}{|c|cc|}
		\hline
		\textbf{Decay Modes}&\textbf{This work}&\textbf{Previous PQCD}\cite{PhysRevD.90.094018}\\
		\hline
		$\mathcal{B}(B_{c}^{-}\rightarrow D^{-}\ell^{+}\ell^{-})$&5.276(595)$\times 10^{-9}$&3.79$^{+1.57}_{-1.31}\times 10^{-9}$\\
		$\mathcal{B}(B_{c}^{+}\rightarrow D^{+}\ell^{+}\ell^{-})$&5.508(609)$\times 10^{-9}$&-\\
		$\mathcal{B}(B_{c}^{-}\rightarrow D^{-}\tau^{+}\tau^{-})$&1.186(84)$\times 10^{-9}$&1.03$^{+0.48}_{-0.38}\times 10^{-9}$\\
		$\mathcal{B}(B_{c}^{+}\rightarrow D^{+}\tau^{+}\tau^{-})$&1.188(86)$\times 10^{-9}$&-\\
		$\mathcal{B}(B_{c}^{-}\rightarrow D^{-}\nu\bar{\nu})$&3.327(384)$\times 10^{-8}$&3.13$^{+1.34}_{-1.08}\times 10^{-8}$\\
		\hline
        \end{tabular}
	\caption{Predictions for the branching fractions of the $B_{c}^{\pm}\rightarrow D^{\pm}$ FCNC transitions with $(l=e,\mu)$ along-with comparison with predictions in existing literature.}
	\label{table:predictions branching fractions rare BcDStar 1}
	\end{table}
        	\begin{table}[htb!]
		\renewcommand{\arraystretch}{1.3}
	\centering
	\begin{tabular}{|c|cc|}
		\hline
		\textbf{Decay Modes}&\textbf{This work}&\textbf{Previous PQCD}\cite{PhysRevD.90.094018}\\
		\hline
		$\mathcal{B}(B_{c}^{-}\rightarrow D^{*-}\ell^{+}\ell^{-})$&1.512(161)$\times 10^{-8}$&1.21$^{+0.55}_{-0.47}\times 10^{-8}$\\
		$\mathcal{B}(B_{c}^{+}\rightarrow D^{*+}\ell^{+}\ell^{-})$&1.537(166)$\times 10^{-8}$&-\\
		$\mathcal{B}(B_{c}^{-}\rightarrow D^{*-}\tau^{+}\tau^{-})$&0.203(70)$\times 10^{-8}$&0.16$^{+0.07}_{-0.06}\times 10^{-8}$\\
		$\mathcal{B}(B_{c}^{+}\rightarrow D^{*+}\tau^{+}\tau^{-})$&0.205(70)$\times 10^{-8}$&-\\
		$\mathcal{B}(B_{c}^{-}\rightarrow D^{*-}\nu\bar{\nu})$&0.980(93)$\times 10^{-7}$&1.10$^{+0.51}_{-0.43}\times 10^{-7}$\\
		\hline
	\end{tabular}
	\caption{Predictions for the branching fractions of the $B_{c}^{\pm}\rightarrow D^{*\pm}$ FCNC transitions with $(l=e,\mu)$, along with a comparison with predictions in existing literature.}
	\label{table:predictions branching fractions rare BcDStar 2}
	\end{table}
	
	In Tables \ref{table:predictions branching fractions rare BcDStar 1} and \ref{table:predictions branching fractions rare BcDStar 2}, the branching fractions of the $\tau$ lepton mode are smaller than the light lepton modes due to phase space suppression. The branching ratio of the decay mode with $\nu\bar{\nu}$ pair is almost an order larger than the corresponding mode with $\ell^{+}\ell^{-}$, the reason for which can be attributed to summation over the three neutrino generations. As for the error estimates, just as in Table \ref{table:predictions branching fractions semileptonic BcDStar}, we also observe a significant improvement in the error estimates, resulting in a much more precise prediction compared to the previous pQCD prediction. In addition to branching fractions, we also calculate the ratio of the tauon and light lepton branching fractions, defined as
	\begin{equation}
		R_{D^{(*)}}^{B_{c}^{-}\rightarrow D^{(*)-}}=\frac{\mathcal{B}(B_{c}^{-}\rightarrow D^{(*)-}\tau^{+}\tau^{-})}{\mathcal{B}(B_{c}^{-}\rightarrow D^{(*)-}\ell^{+}\ell^{-})}.
	\end{equation}
    and present our predictions as follows:
	\begin{equation}
		R_{D}^{B_{c}^{-}\rightarrow D^{-}}=0.225(15),\qquad R_{D^{*}}^{B_{c}^{-}\rightarrow D^{*-}}=0.166(10).
	\end{equation}
		
\end{itemize}

Comparing the prediction of Table \ref{table:predictions branching fractions semileptonic BcDStar} with Tables \ref{table:predictions branching fractions rare BcDStar 1} and \ref{table:predictions branching fractions rare BcDStar 2}, we observe that the branching fractions of the latter table are much smaller than the former. This is primarily because the rare channels mediated by FCNC are forbidden at tree level in SM. They occur only through loop diagrams, such as penguin or box diagrams, making them loop-suppressed compared to the semileptonic modes, which are allowed at tree level in the SM. In addition, the CKM matrix elements in the expression of amplitudes also contribute to further suppression. Further, revisiting Tables \ref{table:predictions branching fractions rare BcDStar 1} and \ref{table:predictions branching fractions rare BcDStar 2}, we observe a difference in the predictions of $\mathcal{B}(B_{c}^{-}\to D^{(*)})$ channels and their respective CP-conjugate modes. This difference hints towards a non-zero value of CP asymmetry. Thus, a prediction of this CP asymmetry observable gathers significance. The observable is defined as
\begin{equation}
	\mathcal{A}_{CP}(q^{2})=\frac{d\mathcal{B}(B_{c}^{-}\rightarrow D_{}^{(*)-}\ell^{+}\ell^{-})/dq^{2}-d\mathcal{B}(B_{c}^{+}\rightarrow D^{(*)+}\ell^{+}\ell^{-})/dq^{2}}{d\mathcal{B}(B_{c}^{-}\rightarrow D^{(*)-}\ell^{+}\ell^{-})/dq^{2}+d\mathcal{B}(B_{c}^{+}\rightarrow D^{(*)+}\ell^{+}\ell^{-})/dq^{2}},
\end{equation}
and the $q^{2}$ averaged values of this observable, integrated over the full physical $q^{2}$ region, are presented in table \ref{table:CP asymmetry Bc to D*}.
\begin{table}[htb!]
	\centering
	\begin{tabular}{|c|c|}
		\hline
		\textbf{Decay Mode}&\textbf{Our predictions}\\
		\hline
		$\langle\mathcal{A}_{CP}\rangle(B_{c}\to D\ell^{+}\ell^{-})$&-0.0211(21)\\
		$\langle\mathcal{A}_{CP}\rangle(B_{c}\to D\tau^{+}\tau^{-})$&-0.0023(2)\\
		$\langle\mathcal{A}_{CP}\rangle(B_{c}\to D^{*}\ell^{+}\ell^{-})$&-0.0119(2)\\
		$\langle\mathcal{A}_{CP}\rangle(B_{c}\to D^{*}\tau^{+}\tau^{-})$&-0.0058(1)\\
		\hline
	\end{tabular}
	\caption{Predictions for CP asymmetry of the respective decay channels.}
	\label{table:CP asymmetry Bc to D*}
\end{table}

From table \ref{table:CP asymmetry Bc to D*}, we can see that CP asymmetry, although very small, still attains a non-zero value. The primary source of this non-zero value arises from the CKM elements $V_{ub}$ and $V_{td}$ in $C_{9,eff}$, which carry a phase term. Furthermore, comparing the first two rows, we can see the lepton mass and reduced phase space for the $\tau$-lepton case, which suppresses the estimate. Also, comparing the values between $D$ and $D^{*}$ modes, we can see the values for the latter modes are smaller than the former ones. This is because the $D$ meson is a pseudoscalar meson, while the $D^{*}$ meson is a vector meson. As a result, the $D^{*}$ meson has longitudinal and transverse polarization, which can have destructive interference between them, resulting in a suppressed total value. This cancellation is absent in $D$ meson, which has just one component.

\subsection{Angular Observables}
\label{subsection:angular observables}

Having predicted the branching fractions of some relevant semileptonic and rare decay modes of the $B_{c}$ meson and the corresponding LFUV observables, we next shift our focus onto calculating the various angular observables concerning the rare $B_{c}^{-}\rightarrow D^{*-}(\rightarrow D^{0}\pi^{-})\ell^{+}\ell^{-}$ cascade channel. A discussion on these observables along with their analytic expressions has already been done in subsection \ref{subsubsection:Angular analysis}.

\begin{figure}[htb!]
	\centering
	\subfloat[\centering]{\includegraphics[width=4.5cm]{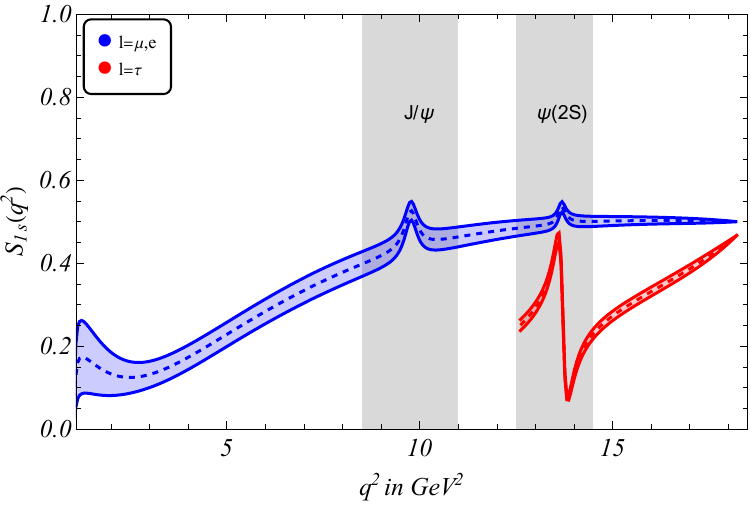}}
	\qquad
	\subfloat[\centering]{\includegraphics[width=4.5cm]{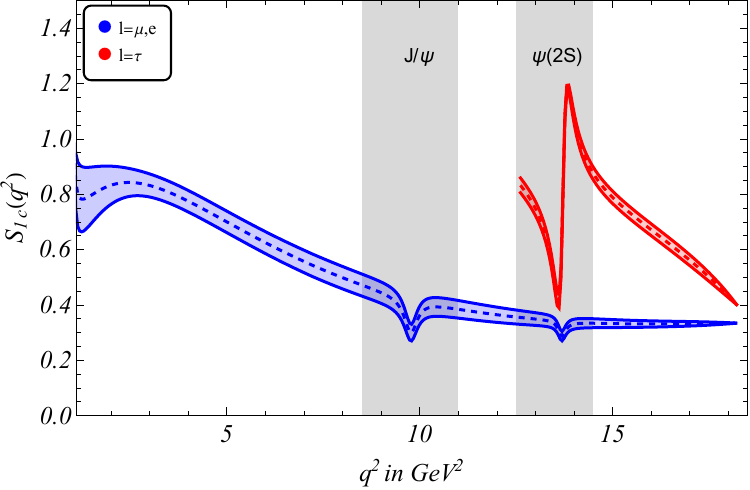}}
	\qquad
	\subfloat[\centering]{\includegraphics[width=4.5cm]{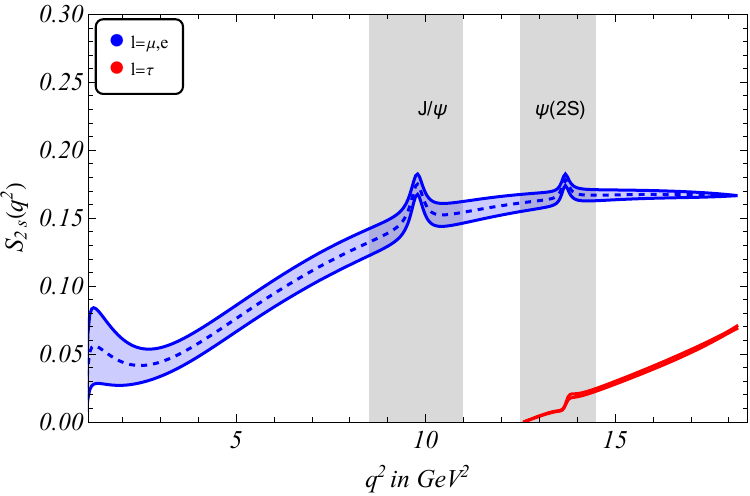}}
	\qquad
	\subfloat[\centering]{\includegraphics[width=4.5cm]{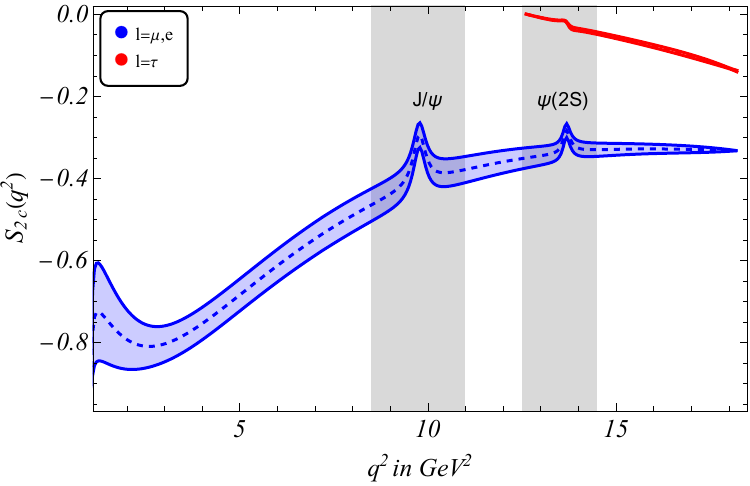}}
	\qquad
	\subfloat[\centering]{\includegraphics[width=4.5cm]{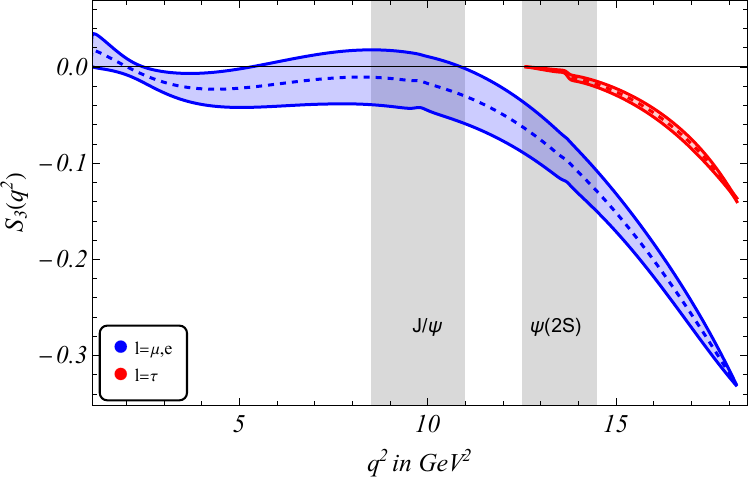}}
	\qquad
	\subfloat[\centering]{\includegraphics[width=4.5cm]{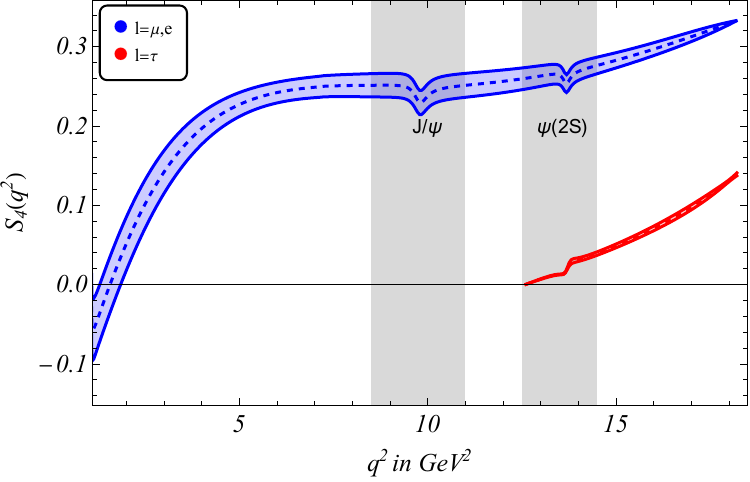}}
	\qquad
	\subfloat[\centering]{\includegraphics[width=4.5cm]{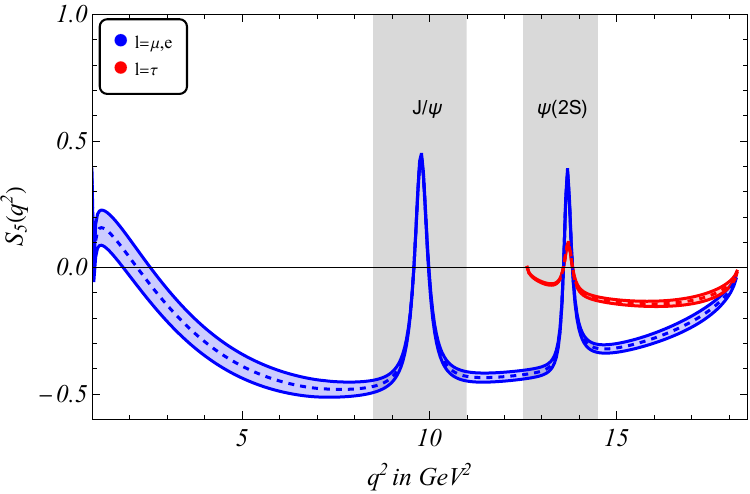}}
	\qquad
	\subfloat[\centering]{\includegraphics[width=4.5cm]{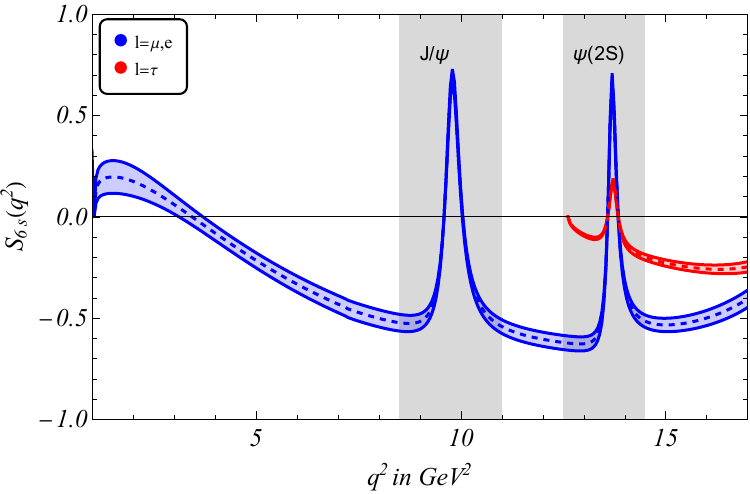}}
	\qquad
	\subfloat[\centering]{\includegraphics[width=4.5cm]{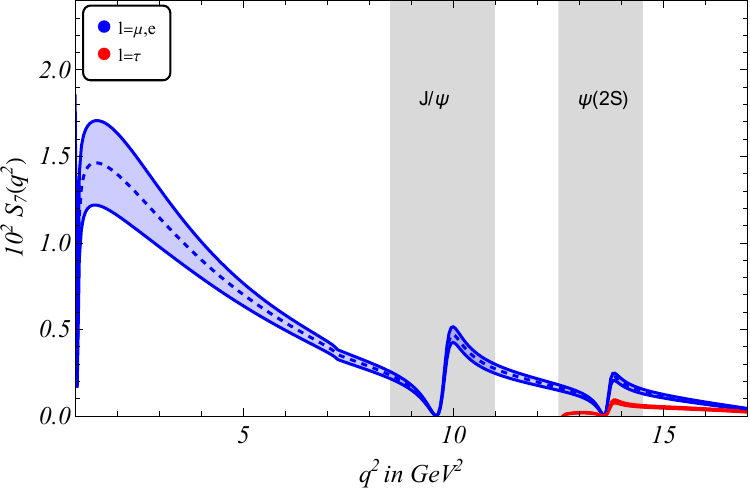}}
	\qquad
	\subfloat[\centering]{\includegraphics[width=4.5cm]{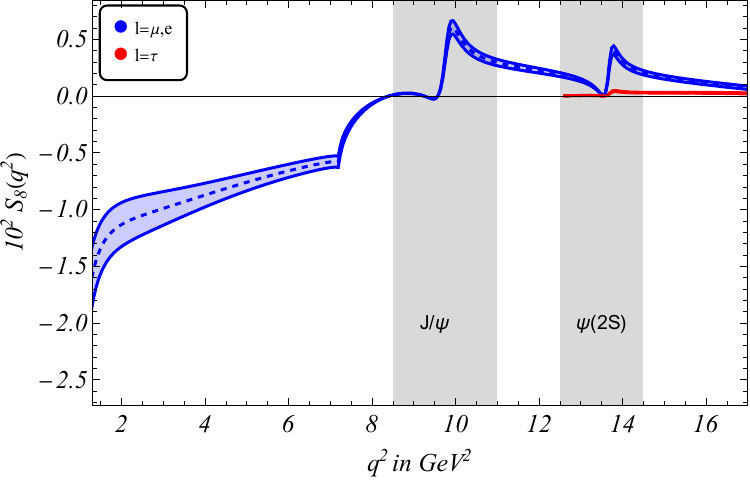}}
	\qquad
	\subfloat[\centering]{\includegraphics[width=4.5cm]{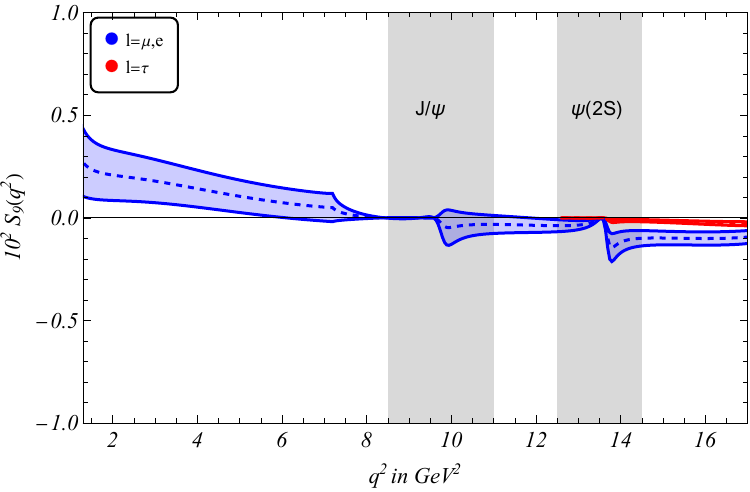}}
	\caption{The $q^{2}$ dependence of normalized CP averaged angular observables $S_{i}$. The blue and the red plots depict our results for the light lepton and $\tau$ lepton modes respectively and the gray bands denote the region of $J/\psi$ and $\psi(2S)$ resonances.}
	\label{fig:CP averaged angular observables}
\end{figure}

We start by presenting our results on the CP average angular coefficients $S_{i}$. In Fig. \ref{fig:CP averaged angular observables} we present $q^{2}$ distribution of each of these angular coefficients.

The blue curves denote the results for light leptons, and the red curves denote those for $\tau$ lepton modes. The gray bands denote the regions where the contributions due to $J/\psi$ and $\psi(2S)$ charmonium states would lie. From Fig.\ref{fig:CP averaged angular observables} we observe that only $S_{5}$ and $S_{6s}$ seems to have zero in the $q^{2}$ range. They cross the zero point at 
\begin{equation}
\begin{split}
S_{5}:q_{0}^{2}=2.212(368)~GeV^{2}, \qquad S_{6s}:q_{0}^{2}=3.489(284)~GeV^{2}.
\end{split}
\end{equation}
This zero point is significant because it denotes the point where the left-handed and right-handed helicities of the transversity amplitudes cancel each other. Compared to $S_{1s}$ to $S_{6s}$, the remaining three observables $S_{7,8,9}$ have a much smaller magnitude. The reason for this can be traced back to analytic expressions in Eqn.\eqref{eqn:transversity amplitudes}, where it can be clearly seen that they are dependent upon the imaginary part of the transversity amplitudes, and hence of the Wilson Coefficients $C_{7}^{eff}$ and $C_{9}^{eff}$, whose imaginary parts, upon checking numerically, is found out to be highly suppressed compared to the real part. From Fig.\ref{fig:CP averaged angular observables}, $S_{8,9}$ also seem to have zeroes, but since the crossing happens within the gray bands, or rather the $J/\psi$ resonance peak, we do not quote the values of $q_{0}^{2}$ for these observables. We also calculate the $q^{2}$- averaged values of these observables, separated into four $q^{2}$ bins, and present the thus obtained values in Table \ref{table:angular observables in bins}. 
\begin{table}[htb!]
	\renewcommand{\arraystretch}{1.3}
	\centering
	\scriptsize
	\begin{tabular}{|c|cc|cc|cc|cc|}
		\hline
		$q^{2}$ bins $(GeV^{2})$& \multicolumn{2}{c|}{$\langle S_{1s}\rangle$} &\multicolumn{2}{c|}{$\langle S_{1c}\rangle$ }&\multicolumn{2}{c|}{$\langle S_{2s}\rangle$}&\multicolumn{2}{c|}{$\langle S_{2c}\rangle$}\\
		\hline
		Lepton Mode:&$l=\mu,e$&$l=\tau$&$l=\mu,e$&$l=\tau$&$l=\mu,e$&$l=\tau$&$l=\mu,e$&$l=\tau$\\
		\hline
		$[1.1,6.0]$&0.192(41)&-&0.748(59)&-&0.064(14)&-&-0.725(55)&-\\
		$[6.0,8.0]$&0.340(18)&-&0.548(42)&-&0.112(10)&-&-0.539(42)&-\\
		$[11.0,12.5]$&0.475(21)&-&0.367(28)&-&0.158(7)&-&-0.365(29)&-\\
		$[15.0,17.0]$&0.503(8)&0.332(11)&0.330(11)&0.669(21)&0.167(3)&0.040(1)&-0.329(11)&-0.079(4)\\
		\hline
		\hline
		$q^{2}$ bins $(GeV^{2})$& \multicolumn{2}{c|}{$\langle S_{3}\rangle$} &\multicolumn{2}{c|}{$\langle S_{4}\rangle$ }&\multicolumn{2}{c|}{$\langle  S_{5}\rangle$}&\multicolumn{2}{c|}{$\langle  S_{6s}\rangle$}\\
		\hline
		Lepton Mode:&$l=\mu,e$&$l=\tau$&$l=\mu,e$&$l=\tau$&$l=\mu,e$&$l=\tau$&$l=\mu,e$&$l=\tau$\\
		\hline
		$[1.1,6.0]$&-0.012(13)&-&0.150(24)&-&-0.211(59)&-&-0.047(34)&-\\
		$[6.0,8.0]$&-0.011(27)&-&0.247(13)&-&-0.472(33)&-&-0.432(43)&-\\
		$[11.0,12.5]$&-0.045(28)&-&0.254(14)&-&-0.430(17)&-&-0.592(34)&-\\
		$[15.0,17.0]$&-0.200(18)&-0.050(5)&0.297(8)&0.071(3)&-0.275(20)&-0.140(10)&-0.490(37)&-0.251(20)\\
		\hline
		\hline
		$q^{2}$ bins $(GeV^{2})$& \multicolumn{2}{c|}{$10^{2}\times\langle S_{7}\rangle$} &\multicolumn{2}{c|}{$10^{2}\times\langle S_{8}\rangle$}&\multicolumn{2}{c|}{$10^{2}\times\langle S_{9}\rangle$}&&\\
		\hline
		Lepton Mode:&$l=\mu,e$&$l=\tau$&$l=\mu,e$&$l=\tau$&$l=\mu,e$&$l=\tau$&&\\
		\hline
		$[1.1,6.0]$&0.976(99)&-&-0.950(99)&-&0.146(105)&-&&\\
		$[6.0,8.0]$&0.387(30)&-&-0.422(37)&-&0.038(48)&-&&\\
		$[11.0,12.5]$&0.209(21)&-&0.237(22)&-&-0.034(36)&-&&\\
		$[15.0,17.0]$&0.074(7)&0.037(3)&0.113(15)&0.026(4)&-0.097(31)&-0.023(8)&&\\
		\hline
	\end{tabular}
	\caption{$q^{2}$ averaged estimates of the various CP averaged angular observables $S_{i}$ in separate $q^{2}$ bins.}
	\label{table:angular observables in bins}
\end{table}

The $q^{2}$ limits of each bin are chosen as follows:
\begin{itemize}
	\item For the first $q^{2}$ bin, the lower limit should ideally be $4 m_{l}^{2}$. However, in our analysis, we have set it at $q^{2}= 1.1~\mathrm {GeV} ^ {2}$. This is because at a small $q^{2}$ value, the relevant decay amplitude is dominated mostly by the photon pole, and hence by just one Wilson Coefficient $C_{7}^{eff}$. Considering this region in the analysis will not add any new information in comparison to what is already available from the analysis of $b\rightarrow d\gamma$ channel. Additionally, including the small $q^{2}$ region might introduce contributions due to light resonances like $\rho,~\omega$ and $\phi$ \cite{Altmannshofer:2008dz}. Hence we avoid this region. The upper limit of $q^{2}=6.0 GeV^{2}$ is taken to avoid the $J/\psi$ resonance peak at $m_{J/\psi}^{2}=9.6~GeV^{2}$ by a safe margin, and also to stay in accordance to experimental conventions.
	\item For the second $q^{2}$ bin, we continue from the first bin, setting the lower limit at $q^{2}=6.0~GeV^{2}$. As for the upper limit, we fix it at $q^{2}= 8.0~\mathrm {GeV} ^ {2}$ to stay below the $J/\psi$ resonance. A point to be noted here is that instead of having one bin in the range $q^{2}\in (1.1,8.0)~GeV^{2}$, we have divided the $q^{2}$ range into two separate bins. This is because the range $6.0-8.0~GeV^{2}$ lies close to the $J/\psi$ resonance, and the non-local $c\bar{c}$ resonance effects are non-negligible in this region, while the range $1.1-6.0~GeV^{2}$ does not suffer from any such contributions. Hence, mixing the two regions will contaminate the clean predictions in the low $q^{2}$ region.
	\item For the third $q^{2}$ bin, the $q^{2}$ range is chosen to lie in the region between the $J/\psi$ and $\psi(2S)$ resonances. For the lower limit, we set it at $q^{2}=11.0~GeV^{2}$, a value conveniently above $J/\psi$ resonance, while for the upper limit, we set it at $q^{2}=12.5~GeV^{2}$, a value safely below the $\psi(2S)$ resonance peak.
	\item For the fourth $q^{2}$ bin, we choose the lower limit at $q^{2}=15.0~GeV^{2}$, a value greater than the $\psi(2S)$ resonance peak. The upper limit of $q^{2}=17.0~GeV^{2}$ is chosen to be in accordance with experimental convention. For $\tau$ lepton modes, however, the lowest kinematic limit $q^{2}=4m_{l}^{2}=12.62$ $GeV^{2}$. Hence, for this case, our results would involve average values only in the fourth bin, since this is the only bin in which the $q^{2}$ range lies within the physical $q^{2}$ limits.
\end{itemize}

Revisiting table \ref{table:angular observables in bins}, there are a few points that require discussion. Among the observables $S_{1s}$ to $S_{6s}$, the observable $S_{3}$ has a smaller central value compared to the other observables. This is primarily due to partial cancellation between the modulus of transversity amplitudes, as can be seen in Eqn.\eqref{eqn:angular observables}. The cancellation becomes less strong as we move from the first bin to the fourth bin, leading to an increase in the absolute value. 

\begin{figure}[htb!]
	\centering
	\subfloat[\centering]{\includegraphics[width=7.0cm]{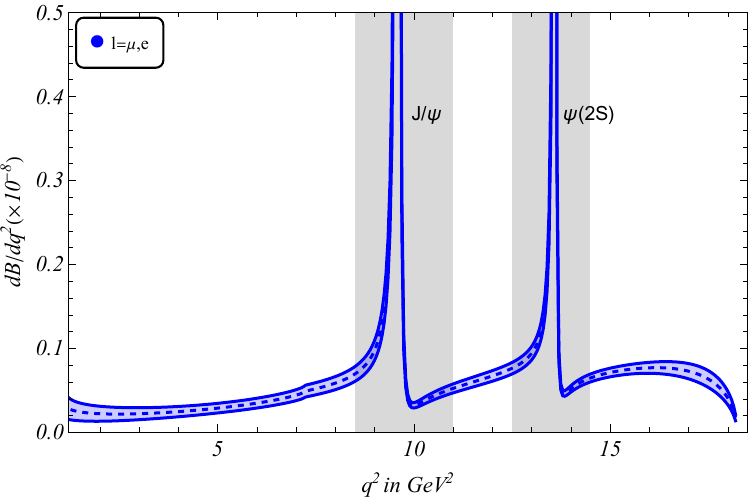}}
	\qquad
	\subfloat[\centering]{\includegraphics[width=7.0cm]{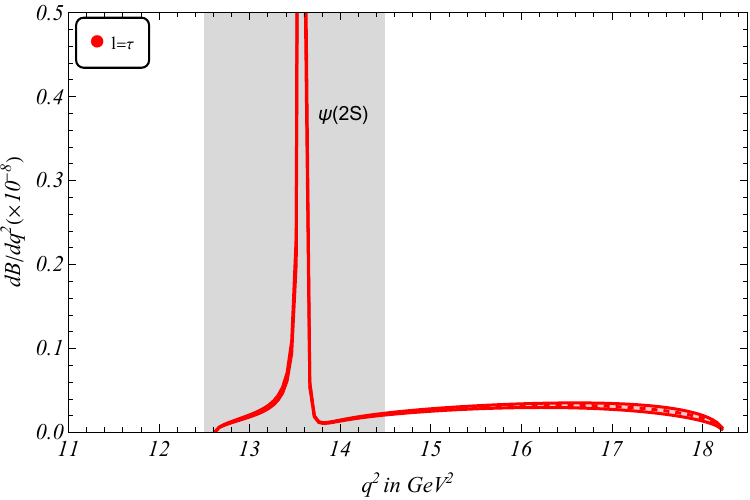}}
	\caption{The $q^{2}$ dependence of differential branching fractions $d\mathcal{B}/dq^{2}$ of $B_{c}^{-}\rightarrow D^{*-}(\rightarrow D^{0}\pi^{-}) \ell^{+}\ell^{-}$. The first plot depicts our result for the light lepton mode, and the second plot depicts our result for the $\tau$ lepton mode, and the gray bands denote the region of $J/\psi$ and $\psi(2S)$ resonances.}
	\label{fig:branching fractions in terms of angular observables}
\end{figure}

Next, we calculate the CP-averaged differential branching ratio by taking the expressions previously mentioned in Eqn.\eqref{eqn:CP averaged branching ratio} and multiplying it with $\tau_{B_{c}}$. The thus obtained $q^{2}$ distribution of the differential branching fractions for $\mu$, $e$ and $\tau$ lepton modes are presented separately in Fig. \ref{fig:branching fractions in terms of angular observables}.\\

Following Fig. \ref{fig:branching fractions in terms of angular observables},
 in Table \ref{table:average decay width in terms of angular observables} we calculate $q^{2}$ averaged values of the branching fractions in the same four $q^{2}$ bins, i.e., $[1.1,6.0]$, $[6.0,8.0]$, $[11.0,12.5]$ and $[15.0,17.0]$ $GeV^{2}$.

\begin{table}[htb!]
	\renewcommand{\arraystretch}{1.3}
	\centering
	\begin{tabular}{|c|c|c|}
		\hline
		$q^{2}$ bins $(GeV^{2})$&$\mathcal{B}(l=e,\mu)$&$\mathcal{B}(l=\tau)$\\
		\hline
		$[1.1,6.0]$&0.130(35)&-\\
		$[6.0,8.0]$&0.092(13)&-\\
		$[11.0,12.5]$&0.095(9)&-\\
		$[15.0,17.0]$&0.151(13)&0.061(4)\\
		\hline
	\end{tabular}
\caption{$q^{2}$ averaged estimates of branching fractions $(\times 10^{-8})$ of $B_{c}^{-}\rightarrow D^{*-}(\rightarrow D^{0}\pi^{-}) \ell^{+}\ell^{-}$ in separate $q^{2}$ bins.}
\label{table:average decay width in terms of angular observables}
\end{table}

We observe a deviation between our results presented in Table \ref{table:average decay width in terms of angular observables} and previous predictions reported in \cite{Ivanov:2024iat}. The primary source of this discrepancy lies in the form factors, specifically $A_{1}$, $A_{2}$ and $V$, which play a dominant role in determining the transversity amplitudes for the light lepton channels. Minor contributions arise from inputs such as $m_{b}$ and $\Lambda_{QCD}$, as well as other hadronic parameters. We have checked our result by comparing $\mathcal{B}(B_{c}^{-}\rightarrow D^{*-}(\rightarrow D^{0}\pi^{-}) \ell^{+}\ell^{-})$ from Table \ref{table:average decay width in terms of angular observables} with an independent calculation of $\mathcal{B}(B_{c}^{-}\rightarrow D^{*-}\ell^{+}\ell^{-})$ using the expression defined in section \ref{subsubsection:decay width and branching fractions} and multiplying it by $\mathcal{B}(D^{*-}\rightarrow D^{0}\pi^{-})$, in each $q^{2}$ bin. We have observed that the two values are in good agreement with each other, thereby validating our results. Furthermore, our analysis allows for a test of lepton flavor universality through the branching fractions in the fourth $q^{2}$ bin, defined as:
\begin{equation}
	R^{\tau\mu}=\frac{\mathcal{B}(B_{c}^{-}\rightarrow D^{*-}(\rightarrow D^{0}\pi^{-}) \tau^{+}\tau^{-}}{\mathcal{B}(B_{c}^{-}\rightarrow D^{*-}(\rightarrow D^{0}\pi^{-}) \ell^{+}\ell^{-}},
\end{equation}
and is found to be $0.408(11)$ in this work.

\begin{figure}[htb!]
	\centering
	\subfloat[\centering]{\includegraphics[width=7.0cm]{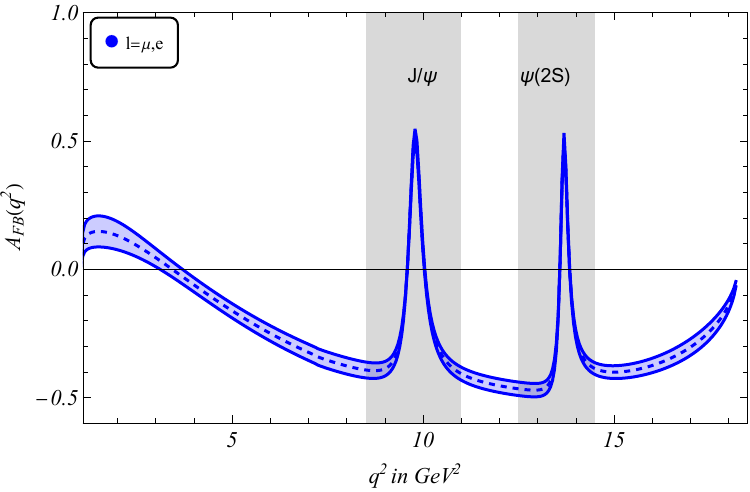}}
	\qquad
	\subfloat[\centering]{\includegraphics[width=7.0cm]{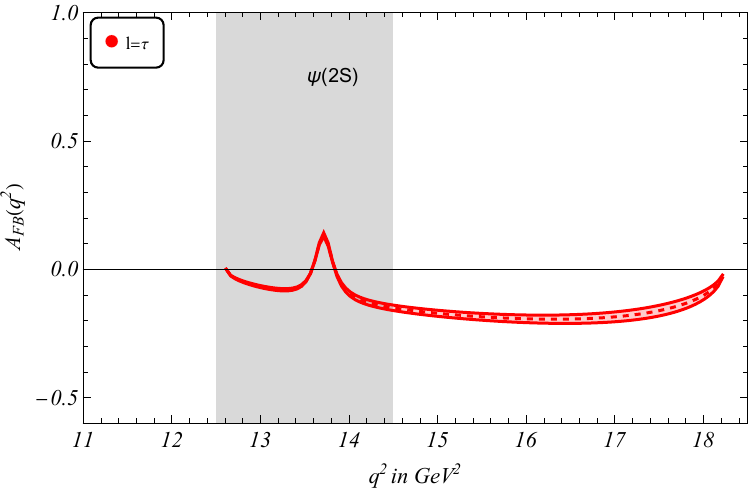}}
	\caption{The $q^{2}$ dependence of lepton forward-backward asymmetry parameter $A_{FB}$ in $B_{c}^{-}\rightarrow D^{*-}(\rightarrow D^{0}\pi^{-}) \ell^{+}\ell^{-}$ channel. The blue plots denote the results for light lepton modes, the red plots denote the results for tau lepton mode, and the gray bands denote the region of $J/\psi$ and $\psi(2S)$ resonances.}
	\label{fig:AFB in terms of angular observables}
\end{figure}

\begin{figure}[htb!]
	\centering
	\subfloat[\centering]{\includegraphics[width=7.0cm]{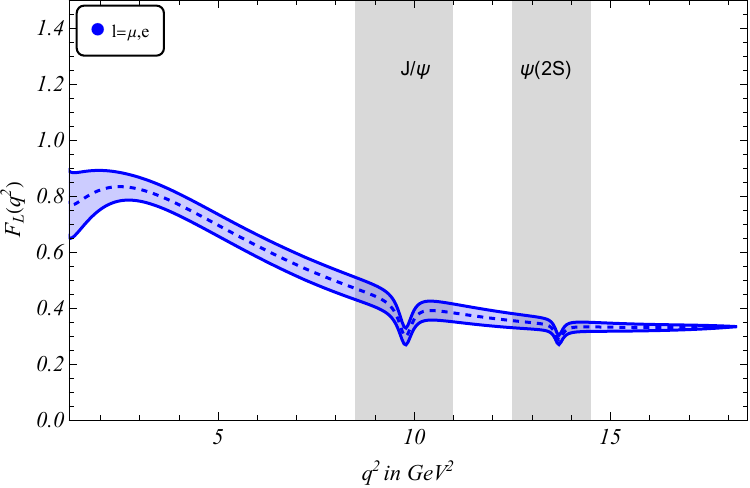}}
	\qquad
	\subfloat[\centering]{\includegraphics[width=7.0cm]{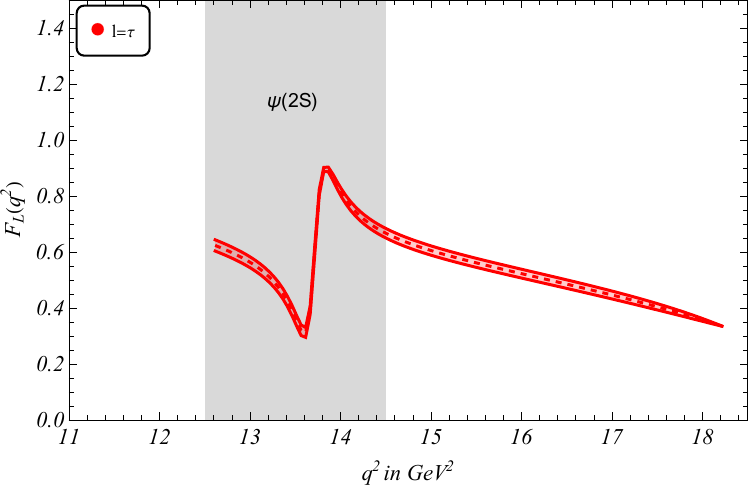}}
	\qquad
	\subfloat[\centering]{\includegraphics[width=7.0cm]{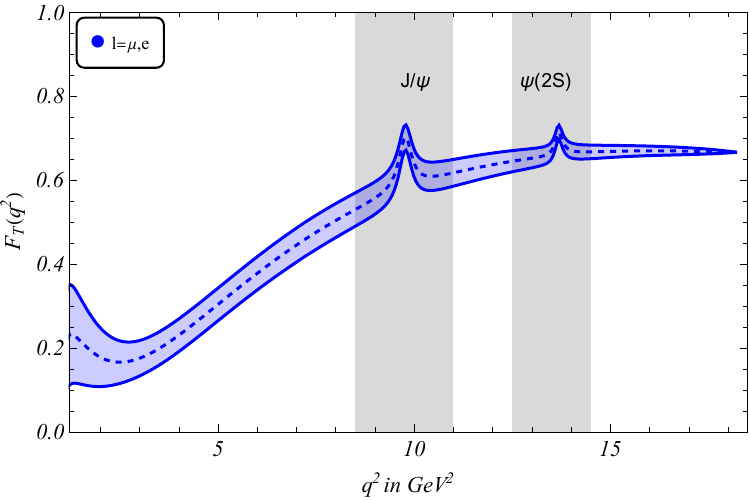}}
	\qquad
	\subfloat[\centering]{\includegraphics[width=7.0cm]{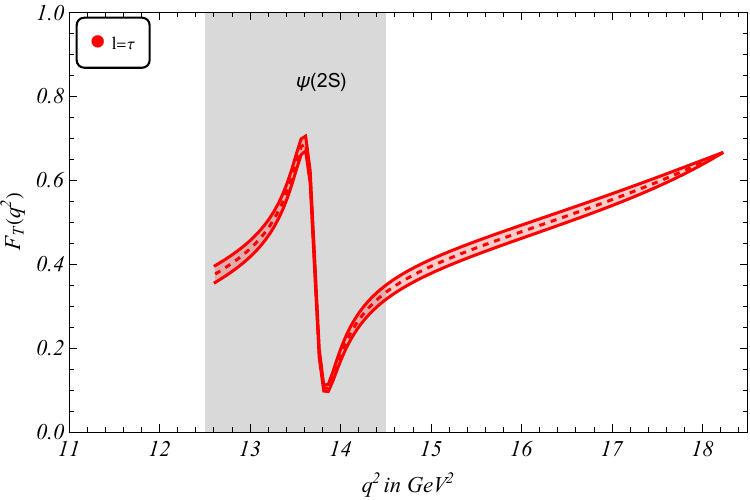}}
	\caption{The $q^{2}$ dependence of $D^{*}$ longitudinal and transverse polarization fractions $F_{L}$ and $F_{T}$ in $B_{c}^{-}\rightarrow D^{*-}(\rightarrow D^{0}\pi^{-}) \ell^{+}\ell^{-}$ channel. The blue plots denote the results for light lepton modes, and the red plots denote the results for tau lepton mode, and the gray bands denote the region of $J/\psi$ and $\psi(2S)$ resonances.}
	\label{fig:FL and FT in terms of angular observables}
\end{figure}

 Having completed the CP-averaged observables and branching fractions, next we turn our attention to some prominent and well-studied observables, namely the forward-backward asymmetry $A_{FB}$, the $D^{*}$ longitudinal and transverse polarization fractions $F_{L}$ and $F_{T}$ respectively. We take the expressions previously discussed in Eqns.\eqref{eqn:forward backward asymmetry} and \eqref{eqn:lepton polarization} in subsection \ref{subsubsection:Angular analysis} and present plots depicting the $q^{2}$ distribution of these observables in Figs.\ref{fig:AFB in terms of angular observables} and \ref{fig:FL and FT in terms of angular observables}.

To facilitate comparison with experimental measurements, we also compute the $q^{2}$ averaged values of these observables. The averaging is performed using the following expression
\begin{equation}
	\langle \mathcal{A}\rangle=\frac{\int_{q_{min}^{2}}^{q_{max}^{2}}\mathcal{A}(q^{2})\left(\frac{d\Gamma}{dq^{2}}+\frac{d\bar{\Gamma}}{dq^{2}}\right)dq^{2}}{\int_{q_{min}^{2}}^{q_{max}^{2}}\left(\frac{d\Gamma}{dq^{2}}+\frac{d\bar{\Gamma}}{dq^{2}}\right)dq^{2}},
\end{equation}
where $\mathcal{A}=A_{FB}$, $F_{L}$ or $F_{T}$. We calculate the average values of these observables in separate $q^{2}$ bins and present them in Table \ref{table:average of physical observables}.
\begin{table}[htb!]
	\renewcommand{\arraystretch}{1.3}
	\centering
	\begin{tabular}{|c|cc|cc|cc|}
		\hline
		$q^{2}$ bins $(GeV^{2})$&\multicolumn{2}{c}{$\langle A_{FB}\rangle$}&\multicolumn{2}{|c|}{$\langle F_{L}\rangle$}&\multicolumn{2}{c|}{$\langle F_{T}\rangle$}\\
		\hline
		Lepton Mode:&$l=\mu,e$&$l=\tau$&$l=\mu,e$&$l=\tau$&$l=\mu,e$&$l=\tau$\\
		\hline
		$[1.1,6.0]$&-0.036(25)&-&0.741(55)&-&0.254(55)&-\\
		$[6.0,8.0]$&-0.326(31)&-&0.550(40)&-&0.452(41)&-\\
		$[11.0,12.5]$&-0.444(26)&-&0.366(29)&-&0.634(29)&-\\
		$[15.0,17.0]$&-0.365(29)&-0.188(15)&0.330(11)&0.522(15)&0.669(10)&0.478(15)\\
		\hline
	\end{tabular}
\caption{$q^{2}$ averaged estimates of forward-backward asymmetry, longitudinal and transverse polarization fractions in separate $q^{2}$ bins.}
\label{table:average of physical observables}
\end{table}

\begin{figure}[htb!]
	\centering
	\subfloat[\centering]{\includegraphics[width=4.5cm]{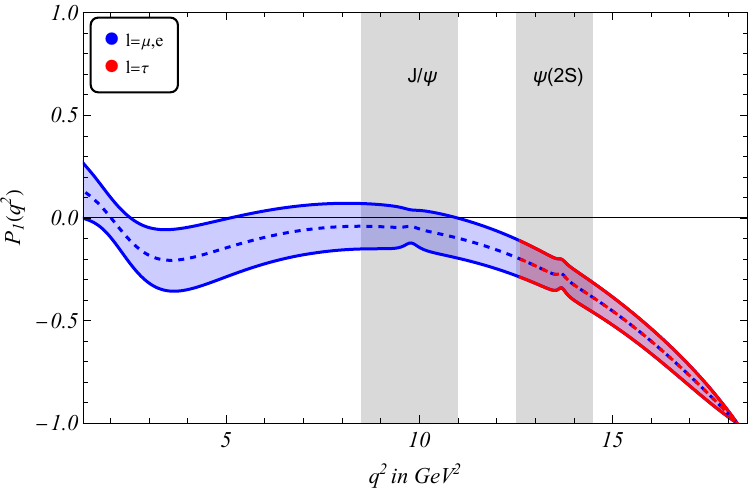}}
	\qquad
	\subfloat[\centering]{\includegraphics[width=4.5cm]{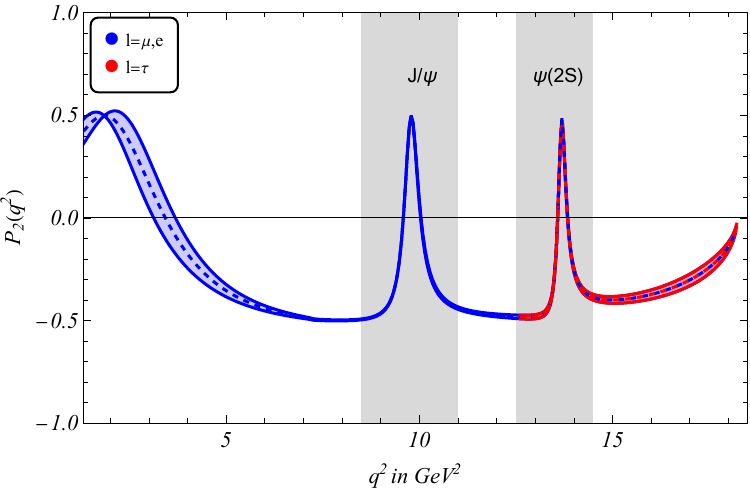}}
	\qquad
	\subfloat[\centering]{\includegraphics[width=4.5cm]{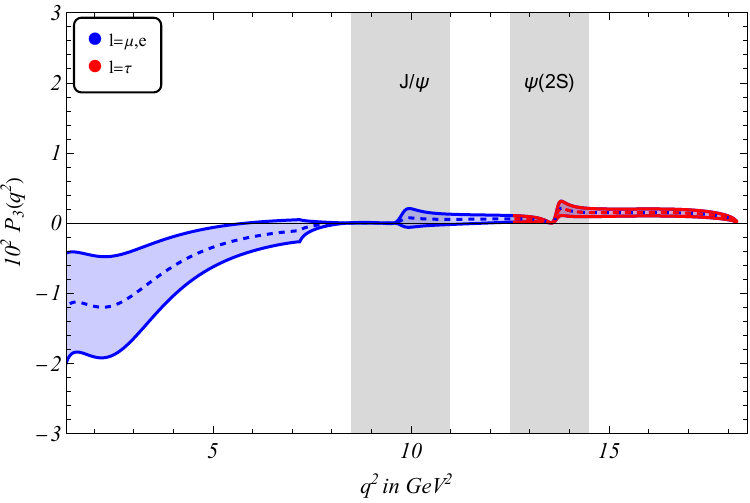}}
	\qquad
	\subfloat[\centering]{\includegraphics[width=4.5cm]{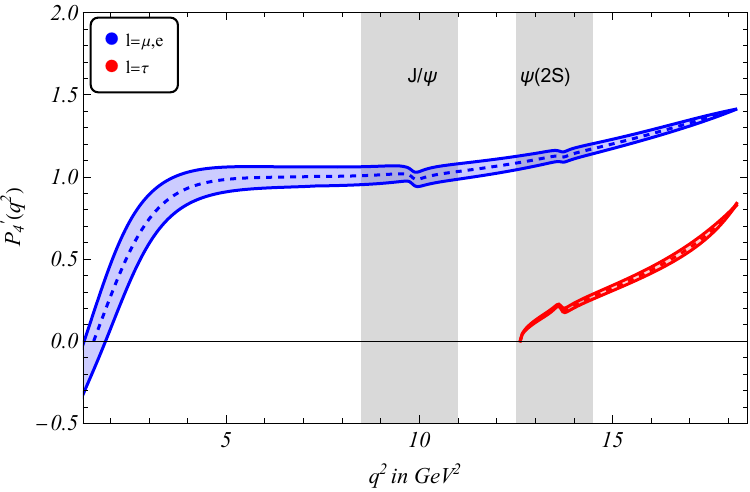}}
	\qquad
	\subfloat[\centering]{\includegraphics[width=4.5cm]{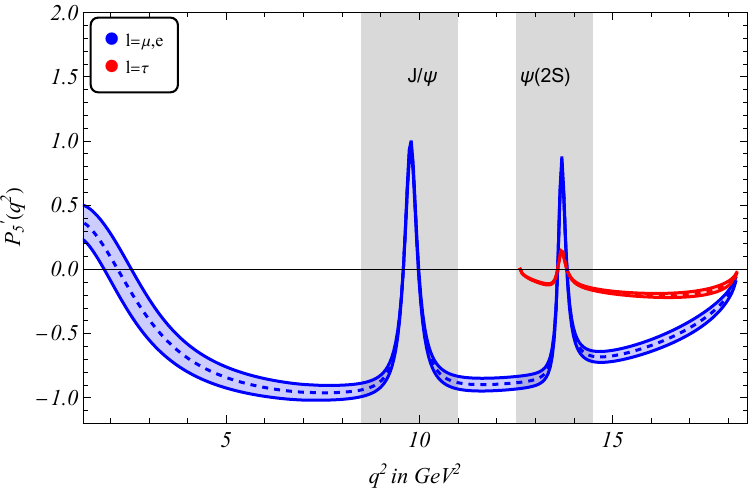}}
	\qquad
	\subfloat[\centering]{\includegraphics[width=4.5cm]{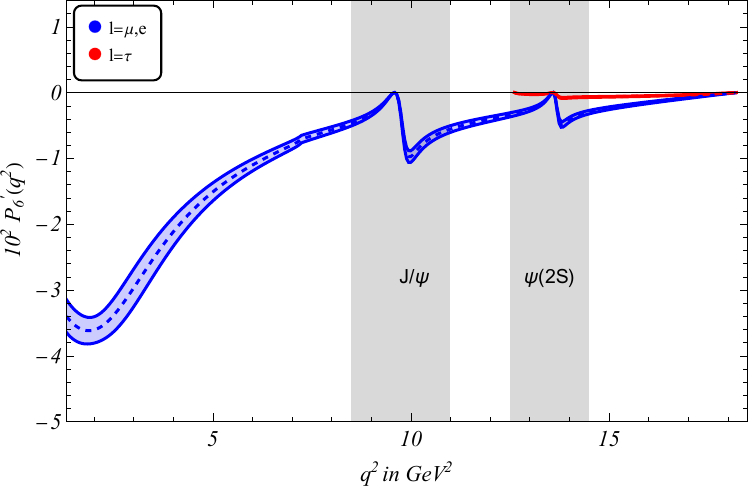}}
	\qquad
	\subfloat[\centering]{\includegraphics[width=4.5cm]{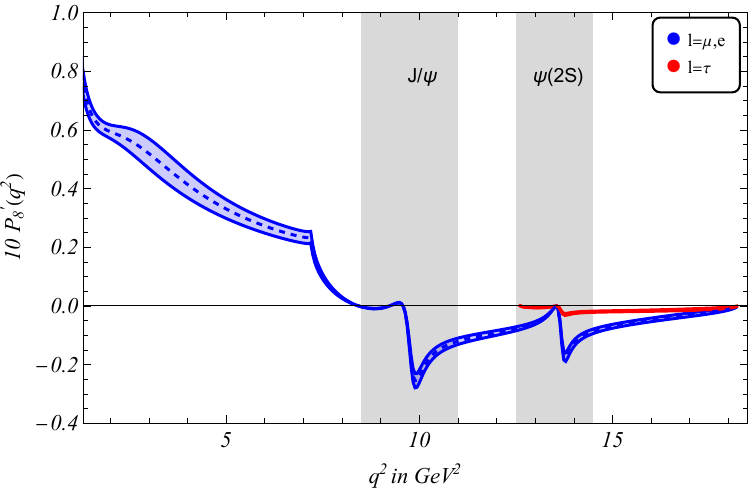}}
	\caption{The $q^{2}$ dependence of clean angular observables $P_{1,2,3}$ and $P_{4,5,6,8}^{'}$. The blue and the red plots depict our results for the light lepton and $\tau$ lepton modes respectively and the gray bands denote the region of $J/\psi$ and $\psi(2S)$ resonances.}
	\label{fig:CP averaged clean angular observables}
\end{figure}

Furthermore, we present predictions for the clean angular observables $P_{1,2,3}$ and $P_{4,5,6,8}^{\prime}$, as defined in Eqn.\eqref{eqn:Clean observables}. These observables are of particular interest due to their reduced sensitivity to hadronic form factors, rendering them comparatively cleaner and less affected by hadronic uncertainties compared to other angular observables. As a result, they serve as powerful probes for exploring potential new physics (NP) effects. The $q^{2}$ dependence of these observables is presented in Fig.\ref{fig:CP averaged clean angular observables}.
In addition, we also present predictions for the average values of these observables in the four $q^{2}$ bins in Table \ref{table:clean observables}. We have verified two of the observables, particularly $P_{2}$ and $P_{5}^{\prime}$ using the relations \cite{LHCb:2020gog}
\begin{equation}
	P_{2}=\frac{2}{3}\cdot\frac{A_{FB}}{1-F_{L}}, \qquad P_{5}^{\prime}=\frac{S_{5}}{\sqrt{F_{L}(1-F_{L})}},
\end{equation}
and have found the results in the table to be in good agreement.

\begin{table}[htb!]
	\renewcommand{\arraystretch}{1.3}
	\centering
	\scriptsize
	\begin{tabular}{|c|cc|cc|cc|cc|}
		\hline
		$q^{2}$ bins $(GeV^{2})$& \multicolumn{2}{c|}{$\langle P_{1}\rangle$} &\multicolumn{2}{c|}{$\langle P_{2}\rangle$ }&\multicolumn{2}{c|}{$10^{2}\times\langle P_{3}\rangle$}&\multicolumn{2}{c|}{}\\
		\hline
		Lepton Mode:&$l=\mu,e$&$l=\tau$&$l=\mu,e$&$l=\tau$&$l=\mu,e$&$l=\tau$&&\\
		\hline
		$[1.1,6.0]$&-0.104(109)&-&-0.134(73)&-&-0.602(411)&-&&\\
		$[6.0,8.0]$&-0.058(116)&-&-0.483(7)&-&-0.080(105)&-&&\\
		$[11.0,12.5]$&-0.154(89)&-&-0.467(8)&-&0.054(61)&-&&\\
		$[15.0,17.0]$&-0.597(64)&-0.623(58)&-0.364(23)&-0.359(23)&0.147(50)&0.147(49)&&\\
		\hline
		\hline
		$q^{2}$ bins $(GeV^{2})$& \multicolumn{2}{c|}{$\langle P_{4}^{'}\rangle$}& \multicolumn{2}{c|}{$\langle P_{5}^{'}\rangle$} &\multicolumn{2}{c|}{$10^{2}\times\langle P_{6}^{'}\rangle$ }&\multicolumn{2}{c|}{$10 \times \langle P_{8}^{'}\rangle$}\\
		\hline
		Lepton Mode:&$l=\mu,e$&$l=\tau$&$l=\mu,e$&$l=\tau$&$l=\mu,e$&$l=\tau$&$l=\mu,e$&$l=\tau$\\
		\hline
		$[1.1,6.0]$&0.745(119)&-&-0.430(79)&-&-4.357(454)&-&0.440(48)&-\\
		$[6.0,8.0]$&1.002(60)&-&-0.770(56)&-&-1.507(122)&-&0.158(14)&-\\
		$[11.0,12.5]$&1.058(43)&-&-0.670(48)&-&-0.854(82)&-&-0.096(8)&-\\
		$[15.0,17.0]$&1.262(24)&0.443(21)&-0.584(48)&-0.200(15)&-0.310(30)&-0.101(10)&-0.048(6)&-0.016(2)\\
		\hline
	\end{tabular}
\caption{$q^{2}$ averaged estimates of clean observables $P_{i}^{(')}$ in separate $q^{2}$ bins.}
\label{table:clean observables}
\end{table}


The results presented in this work provide a comprehensive insight into the angular analysis of the decay $B_{c}^{-}\rightarrow D^{*-}(\rightarrow D^{0}\pi^{-}) \ell^{+}\ell^{-}$ within the SM framework, considering both light and heavy lepton modes. Also, LHCb recently has studied these observables for $B\rightarrow K^{*}\ell^{+}\ell^{-}$ where they have observed a discrepancy of about $3.0~\sigma$ with respect to the SM predictions for $P_{2}$ and $P_{5}^{\prime}$ \cite{LHCb:2020gog} in the third $q^{2}$ bin. Therefore, a study on these observables for our channel in SM is important, as they may play a role when future experimental observations begin to emerge. Any observed deviations from the SM expectations in future data may serve as potential signals of new physics, which can then be investigated through the incorporation of beyond Standard Model scenarios.

\section{Summary and Conclusions}
\label{section:Summary and conclusions}

In this work, we have studied some semileptonic and rare decay modes of the $B_{c}$ meson in the SM. In the framework of pQCD, we have utilized the available lattice inputs on $B\rightarrow D^{(*)}$ and $B_{c}\rightarrow D$ form factors and extracted the shape parameters of the $B_c$ and $D^{(*)}$ meson wave functions, which are sensitive to the pQCD form factors. We obtain the following values of the shape parameters associated with $B_c$ and $B$ meson wave functions:
$$
\omega_{B_{c}} = 1.064(108),\ \ \ \  \omega_{B} = 0.499(140).\ \  
$$
Similarly, for the $D$ and $D^*$ meson wave functions we obtain
$$
  C_{D} = 0.504(52), \ \ \ C_{D^{*}} = 0.492(97),
$$
and 
$$
  \omega_{D} = 0.105(43), \ \ \ \omega_{D^{*}} = 0.099(16).
$$
Furthermore, using these wave functions and the HQSS relations between $B_c\to D$ and $B_c\to D^*$ form factors, we extract the $q^2$ shapes of the $B_c\to D^*$ form factors.

With this information we have then predicted a number of several observables, like branching fractions, lepton flavor universality ratios, and various angular observables, associated with the $B_c^- \rightarrow \bar{D}^{(*)0}\ell^-\bar{\nu}_{\ell}$ and $B_c^- \rightarrow D^{(*)-}\ell^+\ell^-(\nu\bar{\nu})$ decays.

\appendix
\section{Expressions of form factors in PQCD approach}
\label{section:appendix PQCD form factor expressions}
In this appendix, we present the analytical expressions of the form factors already discussed in \ref{subsection:form factors}.
\begin{itemize}
	\item For $B\rightarrow D$ transition the auxiliary form factors $f_{1}(q^{2})$ and $f_{2}(q^{2})$ have the analytic forms \cite{Hu:2019bdf}
\begin{equation}
	\small
\begin{split}
f_{1}(q^{2})=&8\pi m_{B}^{2}C_{F}\int_{0}^{1} dx_{1}dx_{2}\int_{0}^{b_{c}} b_{1}db_{1}b_{2}db_{2}\phi_{B}(x_{1},b_{1})\phi_{D}(x_{2},b_{2})\\
&\Biggl\{[2r(1-rx_{2})]\cdot H_{1}(t_{1})+\biggl[2r(2r_{c}-r)\\
&+x_{1}r\left(-2+2\eta+\sqrt{\eta^{2}-1}-\frac{2\eta}{\sqrt{\eta^{2}-1}}
+\frac{\eta^{2}}{\sqrt{\eta^{2}-1}}\right)\biggr]\cdot H_{2}(t_{2})\Biggr\},
\end{split}
\label{eqn: BD form factors PQCD 1}
\end{equation}
\begin{equation}
	\small
\begin{split}
f_{2}(q^{2})=&8\pi m_{B}^{2}C_{F}\int_{0}^{1} dx_{1}dx_{2}\int_{0}^{b_{c}} b_{1}db_{1}b_{2}db_{2}\phi_{B}(x_{1},b_{1})\phi_{D}(x_{2},b_{2})\\
&\Biggl\{[2-4x_{2}r(1-\eta)]\cdot H_{1}(t_{1})+\left[4r-2r_{c}-x_{1}+\frac{x_{1}}{\sqrt{\eta^{2}-1}}(2-\eta)\right]\cdot H_{2}(t_{2})\Biggr\}.
\end{split}
\label{eqn: BD form factors PQCD}
\end{equation}
\item For $B\rightarrow D^{*}$ transition the axial-vector and vector form factors $A_{0,1,2}(q^{2})$ and $V(q^{2})$ respectively have the analytic forms \cite{Hu:2019bdf}
\begin{equation}
	\small
\begin{split}
A_{0}(q^{2})=&8\pi m_{B}^{2}C_{F}\int_{0}^{1} dx_{1}dx_{2}\int_{0}^{b_{c}} b_{1}db_{1}b_{2}db_{2}\phi_{B}(x_{1},b_{1})\phi_{D^{*}}(x_{2},b_{2})\\
&\Biggl\{[1+r-rx_{2}(2+r-2\eta)]\cdot H_{1}(t_{1})\\
&+\left[r^{2}+r_{c}+\frac{x_{1}}{2}+\frac{\eta x_{1}}{2\sqrt{\eta^{2}-1}}+\frac{rx_{1}}{2\sqrt{\eta^{2}-1}}(1-2\eta(\eta+\sqrt{\eta^{2}-1}))\right]\cdot H_{2}(t_{2})\Biggr\},
\end{split}
\end{equation}
\begin{equation}
\small
\begin{split}
A_{1}(q^{2})=&8\pi m_{B}^{2}C_{F}\int_{0}^{1} dx_{1}dx_{2}\int_{0}^{b_{c}} b_{1}db_{1}b_{2}db_{2}\phi_{B}(x_{1},b_{1})\phi_{D^{*}}(x_{2},b_{2})\frac{r}{1+r}\\
&\Biggl\{2[1+\eta-2rx_{2}+r\eta x_{2}]\cdot H_{1}(t_{1})+\left[2r_{c}+2\eta r-x_{1}\right]\cdot H_{2}(t_{2})\Biggr\},
\end{split}
\end{equation}
\begin{equation}
\small
\begin{split}
A_{2}(q^{2})=&\frac{(1+r)^{2}(\eta-r)}{2r(\eta^{2}-1)}A_{1}(q^{2})\\
&-8\pi m_{B}^{2}C_{F}\int_{0}^{1}dx_{1}dx_{2}\int_{0}^{b_{c}}b_{1}db_{1}b_{2}db_{2}\phi_{B}(x_{1},b_{1})\phi_{D^{*}}(x_{2,b_{2}})\frac{1+r}{\eta^{2}-1}\\
&\times \biggl\lbrace[(1+\eta)(1-r)-rx_{2}(1-2r+\eta(2+r-2\eta))]\cdot H_{1}(t_{1})\\
&+\left[r+r_{c}(\eta-r)-\eta r^{2}+rx_{1}\eta^{2}-\frac{x_{1}}{2}(\eta+r)+x_{1}(\eta r-\frac{1}{2})\sqrt{\eta^{2}-1}\right]\cdot H_{2}(t_{2})\biggr\rbrace,
\end{split}
\end{equation}
\begin{equation}
\small
\begin{split}
V(q^{2})=&8\pi m_{B}^{2}C_{F}\int_{0}^{1}dx_{1}dx_{2}\int_{0}^{b_{c}}b_{1}db_{1}b_{2}db_{2}\phi_{B}(x_{1},b_{1})\phi_{D^{*}}(x_{2,b_{2}})(1+r)\\
&\left\lbrace[1-rx_{2}]\cdot H_{1}(t_{1})+\left[r+\frac{x_{1}}{2\sqrt{\eta^{2}-1}}\right]\cdot H_{2}(t_{2})\right\rbrace.
\end{split}
\label{eqn: BDStar form factors PQCD}
\end{equation}
\item For $B_{c}\rightarrow D$ transition the auxiliary form factors $f_{1}(q^{2})$ and $f_{2}(q^{2})$ and the tensor form factor $F_{T}(q^{2})$ have the following analytic forms \cite{PhysRevD.90.094018}
\begin{equation}
\begin{split}
f_{1}(q^{2})=&16\pi m_{B_{c}}^{2}r C_{F}\int_{0}^{1}dx_{1}dx_{2}\int_{0}^{b_{c}}b_{1}db_{1}b_{2}db_{2}\phi_{B_{c}}(x_{1},b_{1})\phi_{D}(x_{2},b_{2})\\
&\times \biggl\lbrace [1-rx_{2}]\cdot H_{1}(t_{1})-[r+2x_{1}(1-\eta)]\cdot H_{2}(t_{2})\biggr\rbrace,\\
f_{2}(q^{2})=&16\pi m_{B_{c}}^{2} C_{F}\int_{0}^{1}dx_{1}dx_{2}\int_{0}^{b_{c}}b_{1}db_{1}b_{2}db_{2}\phi_{B_{c}}(x_{1},b_{1})\phi_{D}(x_{2},b_{2})\\
&\times \biggl\lbrace [1-2rx_{2}(1-\eta)]\cdot H_{1}(t_{1})+[2r-x_{1}]\cdot H_{2}(t_{2})\biggr\rbrace,\\
F_{T}(q^{2})=&8\pi m_{B_{c}}^{2}(1+r) C_{F}\int_{0}^{1}dx_{1}dx_{2}\int_{0}^{b_{c}}b_{1}db_{1}b_{2}db_{2}\phi_{B_{c}}(x_{1},b_{1})\phi_{D}(x_{2},b_{2})\\
&\times \biggl\lbrace [1-rx_{2}]\cdot H_{1}(t_{1})+[2r-x_{1}]\cdot H_{2}(t_{2})\biggr\rbrace.\\
\end{split}
\label{eqn: BcD form factors PQCD}
\end{equation}
\item For $B_{c}\rightarrow D^{*}$ transition the axial-vector and vector form factors $A_{0,1,2}(q^{2})$ and $V(q^{2})$ respectively have the analytic forms \cite{PhysRevD.90.094018}

\begin{equation}
	\small
\begin{split}
A_{0}(q^{2})=&8\pi m_{B_{c}}^{2}C_{F}\int_{0}^{1} dx_{1}dx_{2}\int_{0}^{b_{c}} b_{1}db_{1}b_{2}db_{2}\phi_{B_{c}}(x_{1},b_{1})\phi_{D^{*}}(x_{2},b_{2})\\
&\Biggl\{[1-rx_{2}(r-2\eta)+r(1-2x_{2})]\cdot H_{1}(t_{1})+\left[r^{2}+x_{1}(1-2r\eta)\right]\cdot H_{2}(t_{2})\Biggr\},\\
A_{1}(q^{2})=&16\pi m_{B_{c}}^{2}C_{F}\frac{r}{1+r} \int_{0}^{1} dx_{1}dx_{2}\int_{0}^{b_{c}} b_{1}db_{1}b_{2}db_{2}\phi_{B_{c}}(x_{1},b_{1})\phi_{D^{*}}(x_{2},b_{2})\\
&\Biggl\{[1+rx_{2}\eta-2rx_{2}+\eta]\cdot H_{1}(t_{1})+\left[r\eta-x_{1}\right]\cdot H_{2}(t_{2})\Biggr\},\\
A_{2}(q^{2})=&\frac{(1+r)^{2}(\eta-r)}{2r(\eta^{2}-1)}A_{1}(q^{2})\\
&-8\pi m_{B_{c}}^{2}C_{F}\frac{1+r}{\eta^{2}-1}\int_{0}^{1}dx_{1}dx_{2}\int_{0}^{b_{c}}b_{1}db_{1}b_{2}db_{2}\phi_{B_{c}}(x_{1},b_{1})\phi_{D^{*}}(x_{2,b_{2}})\\
&\times \biggl\lbrace[\eta(1-r^{2}x_{2})-rx_{2}(1-2\eta^{2}-2r)+(1-r)-r\eta(1+2x_{2})]\cdot H_{1}(t_{1})\\
&+\left[r(1-x_{1}+2x_{1}\eta^{2})-\eta(r^{2}+x_{1})\right]\cdot H_{2}(t_{2})\biggr\rbrace,\\
V(q^{2})=&8\pi m_{B_{c}}^{2}C_{F}\int_{0}^{1}dx_{1}dx_{2}(1+r)\int_{0}^{b_{c}}b_{1}db_{1}b_{2}db_{2}\phi_{B_{c}}(x_{1},b_{1})\phi_{D^{*}}(x_{2,b_{2}})\\
&\biggl\lbrace[1-rx_{2}]\cdot H_{1}(t_{1})+r\cdot H_{2}(t_{2})\biggr\rbrace,
\end{split}
\label{eqn: BcDStar AV form factors PQCD}
\end{equation}
and the tensor form factors $T_{1,2,3}$ have the analytic forms \cite{PhysRevD.90.094018}
\begin{equation}
\begin{split}
T_{1}(q^{2})=&8\pi m_{B_{c}}^{2}C_{F}\int_{0}^{1}dx_{1}dx_{2}\int_{0}^{b_{c}}b_{1}db_{1}b_{2}db_{2}\phi_{B_{c}}(x_{1},b_{1})\phi_{D^{*}}(x_{2,b_{2}})\\
&\biggl\lbrace[1+r(1-x_{2}(2+r-2\eta))]\cdot H_{1}(t_{1})+[r(1-x_{1})]\cdot H_{2}(t_{2})\biggr\rbrace,\\
T_{2}(q^{2})=&16\pi m_{B_{c}}^{2}C_{F}\frac{r}{1-r^{2}}\int_{0}^{1}dx_{1}dx_{2}(1+r)\int_{0}^{b_{c}}b_{1}db_{1}b_{2}db_{2}\phi_{B_{c}}(x_{1},b_{1})\phi_{D^{*}}(x_{2,b_{2}})\\
&\biggl\lbrace[(1-r)(1+\eta)+2rx_{2}(r-\eta)+rx_{2}(2\eta^{2}-r\eta-1)]\cdot H_{1}(t_{1})\\
&+[r(1+x_{1})\eta-r^{2}-x_{1}]\cdot H_{2}(t_{2})\biggr\rbrace,\\
T_{3}(q^{2})=&\frac{r+\eta}{r}\frac{1-r^{2}}{2(\eta^{2}-1)}T_{2}(q^{2})-\frac{1-r^{2}}{\eta^{2}-1}\\
&\times 8\pi m_{B_{c}}^{2}C_{F}\int_{0}^{1}dx_{1}dx_{2}\int_{0}^{b_{c}}b_{1}db_{1}b_{2}db_{2}\phi_{B_{c}}(x_{1},b_{1})\phi_{D^{*}}(x_{2,b_{2}})\\
&\times \biggl\lbrace[1+rx_{2}(\eta-2)+\eta]\cdot H_{1}(t_{1})+\left[x_{1}\eta-r\right]\cdot H_{2}(t_{2})\biggr\rbrace.
\end{split}
\label{eqn: BcDStar Tensor form factors PQCD}
\end{equation}
\end{itemize}

In all the above expressions $r=m/m_{B_{(c)}}$, $r_{b(c)}=m_{b(c)}/m_{B_{(c)}}$, and 
\begin{equation}
\label{eqn:Evolution function}
H_{i}(t_{i})=\alpha_{s}(t_{i})h_{i}(x_{1},x_{2},b_{1},b_{2})\exp[-S_{ab}(t_{i})],
\end{equation}
with $\alpha_{s}(t)$, $h_{i}(x_{1},x_{2},b_{1},b_{2})$ and $S_{ab}(t)$  representing the strong coupling constant evaluated at scale $t$, the hard kernel and the Sudakov factor respectively. The detailed expressions for these terms has been presented in the following appendix.
\section{Scales and relevant functions in the hard kernel}
\label{section:appendix hard functions}

In this appendix, we present analytic expressions for the hard functions and scales that were introduced in the previous appendix. The hard kernel $h_{i}$ comes from the Fourier transform of virtual quark and gluon propagators
\begin{equation}
\begin{split}
h_{1}(x_{1},x_{2},b_{1},b_{2})=&K_{0}(\beta_{1}b_{1})\left[\theta(b_{1}-b_{2})I_{0}(\alpha_{1}b_{2})K_{0}(\alpha_{1}b_{1})+\theta(b_{2}-b_{1})I_{0}(\alpha_{1}b_{1})K_{0}(\alpha_{1}b_{2})\right]S_{t}(x_{2}),\\
h_{2}(x_{1},x_{2},b_{1},b_{2})=&K_{0}(\beta_{2}b_{2})\left[\theta(b_{1}-b_{2})I_{0}(\alpha_{2}b_{2})K_{0}(\alpha_{2}b_{1})+\theta(b_{2}-b_{1})I_{0}(\alpha_{2}b_{1})K_{0}(\alpha_{2}b_{2})\right]S_{t}(x_{1}),
\end{split}
\end{equation}
where $K_{0}$ and $I_{0}$ are the modified Bessel functions, and
\begin{equation}
\begin{split}
\alpha_{1}=&m_{B}\sqrt{x_{2}r\eta^{+}},\\
\alpha_{2}=&m_{B}\sqrt{x_{1}r\eta^{+}-r^{2}+r_{c}^{2}},\\
\beta_{1,2}=&m_{B}\sqrt{x_{1}x_{2}r\eta^{+}},
\end{split}
\end{equation}
for form factors of $B\rightarrow D^{(*)}$ form factors in Eqns.\eqref{eqn: BD form factors PQCD 1}-\eqref{eqn: BDStar form factors PQCD}, taken from \cite{Hu:2019bdf,Wang:2012lrc}, and
\begin{equation}
\begin{split}
\alpha_{1}=&m_{B_{c}}\sqrt{2rx_{2}\eta+r_{b}^{2}-1-r^{2}x_{2}^{2}},\\
\alpha_{2}=&m_{B_{c}}\sqrt{rx_{1}\eta^{+}+r_{u(d)}^{2}-r^{2}},\\
\beta_{1,2}=&m_{B_{c}}\sqrt{x_{1}x_{2}r\eta^{+}-r^{2}x_{2}^{2}},
\end{split}
\end{equation}
for form factors of $B_{c}\rightarrow D^{(*)}$ decays shown in Eqns. \eqref{eqn: BcD form factors PQCD}-\eqref{eqn: BcDStar Tensor form factors PQCD}, taken from \cite{Hu:2019qcn}. In addition to the hard kernels in Eqn.\eqref{eqn:Evolution function}, the Sudakov factors $S_{ab}(t)$ evaluated in the modified PQCD framework have been taken from \cite{PhysRevD.97.113001}. The hard scale $t$ is chosen to be the maximum of the virtuality of internal momentum transition in the hard amplitudes
\begin{equation}
t_{1}=\text{max}(\alpha_{1},1/b_{1},1/b_{2}), \qquad t_{2}=\text{max}(\alpha_{2},1/b_{1},1/b_{2}),
\end{equation}
and the jet function $S_{t}(x)$ has the same form as Eqn \eqref{eqn:jet function}.
\section{Synthetic data for form factors}
\label{section:appendix synthetic data of form factors}

In this appendix, we present the synthetic data for the $B_{c}\rightarrow D^{(*)}$ form factors at discrete $w$ values, or more fundamentally, $q^{2}$ values. These synthetic data have been used in the respective chi-square minimizations that we have done in this work.

\begin{itemize}

\item In Table \ref{table:input BcD} we present lattice data for $B_{c}\rightarrow D$ form factors at $w=1.0,~1.15$ and $1.3$. These data points have been used to extract the pole expansion parameters in Table \ref{table:Soft functions at w values}.
	\begin{table}[htb!]
		\renewcommand{\arraystretch}{1.3}
	\centering
	\footnotesize
	\begin{tabular}{|c|c|cccccc|}
		\hline
		\textbf{Form Factors} & \textbf{Value} &  &    & \textbf{Correlation} & &  &  \\ 
		\cline{3-8}
		\textbf{at} $\boldsymbol{w}$ & \textbf{from HPQCD}&$F_{+}(1.0)$ &$F_{+}(1.15)$ & $F_{+}(1.3)$&$F_{0}(1.0)$ & $F_{0}(1.15)$& $F_{0}(1.3)$\\
		\hline 
		$F_{+}(1.0)$ & 1.454(254)  &1.0 & 0.926 & 0.776 & -0.001 & 0.059 & 0.009 \\ 
		
		$F_{+}(1.15)$ & 0.851(120)  & & 1.0 & 0.951 & 0.039 & 0.086 & 0.081 \\ 
		
		$F_{+}(1.3)$ & 0.549(164)  & &  & 1.0 & 0.101 & 0.137 & 0.200 \\ 
		
		$F_{0}(1.0)$ & 0.612(33)  & &  &  & 1.0 & 0.877 & 0.746 \\ 
		
		$F_{0}(1.15)$ & 0.417(23)  & &  &  &  & 1.0 & 0.914 \\ 
		
		$F_{0}(1.3)$ & 0.315(20)  & &  &  &  &  & 1.0 \\ 
		\hline
	\end{tabular} 
	\captionof{table}{HPQCD data for $B_{c}\rightarrow D$ form factors, along-with their correlation}	
	\label{table:input BcD}
\end{table}
\item In Tables \ref{table: inputs BcD FT}, \ref{table: inputs BcDStar A012V} and \ref{table: inputs BcDStar T123}  we present synthetic data for rest of the $B_{c}\rightarrow D^{(*)}$ form factors generated using the soft function parameters extracted in Table \ref{table:Soft function parameters}. For the $B_{c}\rightarrow D^{*}$ form factors, we have taken each form factor uncorrelated to the other ones. This was done because all the form factors were derived from the same universal functions $\Sigma_{1}$ and $\Sigma_{2}$, making the full correlation matrix positive semidefinite, leading to numerical instability during the chi-square minimization. To avoid this, we have considered all the form factors to be uncorrelated.  

\begin{table}[htb!]
	\renewcommand{\arraystretch}{1.3}
	\centering
	\footnotesize
	\begin{tabular}{|c|c|ccc|}
		\hline
		\textbf{Form}&\textbf{Value}&&\textbf{Correlation}&\\
		\cline{3-5}
		\textbf{Factors}&\textbf{Obtained}&$F_{T}(1.0)$&$F_{T}(1.15)$&$F_{T}(1.3)$\\
		\hline
		$F_{T}(1.0)$&1.951(470)&1.0&0.924&0.792\\
		$F_{T}(1.15)$&1.184(253)&&1.0&0.964\\
		$F_{T}(1.3)$&0.794(173)&&&1.0\\
		\hline
	\end{tabular}
	\caption{Synthetic data for $B_{c}\rightarrow D$ tensor form factor $F_{T}$ obtained using soft function parameters extracted in this work.}
	\label{table: inputs BcD FT}
\end{table}

\begin{table}[htb!]
	\renewcommand{\arraystretch}{1.3}
	\centering
	\scriptsize
	\begin{tabular}{|c|c|ccc|c|c|ccc|}
		\hline
		\textbf{Form}&\textbf{Value}&&\textbf{Correlation}&&\textbf{Form}&\textbf{Value}&&\textbf{Correlation}&\\
		\cline{3-5} \cline{8-10}
		\textbf{Factors}&\textbf{Obtained}&$A_{0}(1.0)$&$A_{0}(1.15)$&$A_{0}(1.3)$&\textbf{Factors}&\textbf{Obtained}&$A_{1}(1.0)$&$A_{1}(1.15)$&$A_{1}(1.3)$\\
		\hline
		$A_{0}(1.0)$&2.350(259)&1.0&0.848&-0.071&$A_{1}(1.0)$&0.624(34)&1.0&0.807&0.605\\
		$A_{0}(1.15)$&1.231(79)&&1.0&0.338&$A_{1}(1.15)$&0.397(28)&&1.0&0.925\\
		$A_{0}(1.3)$&0.694(50)&&&1.0&$A_{1}(1.3)$&0.278(30)&&&1.0\\
		\hline
			\textbf{Form}&\textbf{Value}&&\textbf{Correlation}&&\textbf{Form}&\textbf{Value}&&\textbf{Correlation}&\\
			\cline{3-5} \cline{8-10}
		\textbf{Factors}&\textbf{Obtained}&$A_{2}(1.0)$&$A_{2}(1.15)$&$A_{2}(1.3)$&\textbf{Factors}&\textbf{Obtained}&$V(1.0)$&$V(1.15)$&$V(1.3)$\\
		\hline
		$A_{2}(1.0)$&0.742(179)&1.0&0.924&0.792&$V(1.0)$&2.059(496)&1.0&0.925&0.792\\
		$A_{2}(1.15)$&0.450(96)&&1.0&0.964&$V(1.15)$&1.249(267)&&1.0&0.964\\
		$A_{2}(1.3)$&0.302(66)&&&1.0&$V(1.3)$&0.838(173)&&&1.0\\
		\hline
	\end{tabular}
	\caption{Synthetic data for $B_{c}\rightarrow D^{*}$ axial-vector and vector form factors $A_{0,1,2}$ and $V$ obtained using soft function parameters extracted in this work.}
	\label{table: inputs BcDStar A012V}
\end{table}

\begin{table}[htb!]
	\renewcommand{\arraystretch}{1.3}
	\centering
	\scriptsize
	\begin{tabular}{|c|c|ccc|c|c|ccc|}
		\hline
		\textbf{Form}&\textbf{Value}&&\textbf{Correlation}&&\textbf{Form}&\textbf{Value}&&\textbf{Correlation}&\\
		\cline{3-5} \cline{8-10}
		\textbf{Factors}&\textbf{Obtained}&$T_{1}(1.0)$&$T_{1}(1.15)$&$T_{1}(1.3)$&\textbf{Factors}&\textbf{Obtained}&$T_{2}(1.0)$&$T_{2}(1.15)$&$T_{2}(1.3)$\\
		\hline
		$T_{1}(1.0)$&1.473(255)&1.0&0.927&0.775&$T_{2}(1.0)$&0.624(34)&1.0&0.862&0.710\\
		$T_{1}(1.15)$&0.861(120)&&1.0&0.951&$T_{2}(1.15)$&0.439(23)&&1.0&0.899\\
		$T_{1}(1.3)$&0.555(70)&&&1.0&$T_{2}(1.3)$&0.339(20)&&&1.0\\
		\hline
		\textbf{Form}&\textbf{Value}&&\textbf{Correlation}&&&&&&\\
		\cline{3-5}
		\textbf{Factors}&\textbf{Obtained}&$T_{3}(1.0)$&$T_{3}(1.15)$&$T_{3}(1.3)$&&&&&\\
		\hline
		$T_{3}(1.0)$&1.647(499)&1.0&0.923&0.795&&&&&\\
		$T_{3}(1.15)$&1.033(287)&&1.0&0.966&&&&&\\
		$T_{3}(1.3)$&0.715(209)&&&1.0&&&&&\\
		\hline
	\end{tabular}
	\caption{Synthetic data for $B_{c}\rightarrow D^{*}$ tensor form factors $T_{1,2,3}$ obtained using soft function parameters extracted in this work.}
	\label{table: inputs BcDStar T123}
\end{table}
\end{itemize}

\section{Correlation matrices}
\label{section:appendix correlation matrices}
In this appendix, we present the correlation matrices obtained between the extracted parameters in each of the chi-square optimizations.

\begin{itemize}
	\item In Table \ref{table: correlation LCDA BcD} we present the correlation matrix between the LCDA shape parameters of $B$, $B_{c}$, and $D^{(*)}$ mesons, whose values we had extracted and showcased in Table \ref{table:extracted value of Bc and D shape parameters}.
\begin{table}[htbp]
	\centering
	\setlength{\tabcolsep}{1.5pt} 
	\renewcommand{\arraystretch}{1.3}
	\scriptsize 

	\resizebox{\textwidth}{!}{%
		
		\begin{tabular}{|c|ccccccccccccccc|}
			\hline
			 & $\omega_{B_{c}}$ & $\omega_{B}$ & $C_{D}$ & $C_{D^{*}}$ & $\omega_{D}$ & $\omega_{D^{*}}$ & $m_{b}$ & $m_{c}$ & $\delta_{f_{1}}^{B\to D}$ & $\delta_{f_{2}}^{B\to D}$ & $\delta_{A_{0}}^{B\to D^{*}}$ & $\delta_{A_{1}}^{B\to D^{*}}$ & $\delta_{V}^{B\to D^{*}}$ & $\delta_{f_{1}}^{B_{c}\to D}$ & $\delta_{f_{2}}^{B_{c}\to D}$ \\
			\hline 
			
			$\omega_{B_{c}}$ & 1.0 & 0.165 & 0.264 & -0.141 & -0.162 & 0.048 & 0.050 & -0.005 & -0.050 & -0.029 & 0.040 & -0.060 & -0.005 & 0.004 & -0.005 \\
			
			$\omega_{B}$ & & 1.0 & -0.121 & -0.085 & -0.108 & -0.190 & 0.026 & -0.110 & 0.025 & -0.039 & 0.014 & 0.069 & 0.100 & -0.075 & 0.100 \\
			
			$C_{D}$ & & & 1.0 & -0.133 & 0.014 & 0.563 & 0.051 & 0.424 & -0.312 & -0.073 & 0.075 & -0.293 & -0.392 & 0.294 & -0.393 \\
			
			$C_{D^{*}}$ & & & & 1.0 & 0.751 & -0.090 & -0.129 & 0.026 & -0.169 & -0.032 & 0.089 & 0.023 & -0.139 & 0.104 & -0.139 \\
			
			$\omega_{D}$ & & & & & 1.0 & 0.168 & -0.057 & -0.014 & -0.329 & -0.090 & 0.262 & -0.092 & -0.271 & 0.203 & -0.271 \\
			
			$\omega_{D^{*}}$ & & & & & & 1.0 & 0.161 & 0.281 & -0.588 & -0.088 & 0.454 & -0.153 & -0.715 & 0.535 & -0.715 \\
			
			$m_{b}$ & & & & & & & 1.0 & 0.254 & 0.104 & 0.044 & 0.008 & 0.330 & -0.139 & 0.103 & -0.138 \\
			
			$m_{c}$ & & & & & & & & 1.0 & 0.010 & 0.137 & 0.149 & -0.061 & -0.449 & 0.335 & -0.449 \\
			
			$\delta_{f_{1}}^{B\to D}$ & & & & & & & & & 1.0 & -0.037 & -0.529 & 0.466 & 0.757 & -0.567 & 0.758 \\
			
			$\delta_{f_{2}}^{B\to D}$ & & & & & & & & & & 1.0 & -0.190 & 0.099 & 0.091 & -0.068 & 0.089 \\
			
			$\delta_{A_{0}}^{B\to D^{*}}$ & & & & & & & & & & & 1.0 & -0.294 & -0.549 & 0.412 & -0.551 \\
			
			$\delta_{A_{1}}^{B\to D^{*}}$ & & & & & & & & & & & & 1.0 & 0.390 & -0.293 & 0.392 \\

            $\delta_{V}^{B\to D^{*}}$ & & & & & & & & & & & & & 1.0 & -0.749 & 1.000 \\

            $\delta_{f_{1}}^{B_{c}\to D}$ & & & & & & & & & & & & & & 1.0 & -0.746 \\

            $\delta_{f_{2}}^{B_{c}\to D}$ & & & & & & & & & & & & & & & 1.0 \\
			\hline
			
		\end{tabular} 
		}
		\caption{Correlation matrix between extracted LCDA shape parameters of $B_{c}$, $B$ and $D^{(*)}$ mesons along with other nuisance parameters.}
		\label{table: correlation LCDA BcD}
\end{table}

\item In Table \ref{table:correlation form factors BcDStar} we present the correlation matrix between the nine $B_{c}\rightarrow D^{(*)}$ semileptonic form factors at $q^{2}=0$ predicted in Table \ref{table:form factor predictions BcDStar}.

\begin{table}[htb!]
	\centering
	\setlength{\tabcolsep}{6pt} 
	\renewcommand{\arraystretch}{1.3}
	
	\begin{tabular}{|c|ccccccccc|}
		\hline
		& $F_{+}(0)$ & $F_{T}(0)$ & $A_{0}(0)$ & $A_{1}(0)$ & $A_{2}(0)$ & $V(0)$ & $T_{1}(0)$ & $T_{2}(0)$ & $T_{3}(0)$ \\
		\hline 
		$F_{+}(0)$ & 1.0 & 0.456 & 0.379 & 0.394 & 0.402 & 0.410 & 0.393 & 0.393 & 0.453 \\ 
		$F_{T}(0)$ & & 1.0 & 0.295 & 0.307 & 0.313 & 0.320 & 0.306 & 0.306 & 0.353 \\ 
		$A_{0}(0)$ & & & 1.0 & 0.273 & 0.278 & 0.284 & 0.273 & 0.273 & 0.312 \\ 
		$A_{1}(0)$ & & & & 1.0 & 0.288 & 0.294 & 0.282 & 0.282 & 0.322 \\ 
		$A_{2}(0)$ & & & & & 1.0 & 0.300 & 0.287 & 0.287 & 0.329 \\ 
		$V(0)$ & & & & & & 1.0 & 0.293 & 0.293 & 0.336 \\ 
		$T_{1}(0)$ & & & & & & & 1.0 & 0.281 & 0.322 \\ 
		$T_{2}(0)$ & & & & & & & & 1.0 & 0.322 \\ 
		$T_{3}(0)$ & & & & & & & & & 1.0 \\ 
		\hline
	\end{tabular} 
	
	\caption{Correlation Matrix between predicted $B_{c}\rightarrow D^{(*)}$ semileptonic form factors at $q^{2}=0$ using modified pQCD.}
	\label{table:correlation form factors BcDStar}
\end{table}

\item In Table \ref{table:correlation BCL parameters BcD} we present the correlation matrix between the BCL parameters of $B_{c}\rightarrow D$ form factors extracted in Table \ref{table:BCL Bc to D}.

\begin{table}[htb!]
	\centering
	\renewcommand{\arraystretch}{1.3}
	\setlength{\tabcolsep}{3pt}
	\begin{tabular}{|c|cccccccc|}
		\hline
		& $a_{0}^{F_{+}}$ & $a_{1}^{F_{+}}$ & $a_{2}^{F_{+}}$ & $a_{1}^{F_{0}}$ & $a_{2}^{F_{0}}$ & $a_{0}^{F_{T}}$ & $a_{1}^{F_{T}}$ & $a_{2}^{F_{T}}$ \\
		\hline
		$a_{0}^{F_{+}}$ & 1.0 & -0.771 & -0.317 & 0.126 & 0.069 & 0.033 & -0.003 & 0.000 \\
		$a_{1}^{F_{+}}$ & & 1.0 & -0.014 & 0.272 & 0.219 & 0.059 & -0.006 & 0.000 \\
		$a_{2}^{F_{+}}$ & & & 1.0 & 0.253 & 0.353 & 0.054 & -0.005 & 0.000 \\
		$a_{1}^{F_{0}}$ & & & & 1.0 & 0.692 & 0.166 & -0.017 & 0.000 \\
		$a_{2}^{F_{0}}$ & & & & & 1.0 & 0.147 & -0.015 & 0.000 \\
		$a_{0}^{F_{T}}$ & & & & & & 1.0 & -0.502 & -0.163 \\
		$a_{1}^{F_{T}}$ & & & & & & & 1.0 & -0.598 \\
		$a_{2}^{F_{T}}$ & & & & & & & & 1.0 \\
		\hline
	\end{tabular}
	\caption{Correlation between extracted BCL parameters of $B_{c}\rightarrow D$ form factors.}
	\label{table:correlation BCL parameters BcD}
\end{table}

\item In Table \ref{table:correlation BCL parameters BcDStar} we present the correlation matrix between the BCL parameters of $B_{c}\rightarrow D^{*}$ form factors extracted in Table \ref{table:BCL Bc to D*}.
\begin{table}[htb!]
	\centering
	\setlength{\tabcolsep}{1.5pt} 
	\renewcommand{\arraystretch}{1.3}
	\scriptsize 

	\resizebox{\textwidth}{!}{%
	\begin{tabular}{|c|ccccccccccccccccccc|}
		\hline
        &$a_{0}^{A_{0}}$&$a_{1}^{A_{0}}$&$a_{2}^{A_{0}}$&$a_{0}^{A_{1}}$&$a_{1}^{A_{1}}$&$a_{2}^{A_{1}}$&$a_{1}^{A_{2}}$&$a_{2}^{A_{2}}$&$a_{0}^{V}$&$a_{1}^{V}$&$a_{2}^{V}$&$a_{0}^{T_{1}}$&$a_{1}^{T_{1}}$&$a_{2}^{T_{1}}$&$a_{1}^{T_{2}}$&$a_{2}^{T_{2}}$&$a_{0}^{T_{3}}$&$a_{1}^{T_{3}}$&$a_{2}^{T_{3}}$\\
		\hline
		$a_{0}^{A_{0}}$ & 1.0 & 0.295 & -0.206 & 0.016 & 0.229 & 0.278 & 0.031 & 0.041 & 0.040 & 0.085 & 0.075 & 0.089 & 0.063 & 0.070 & 0.162 & 0.153 & -0.003 & 0.064 & 0.059 \\
		$a_{1}^{A_{0}}$ & & 1.0 & -0.295 & -0.002 & 0.043 & 0.054 & -0.237 & -0.163 & 0.005 & 0.009 & 0.008 & 0.010 & 0.007 & 0.008 & 0.019 & 0.018 & 0.000 & 0.008 & 0.007 \\
		$a_{2}^{A_{0}}$ & & & 1.0 & 0.013 & 0.240 & 0.292 & -0.125 & -0.068 & 0.040 & 0.084 & 0.074 & 0.089 & 0.063 & 0.070 & 0.162 & 0.152 & -0.003 & 0.064 & 0.058 \\
		$a_{0}^{A_{1}}$ & & & & 1.0 & 0.185 & -0.456 & 0.018 & 0.189 & 0.003 & 0.005 & 0.005 & 0.006 & 0.004 & 0.004 & 0.010 & 0.010 & -0.000 & 0.004 & 0.004 \\
		$a_{1}^{A_{1}}$ & & & & & 1.0 & 0.705 & 0.565 & 0.756 & 0.035 & 0.074 & 0.065 & 0.078 & 0.055 & 0.061 & 0.142 & 0.134 & -0.002 & 0.056 & 0.051 \\
		$a_{2}^{A_{1}}$ & & & & & & 1.0 & 0.525 & 0.478 & 0.043 & 0.090 & 0.079 & 0.094 & 0.067 & 0.074 & 0.172 & 0.162 & -0.003 & 0.068 & 0.062 \\
		$a_{1}^{A_{2}}$ & & & & & & & 1.0 & 0.531 & 0.007 & 0.020 & 0.016 & 0.019 & 0.014 & 0.015 & 0.034 & 0.032 & -0.001 & 0.014 & 0.012 \\
		$a_{2}^{A_{2}}$ & & & & & & & & 1.0 & -0.153 & 0.368 & -0.130 & -0.015 & 0.062 & -0.020 & 0.024 & 0.017 & -0.001 & 0.014 & 0.013 \\
		$a_{0}^{V}$ & & & & & & & & & 1.0 & -0.387 & -0.610 & 0.559 & -0.750 & 0.579 & 0.214 & 0.289 & -0.000 & 0.010 & 0.009 \\
		$a_{1}^{V}$ & & & & & & & & & & 1.0 & -0.141 & -0.057 & 0.141 & -0.067 & 0.023 & 0.008 & -0.001 & 0.021 & 0.019 \\
		$a_{2}^{V}$ & & & & & & & & & & & 1.0 & -0.412 & 0.628 & -0.437 & -0.105 & -0.170 & -0.001 & 0.018 & 0.017 \\
		$a_{0}^{T_{1}}$ & & & & & & & & & & & & 1.0 & -0.221 & -0.094 & 0.300 & 0.576 & -0.003 & 0.032 & 0.016 \\
		$a_{1}^{T_{1}}$ & & & & & & & & & & & & & 1.0 & -0.314 & 0.410 & 0.313 & 0.002 & 0.002 & 0.020 \\
		$a_{2}^{T_{1}}$ & & & & & & & & & & & & & & 1.0 & 0.570 & 0.337 & 0.005 & -0.008 & 0.026 \\
		$a_{1}^{T_{2}}$ & & & & & & & & & & & & & & & 1.0 & 0.873 & 0.004 & 0.014 & 0.047 \\
		$a_{2}^{T_{2}}$ & & & & & & & & & & & & & & & & 1.0 & -0.067 & 0.348 & -0.088 \\
		$a_{0}^{T_{3}}$ & & & & & & & & & & & & & & & & & 1.0 & -0.203 & -0.812 \\
		$a_{1}^{T_{3}}$ & & & & & & & & & & & & & & & & & & 1.0 & -0.291 \\
		$a_{2}^{T_{3}}$ & & & & & & & & & & & & & & & & & & & 1.0 \\
		\hline
	\end{tabular}
	}
	\caption{Correlation between extracted BCL parameters of $B_{c}\rightarrow D^{*}$ form factors.}
	\label{table:correlation BCL parameters BcDStar}
\end{table}

\end{itemize}

\bibliographystyle{JHEP}
\bibliography{biblio}

@article{CDF:1998ihx,
	author = "Abe, F. and others",
	collaboration = "CDF",
	title = "{Observation of the $B_c$ meson in $p\bar{p}$ collisions at $\sqrt{s} = 1.8$ TeV}",
	eprint = "hep-ex/9805034",
	archivePrefix = "arXiv",
	reportNumber = "FERMILAB-PUB-98-157-E, CDF-PUB-BOTTOM-CDFR-4496, LS-6941",
	doi = "10.1103/PhysRevLett.81.2432",
	journal = "Phys. Rev. Lett.",
	volume = "81",
	pages = "2432--2437",
	year = "1998"
}

@article{PepeAltarelli:2008yyl,
	author = "Pepe Altarelli, Monica and Teubert, Frederic",
	editor = "Kane, Gordon and Pierce, Aaron",
	title = "{$B$ Physics at LHCb}",
	eprint = "0802.1901",
	archivePrefix = "arXiv",
	primaryClass = "hep-ph",
	doi = "10.1142/S0217751X08042791",
	journal = "Int. J. Mod. Phys. A",
	volume = "23",
	pages = "5117--5136",
	year = "2008"
}

@article{CDF:1999uew,
	author = "Affolder, T. and others",
	collaboration = "CDF",
	title = "{Search for the flavor-changing neutral current decays $B^+ \to \mu^+ \mu^- K^+$ and $B^0 \to \mu^+ \mu^- K^{*0}$}",
	eprint = "hep-ex/9905004",
	archivePrefix = "arXiv",
	reportNumber = "FERMILAB-PUB-99-138-E",
	doi = "10.1103/PhysRevLett.83.3378",
	journal = "Phys. Rev. Lett.",
	volume = "83",
	pages = "3378--3383",
	year = "1999"
}

@article{Belle:2001oey,
	author = "Abe, Kazuo and others",
	collaboration = "Belle",
	title = "{Observation of the decay $B \to K \ell^{+} \ell^{-}$}",
	eprint = "hep-ex/0109026",
	archivePrefix = "arXiv",
	reportNumber = "KEK-PREPRINT-2001-118, BELLE-PREPRINT-2001-13, DPNU-01-30",
	doi = "10.1103/PhysRevLett.88.021801",
	journal = "Phys. Rev. Lett.",
	volume = "88",
	pages = "021801",
	year = "2002"
}

@article{Belle:2003ivt,
	author = "Ishikawa, A. and others",
	editor = "Cheung, H. W. K. and Pratt, T. S.",
	collaboration = "Belle",
	title = "{Observation of B ---\ensuremath{>} K* l+ l-}",
	eprint = "hep-ex/0308044",
	archivePrefix = "arXiv",
	reportNumber = "BELLE-CONF-0351",
	doi = "10.1103/PhysRevLett.91.261601",
	journal = "Phys. Rev. Lett.",
	volume = "91",
	pages = "261601",
	year = "2003"
}

@article{Belle:2009zue,
	author = "Wei, J. -T. and others",
	collaboration = "Belle",
	title = "{Measurement of the Differential Branching Fraction and Forward-Backward Asymmetry for $B \to K^{(*)}\ell^+\ell^-$}",
	eprint = "0904.0770",
	archivePrefix = "arXiv",
	primaryClass = "hep-ex",
	reportNumber = "BELLE-PREPRINT-2009-7, KEK-PREPRINT-2008-56",
	doi = "10.1103/PhysRevLett.103.171801",
	journal = "Phys. Rev. Lett.",
	volume = "103",
	pages = "171801",
	year = "2009"
}

@article{Belle:2016fev,
	author = "Wehle, S. and others",
	collaboration = "Belle",
	title = "{Lepton-Flavor-Dependent Angular Analysis of $B\to K^\ast \ell^+\ell^-$}",
	eprint = "1612.05014",
	archivePrefix = "arXiv",
	primaryClass = "hep-ex",
	reportNumber = "BELLE-PREPRINT-2016-15, KEK-PREPRINT-2016-54",
	doi = "10.1103/PhysRevLett.118.111801",
	journal = "Phys. Rev. Lett.",
	volume = "118",
	number = "11",
	pages = "111801",
	year = "2017"
}

@article{BELLE:2019xld,
	author = "Choudhury, S. and others",
	collaboration = "BELLE",
	title = "{Test of lepton flavor universality and search for lepton flavor violation in $B \rightarrow K\ell \ell$ decays}",
	eprint = "1908.01848",
	archivePrefix = "arXiv",
	primaryClass = "hep-ex",
	reportNumber = "BELLE-CONF-1904, Belle Preprint 2020-11, KEK Preprint 2020-12",
	doi = "10.1007/JHEP03(2021)105",
	journal = "JHEP",
	volume = "03",
	pages = "105",
	year = "2021"
}

@article{Belle:2019oag,
	author = "Abdesselam, A. and others",
	collaboration = "Belle",
	title = "{Test of Lepton-Flavor Universality in ${B\to K^\ast\ell^+\ell^-}$ Decays at Belle}",
	eprint = "1904.02440",
	archivePrefix = "arXiv",
	primaryClass = "hep-ex",
	reportNumber = "BELLE-CONF-1901, Belle Preprint 2020-14, KEK Preprint 2020-16",
	doi = "10.1103/PhysRevLett.126.161801",
	journal = "Phys. Rev. Lett.",
	volume = "126",
	number = "16",
	pages = "161801",
	year = "2021"
}

@article{BaBar:2003szi,
	author = "Aubert, Bernard and others",
	collaboration = "BaBar",
	title = "{Evidence for the rare decay $B \to K^* \ell^+ \ell^-$ and measurement of the $B \to K \ell^+ \ell^-$ branching fraction}",
	eprint = "hep-ex/0308042",
	archivePrefix = "arXiv",
	reportNumber = "SLAC-PUB-10132, BABAR-PUB-03-021",
	doi = "10.1103/PhysRevLett.91.221802",
	journal = "Phys. Rev. Lett.",
	volume = "91",
	pages = "221802",
	year = "2003"
}

@article{BaBar:2008jdv,
	author = "Aubert, Bernard and others",
	collaboration = "BaBar",
	title = "{Direct CP, Lepton Flavor and Isospin Asymmetries in the Decays $B \to K^{(*)} \ell^{+} \ell^{-}$}",
	eprint = "0807.4119",
	archivePrefix = "arXiv",
	primaryClass = "hep-ex",
	reportNumber = "BABAR-PUB-08-022, SLAC-PUB-13322",
	doi = "10.1103/PhysRevLett.102.091803",
	journal = "Phys. Rev. Lett.",
	volume = "102",
	pages = "091803",
	year = "2009"
}

@article{BaBar:2012mrf,
	author = "Lees, J. P. and others",
	collaboration = "BaBar",
	title = "{Measurement of Branching Fractions and Rate Asymmetries in the Rare Decays $B \to K^{(*)} l^+ l^-$}",
	eprint = "1204.3933",
	archivePrefix = "arXiv",
	primaryClass = "hep-ex",
	reportNumber = "SLAC-PUB-14957, BABAR-PUB-12-002",
	doi = "10.1103/PhysRevD.86.032012",
	journal = "Phys. Rev. D",
	volume = "86",
	pages = "032012",
	year = "2012"
}

@article{CMS:2015bcy,
	author = "Khachatryan, Vardan and others",
	collaboration = "CMS",
	title = "{Angular analysis of the decay $B^0 \to K^{*0} \mu^+ \mu^-$ from pp collisions at $\sqrt  s = 8$ TeV}",
	eprint = "1507.08126",
	archivePrefix = "arXiv",
	primaryClass = "hep-ex",
	reportNumber = "CMS-BPH-13-010, CERN-PH-EP-2015-178",
	doi = "10.1016/j.physletb.2015.12.020",
	journal = "Phys. Lett. B",
	volume = "753",
	pages = "424--448",
	year = "2016"
}

@article{LHCb:2012juf,
	author = "Aaij, R and others",
	collaboration = "LHCb",
	title = "{Differential branching fraction and angular analysis of the $B^{+} \rightarrow K^{+}\mu^{+}\mu^{-}$ decay}",
	eprint = "1209.4284",
	archivePrefix = "arXiv",
	primaryClass = "hep-ex",
	reportNumber = "LHCB-PAPER-2012-024, CERN-PH-EP-2012-263",
	doi = "10.1007/JHEP02(2013)105",
	journal = "JHEP",
	volume = "02",
	pages = "105",
	year = "2013"
}

@article{LHCb:2014vgu,
	author = "Aaij, Roel and others",
	collaboration = "LHCb",
	title = "{Test of lepton universality using $B^{+}\rightarrow K^{+}\ell^{+}\ell^{-}$ decays}",
	eprint = "1406.6482",
	archivePrefix = "arXiv",
	primaryClass = "hep-ex",
	reportNumber = "CERN-PH-EP-2014-140, LHCB-PAPER-2014-024",
	doi = "10.1103/PhysRevLett.113.151601",
	journal = "Phys. Rev. Lett.",
	volume = "113",
	pages = "151601",
	year = "2014"
}

@article{LHCb:2017avl,
	author = "Aaij, R. and others",
	collaboration = "LHCb",
	title = "{Test of lepton universality with $B^{0} \rightarrow K^{*0}\ell^{+}\ell^{-}$ decays}",
	eprint = "1705.05802",
	archivePrefix = "arXiv",
	primaryClass = "hep-ex",
	reportNumber = "LHCB-PAPER-2017-013, CERN-EP-2017-100",
	doi = "10.1007/JHEP08(2017)055",
	journal = "JHEP",
	volume = "08",
	pages = "055",
	year = "2017"
}

@article{LHCb:2021trn,
	author = "Aaij, Roel and others",
	collaboration = "LHCb",
	title = "{Test of lepton universality in beauty-quark decays}",
	eprint = "2103.11769",
	archivePrefix = "arXiv",
	primaryClass = "hep-ex",
	reportNumber = "LHCb-PAPER-2021-004, CERN-EP-2021-042",
	doi = "10.1038/s41567-023-02095-3",
	journal = "Nature Phys.",
	volume = "18",
	number = "3",
	pages = "277--282",
	year = "2022",
	note = "[Addendum: Nature Phys. 19, (2023)]"
}

@article{LHCb:2013ghj,
	author = "Aaij, R and others",
	collaboration = "LHCb",
	title = "{Measurement of Form-Factor-Independent Observables in the Decay $B^{0} \to K^{*0} \mu^+ \mu^-$}",
	eprint = "1308.1707",
	archivePrefix = "arXiv",
	primaryClass = "hep-ex",
	reportNumber = "LHCB-PAPER-2013-037, CERN-PH-EP-2013-146",
	doi = "10.1103/PhysRevLett.111.191801",
	journal = "Phys. Rev. Lett.",
	volume = "111",
	pages = "191801",
	year = "2013"
}

@article{Azizi:2008vv,
	author = "Azizi, K. and Falahati, F. and Bashiry, V. and Zebarjad, S. M.",
	title = "{Analysis of the Rare B(c) ---\ensuremath{>} D*(s,d) l+ l- Decays in QCD}",
	eprint = "0806.0583",
	archivePrefix = "arXiv",
	primaryClass = "hep-ph",
	doi = "10.1103/PhysRevD.77.114024",
	journal = "Phys. Rev. D",
	volume = "77",
	pages = "114024",
	year = "2008"
}

@article{Kiselev:2002vz,
	author = "Kiselev, V. V.",
	title = "{Exclusive decays and lifetime of $B_c$ meson in QCD sum rules}",
	eprint = "hep-ph/0211021",
	archivePrefix = "arXiv",
	month = "11",
	year = "2002"
}

@article{Geng:2001vy,
	author = "Geng, C. Q. and Hwang, Chien-Wen and Liu, C. C.",
	title = "{Study of rare $B^+_{c} \to D_{d,s}$ + lepton anti-lepton decays}",
	eprint = "hep-ph/0110376",
	archivePrefix = "arXiv",
	doi = "10.1103/PhysRevD.65.094037",
	journal = "Phys. Rev. D",
	volume = "65",
	pages = "094037",
	year = "2002"
}

@article{PhysRevD.90.094018,
	title = {Semileptonic decays ${B}_{c}^{+}\ensuremath{\rightarrow}{D}_{(s)}^{(*)}({l}^{+}{\ensuremath{\nu}}_{l},{l}^{+}{l}^{\ensuremath{-}},\ensuremath{\nu}\overline{\ensuremath{\nu}})$ in the perturbative QCD approach},
	author = {Wang, Wen-Fei and Yu, Xin and L\"u, Cai-Dian and Xiao, Zhen-Jun},
	journal = {Phys. Rev. D},
	volume = {90},
	issue = {9},
	pages = {094018},
	numpages = {9},
	year = {2014},
	month = {Nov},
	publisher = {American Physical Society},
	doi = {10.1103/PhysRevD.90.094018},
	url = {https://link.aps.org/doi/10.1103/PhysRevD.90.094018}
}

@article{Rui:2014tpa,
	author = "Rui, Zhou and Zou, Zhi-Tian",
	title = "{S-wave ground state charmonium decays of $B_c$ mesons in the perturbative QCD approach}",
	eprint = "1407.5550",
	archivePrefix = "arXiv",
	primaryClass = "hep-ph",
	doi = "10.1103/PhysRevD.90.114030",
	journal = "Phys. Rev. D",
	volume = "90",
	number = "11",
	pages = "114030",
	year = "2014"
}

@article{Rui:2018kqr,
	author = "Rui, Zhou and Zhang, Jie and Zhang, Li-Li",
	title = "{Semileptonic decays of $B_c$ meson to $P$-wave charmonium states}",
	eprint = "1806.00796",
	archivePrefix = "arXiv",
	primaryClass = "hep-ph",
	doi = "10.1103/PhysRevD.98.033007",
	journal = "Phys. Rev. D",
	volume = "98",
	number = "3",
	pages = "033007",
	year = "2018"
}

@article{Rui:2017pre,
	author = "Rui, Zhou",
	title = "{Probing the $P$-wave charmonium decays of $B_c$ meson}",
	eprint = "1712.08928",
	archivePrefix = "arXiv",
	primaryClass = "hep-ph",
	doi = "10.1103/PhysRevD.97.033001",
	journal = "Phys. Rev. D",
	volume = "97",
	number = "3",
	pages = "033001",
	year = "2018"
}

@article{Hu:2019bdf,
	author = "Hu, Xue-Qing and Jin, Su-Ping and Xiao, Zhen-Jun",
	title = "{Semileptonic decays $B/B_s \to (D^{(*)},D_s^{(*)}) l \nu_l$ in the PQCD approach with the lattice QCD input}",
	eprint = "1912.03981",
	archivePrefix = "arXiv",
	primaryClass = "hep-ph",
	doi = "10.1088/1674-1137/44/5/053102",
	journal = "Chin. Phys. C",
	volume = "44",
	number = "5",
	pages = "053102",
	year = "2020"
}

@article{ParticleDataGroup:2022pth,
	author = "Workman, R. L. and others",
	collaboration = "Particle Data Group",
	title = "{Review of Particle Physics}",
	doi = "10.1093/ptep/ptac097",
	journal = "PTEP",
	volume = "2022",
	pages = "083C01",
	year = "2022"
}

@article{Buchalla:1995vs,
	author = "Buchalla, Gerhard and Buras, Andrzej J. and Lautenbacher, Markus E.",
	title = "{Weak decays beyond leading logarithms}",
	eprint = "hep-ph/9512380",
	archivePrefix = "arXiv",
	reportNumber = "SLAC-PUB-7009, SLAC-PUB-95-7009, MPI-PH-95-104, TUM-T31-100-95, FERMILAB-PUB-95-305-T",
	doi = "10.1103/RevModPhys.68.1125",
	journal = "Rev. Mod. Phys.",
	volume = "68",
	pages = "1125--1144",
	year = "1996"
}

@article{Wang:2012ab,
	author = "Wang, Wen-Fei and Xiao, Zhen-Jun",
	title = "{The semileptonic decays $B/B_s \to (\pi, K)(\ell^+\ell^-,\ell\nu,\nu\bar{\nu})$ in the perturbative QCD approach beyond the leading-order}",
	eprint = "1207.0265",
	archivePrefix = "arXiv",
	primaryClass = "hep-ph",
	doi = "10.1103/PhysRevD.86.114025",
	journal = "Phys. Rev. D",
	volume = "86",
	pages = "114025",
	year = "2012"
}

@article{Biswas:2021cyd,
	author = "Biswas, Aritra and Nandi, Soumitra",
	title = "{A closer look at observables from exclusive semileptonic $B\to(\pi,\rho)\ell\nu_{\ell}$ decays}",
	eprint = "2105.01732",
	archivePrefix = "arXiv",
	primaryClass = "hep-ph",
	doi = "10.1007/JHEP09(2021)127",
	journal = "JHEP",
	volume = "09",
	pages = "127",
	year = "2021"
}

@article{Sakaki:2013bfa,
	author = "Sakaki, Yasuhito and Tanaka, Minoru and Tayduganov, Andrey and Watanabe, Ryoutaro",
	title = "{Testing leptoquark models in $\bar B \to D^{(*)} \tau \bar\nu$}",
	eprint = "1309.0301",
	archivePrefix = "arXiv",
	primaryClass = "hep-ph",
	reportNumber = "OU-HET-791, KEK-TH-1660, OU-HET 791",
	doi = "10.1103/PhysRevD.88.094012",
	journal = "Phys. Rev. D",
	volume = "88",
	number = "9",
	pages = "094012",
	year = "2013"
}

@article{Ali:1999mm,
	author = "Ali, Ahmed and Ball, Patricia and Handoko, L. T. and Hiller, G.",
	title = "{A Comparative study of the decays $B \to$ ($K$, $K^{*)} \ell^+ \ell^-$ in standard model and supersymmetric theories}",
	eprint = "hep-ph/9910221",
	archivePrefix = "arXiv",
	reportNumber = "SLAC-PUB-8269, DESY-99-146, CERN-TH-99-298, LNF-99-026-P",
	doi = "10.1103/PhysRevD.61.074024",
	journal = "Phys. Rev. D",
	volume = "61",
	pages = "074024",
	year = "2000"
}

@article{Barakat:2001ef,
	author = "Barakat, Thabit",
	title = "{Fourth generation effects on the rare B ---\ensuremath{>} K* neutrino anti-neutrino decay}",
	eprint = "hep-ph/0105116",
	archivePrefix = "arXiv",
	doi = "10.1088/1367-2630/4/1/325",
	journal = "New J. Phys.",
	volume = "4",
	pages = "25",
	year = "2002"
}

@article{Altmannshofer:2008dz,
	author = "Altmannshofer, Wolfgang and Ball, Patricia and Bharucha, Aoife and Buras, Andrzej J. and Straub, David M. and Wick, Michael",
	title = "{Symmetries and Asymmetries of $B \to K^{*} \mu^{+} \mu^{-}$ Decays in the Standard Model and Beyond}",
	eprint = "0811.1214",
	archivePrefix = "arXiv",
	primaryClass = "hep-ph",
	reportNumber = "IPPP-08-58, DCPT-08-116, TUM-HEP-696-08",
	doi = "10.1088/1126-6708/2009/01/019",
	journal = "JHEP",
	volume = "01",
	pages = "019",
	year = "2009"
}

@article{Soni:2020bvu,
	author = "Soni, Nakul R. and Issadykov, Aidos and Gadaria, Akshay N. and Patel, Janaki J. and Pandya, Jignesh N.",
	title = "{Rare $b \rightarrow d$ decays in covariant confined quark model}",
	eprint = "2008.07202",
	archivePrefix = "arXiv",
	primaryClass = "hep-ph",
	doi = "10.1140/epja/s10050-022-00685-y",
	journal = "Eur. Phys. J. A",
	volume = "58",
	number = "3",
	pages = "39",
	year = "2022"
}

@article{Chen:2001zc,
	author = "Chen, Chuan-Hung and Geng, C. Q.",
	title = "{Baryonic rare decays of Lambda(b) ---\ensuremath{>} Lambda lepton+ lepton-}",
	eprint = "hep-ph/0106193",
	archivePrefix = "arXiv",
	doi = "10.1103/PhysRevD.64.074001",
	journal = "Phys. Rev. D",
	volume = "64",
	pages = "074001",
	year = "2001"
}

@article{Li:2023mrj,
	author = "Li, Yu-Shuai and Liu, Xiang",
	title = "{Angular distribution of the FCNC process Bc\textrightarrow{}Ds*(\textrightarrow{}Ds\ensuremath{\pi})\ensuremath{\ell}+\ensuremath{\ell}-}",
	eprint = "2309.08191",
	archivePrefix = "arXiv",
	primaryClass = "hep-ph",
	doi = "10.1103/PhysRevD.108.093005",
	journal = "Phys. Rev. D",
	volume = "108",
	number = "9",
	pages = "093005",
	year = "2023"
}

@article{Jin:2020jtu,
	author = "Jin, Su-Ping and Hu, Xue-Qing and Xiao, Zhen-Jun",
	title = "{Study of $B_s\to K^{(*)}\ell^+ \ell^-$ decays in the PQCD factorization approach with lattice QCD input}",
	eprint = "2003.12226",
	archivePrefix = "arXiv",
	primaryClass = "hep-ph",
	doi = "10.1103/PhysRevD.102.013001",
	journal = "Phys. Rev. D",
	volume = "102",
	number = "1",
	pages = "013001",
	year = "2020"
}

@article{Nayek:2018rcq,
	author = "Nayek, P. and Maji, P. and Sahoo, S.",
	title = "{Study of semileptonic decays $B \to \pi l^+ l^-$ and $B \to \rho l^+ l^-$ in nonuniversal Z' model}",
	eprint = "1811.09991",
	archivePrefix = "arXiv",
	primaryClass = "hep-ph",
	doi = "10.1103/PhysRevD.99.013005",
	journal = "Phys. Rev. D",
	volume = "99",
	number = "1",
	pages = "013005",
	year = "2019"
}

@article{Matias:2012xw,
	author = "Matias, Joaquim and Mescia, Federico and Ramon, Marc and Virto, Javier",
	title = "{Complete Anatomy of $\bar{B}_d -> \bar{K}^{* 0} (-> K \pi)l^+l^-$ and its angular distribution}",
	eprint = "1202.4266",
	archivePrefix = "arXiv",
	primaryClass = "hep-ph",
	reportNumber = "UAB-FT-706, ICCUB-12-076, ECM-UB-68",
	doi = "10.1007/JHEP04(2012)104",
	journal = "JHEP",
	volume = "04",
	pages = "104",
	year = "2012"
}

@article{Descotes-Genon:2013vna,
	author = "Descotes-Genon, Sebastien and Hurth, Tobias and Matias, Joaquim and Virto, Javier",
	title = "{Optimizing the basis of $B\to K^*ll$ observables in the full kinematic range}",
	eprint = "1303.5794",
	archivePrefix = "arXiv",
	primaryClass = "hep-ph",
	reportNumber = "UAB-FT-732, LPT-ORSAY-13-24, MITP-13-020",
	doi = "10.1007/JHEP05(2013)137",
	journal = "JHEP",
	volume = "05",
	pages = "137",
	year = "2013"
}

@article{Biswas:2022lhu,
	author = "Biswas, Aritra and Nandi, Soumitra and Patra, Sunando Kumar and Ray, Ipsita",
	title = "{Study of the b \textrightarrow{} d\ensuremath{\ell}\ensuremath{\ell} transitions in the Standard Model and test of New Physics sensitivities}",
	eprint = "2208.14463",
	archivePrefix = "arXiv",
	primaryClass = "hep-ph",
	doi = "10.1007/JHEP03(2023)247",
	journal = "JHEP",
	volume = "03",
	pages = "247",
	year = "2023"
}

@article{PhysRevD.97.113001,
	title = {Improved perturbative QCD formalism for ${B}_{c}$ meson decays},
	author = {Liu, Xin and Li, Hsiang-nan and Xiao, Zhen-Jun},
	journal = {Phys. Rev. D},
	volume = {97},
	issue = {11},
	pages = {113001},
	numpages = {9},
	year = {2018},
	month = {Jun},
	publisher = {American Physical Society},
	doi = {10.1103/PhysRevD.97.113001},
	url = {https://link.aps.org/doi/10.1103/PhysRevD.97.113001}
	}

@article{Kurimoto:2001zj,
	author = "Kurimoto, T. and Li, Hsiang-nan and Sanda, A. I.",
	title = "{Leading power contributions to B ---\ensuremath{>} pi, rho transition form-factors}",
	eprint = "hep-ph/0105003",
	archivePrefix = "arXiv",
	reportNumber = "NCKU-HEP-01-01, DPNU-00-13",
	doi = "10.1103/PhysRevD.65.014007",
	journal = "Phys. Rev. D",
	volume = "65",
	pages = "014007",
	year = "2002"
}

@article{Colangelo:1995jv,
	author = "Colangelo, P. and De Fazio, F. and Santorelli, Pietro and Scrimieri, E.",
	title = "{QCD sum rule analysis of the decays $B \to K \ell^{+} \ell^{-}$ and $B \to K^{*} \ell^{+} \ell^{-}$}",
	eprint = "hep-ph/9510403",
	archivePrefix = "arXiv",
	reportNumber = "BARI-TH-95-206, DSF-T-95-42",
	doi = "10.1103/PhysRevD.53.3672",
	journal = "Phys. Rev. D",
	volume = "53",
	pages = "3672--3686",
	year = "1996",
	note = "[Erratum: Phys.Rev.D 57, 3186 (1998)]"
}

@article{Li:1994iu,
	author = "Li, Hsiang-nan and Yu, Hoi-Lai",
	title = "{Perturbative QCD analysis of B meson decays}",
	eprint = "hep-ph/9411308",
	archivePrefix = "arXiv",
	reportNumber = "CCUTH-94-04, IP-ASTP-12-94",
	doi = "10.1103/PhysRevD.53.2480",
	journal = "Phys. Rev. D",
	volume = "53",
	pages = "2480--2490",
	year = "1996"
}

@article{Kurimoto:2002sb,
	author = "Kurimoto, T. and Li, Hsiang-nan and Sanda, A. I.",
	title = "{B ---\ensuremath{>} D(*) form-factors in perturbative QCD}",
	eprint = "hep-ph/0210289",
	archivePrefix = "arXiv",
	reportNumber = "DPNU-00-13",
	doi = "10.1103/PhysRevD.67.054028",
	journal = "Phys. Rev. D",
	volume = "67",
	pages = "054028",
	year = "2003"
}

@article{Li:2001ay,
	author = "Li, Hsiang-nan",
	title = "{Threshold resummation for exclusive B meson decays}",
	eprint = "hep-ph/0102013",
	archivePrefix = "arXiv",
	reportNumber = "NCKU-HEP-01-02",
	doi = "10.1103/PhysRevD.66.094010",
	journal = "Phys. Rev. D",
	volume = "66",
	pages = "094010",
	year = "2002"
}

@article{Nagashima:2002ia,
	author = "Nagashima, Makiko and Li, Hsiang-nan",
	title = "{k(T) factorization of exclusive processes}",
	eprint = "hep-ph/0210173",
	archivePrefix = "arXiv",
	reportNumber = "OCHA-PP-193",
	doi = "10.1103/PhysRevD.67.034001",
	journal = "Phys. Rev. D",
	volume = "67",
	pages = "034001",
	year = "2003"
}

@article{Li:1999kna,
	author = "Li, Hsiang-nan and Melic, Blazenka",
	title = "{Determination of heavy meson wave functions from B decays}",
	eprint = "hep-ph/9902205",
	archivePrefix = "arXiv",
	reportNumber = "NCKU-HEP-99-01, IRB-TH-1-99",
	doi = "10.1007/s100520050665",
	journal = "Eur. Phys. J. C",
	volume = "11",
	pages = "695--702",
	year = "1999"
}

@article{Xiao:2011tx,
	author = "Xiao, Zhen-Jun and Wang, Wen-Fei and Fan, Ying-ying",
	title = "{Revisiting the pure annihilation decays $B_s\to \pi^+ \pi^-$ and $B^0 \to K^+ K^-$: the data and the pQCD predictions}",
	eprint = "1111.6264",
	archivePrefix = "arXiv",
	primaryClass = "hep-ph",
	doi = "10.1103/PhysRevD.85.094003",
	journal = "Phys. Rev. D",
	volume = "85",
	pages = "094003",
	year = "2012"
}

@article{Li:2008ts,
	author = "Li, Run-Hui and Lu, Cai-Dian and Zou, Hao",
	title = "{The B(B(s)) ---\ensuremath{>} D(s) P, D(s) V, D*(s) P and D*(s) V decays in the perturbative QCD approach}",
	eprint = "0803.1073",
	archivePrefix = "arXiv",
	primaryClass = "hep-ph",
	doi = "10.1103/PhysRevD.78.014018",
	journal = "Phys. Rev. D",
	volume = "78",
	pages = "014018",
	year = "2008"
}

@article{McNeile:2012qf,
	author = "McNeile, C. and Davies, C. T. H. and Follana, E. and Hornbostel, K. and Lepage, G. P.",
	title = "{Heavy meson masses and decay constants from relativistic heavy quarks in full lattice QCD}",
	eprint = "1207.0994",
	archivePrefix = "arXiv",
	primaryClass = "hep-lat",
	doi = "10.1103/PhysRevD.86.074503",
	journal = "Phys. Rev. D",
	volume = "86",
	pages = "074503",
	year = "2012"
}

@article{Lubicz:2017asp,
	author = "Lubicz, V. and Melis, A. and Simula, S.",
	collaboration = "ETM",
	title = "{Masses and decay constants of D*$_{(s)}$ and B*$_{(s)}$ mesons with N$_f = 2 + 1 + 1$ twisted mass fermions}",
	eprint = "1707.04529",
	archivePrefix = "arXiv",
	primaryClass = "hep-lat",
	reportNumber = "PREPRINT-RM3-TH-17-8",
	doi = "10.1103/PhysRevD.96.034524",
	journal = "Phys. Rev. D",
	volume = "96",
	number = "3",
	pages = "034524",
	year = "2017"
}

@article{Na:2015kha,
	author = "Na, Heechang and Bouchard, Chris M. and Lepage, G. Peter and Monahan, Chris and Shigemitsu, Junko",
	collaboration = "HPQCD",
	title = "{$B \rightarrow D l \nu$ form factors at nonzero recoil and extraction of $|V_{cb}|$}",
	eprint = "1505.03925",
	archivePrefix = "arXiv",
	primaryClass = "hep-lat",
	doi = "10.1103/PhysRevD.93.119906",
	journal = "Phys. Rev. D",
	volume = "92",
	number = "5",
	pages = "054510",
	year = "2015",
	note = "[Erratum: Phys.Rev.D 93, 119906 (2016)]"
}

@article{FermilabLattice:2021cdg,
	author = "Bazavov, A. and others",
	collaboration = "Fermilab Lattice, MILC, Fermilab Lattice, MILC",
	title = "{Semileptonic form factors for $B\rightarrow D^*\ell \nu $ at nonzero recoil from $2+1$-flavor lattice QCD: Fermilab Lattice~and~MILC~Collaborations}",
	eprint = "2105.14019",
	archivePrefix = "arXiv",
	primaryClass = "hep-lat",
	reportNumber = "FERMILAB-PUB-21-261-T~, FERMILAB-PUB-21/261-T",
	doi = "10.1140/epjc/s10052-022-10984-9",
	journal = "Eur. Phys. J. C",
	volume = "82",
	number = "12",
	pages = "1141",
	year = "2022",
	note = "[Erratum: Eur.Phys.J.C 83, 21 (2023)]"
}

@article{Bourrely:2008za,
	author = "Bourrely, Claude and Caprini, Irinel and Lellouch, Laurent",
	title = "{Model-independent description of B ---\ensuremath{>} pi l nu decays and a determination of |V(ub)|}",
	eprint = "0807.2722",
	archivePrefix = "arXiv",
	primaryClass = "hep-ph",
	reportNumber = "CPT-P36-2007",
	doi = "10.1103/PhysRevD.82.099902",
	journal = "Phys. Rev. D",
	volume = "79",
	pages = "013008",
	year = "2009",
	note = "[Erratum: Phys.Rev.D 82, 099902 (2010)]"
}

@article{Boyd:1995sq,
	author = "Boyd, C. Glenn and Grinstein, Benjamin and Lebed, Richard F.",
	title = "{Model independent determinations of anti-B ---\ensuremath{>} D (lepton), D* (lepton) anti-neutrino form-factors}",
	eprint = "hep-ph/9508211",
	archivePrefix = "arXiv",
	reportNumber = "UCSD-PTH-95-11",
	doi = "10.1016/0550-3213(95)00653-2",
	journal = "Nucl. Phys. B",
	volume = "461",
	pages = "493--511",
	year = "1996"
}

@article{Boyd:1995cf,
	author = "Boyd, C. Glenn and Grinstein, Benjamin and Lebed, Richard F.",
	title = "{Model independent extraction of |V(cb)| using dispersion relations}",
	eprint = "hep-ph/9504235",
	archivePrefix = "arXiv",
	reportNumber = "UCSD-PTH-95-03",
	doi = "10.1016/0370-2693(95)00480-9",
	journal = "Phys. Lett. B",
	volume = "353",
	pages = "306--312",
	year = "1995"
}

@article{Boyd:1997kz,
	author = "Boyd, C. Glenn and Grinstein, Benjamin and Lebed, Richard F.",
	title = "{Precision corrections to dispersive bounds on form-factors}",
	eprint = "hep-ph/9705252",
	archivePrefix = "arXiv",
	reportNumber = "CMU-HEP-97-07A, UCSD-PTH-97-12",
	doi = "10.1103/PhysRevD.56.6895",
	journal = "Phys. Rev. D",
	volume = "56",
	pages = "6895--6911",
	year = "1997"
}

@article{Bordone:2019guc,
	author = "Bordone, Marzia and Gubernari, Nico and van Dyk, Danny and Jung, Martin",
	title = "{Heavy-Quark expansion for ${{\bar{B}}_s\rightarrow D^{(*)}_s}$ form factors and unitarity bounds beyond the ${SU(3)_F}$ limit}",
	eprint = "1912.09335",
	archivePrefix = "arXiv",
	primaryClass = "hep-ph",
	reportNumber = "EOS-2019-04, P3H-19-050, SI-HEP-2019-20, TUM-HEP 1241/19",
	doi = "10.1140/epjc/s10052-020-7850-9",
	journal = "Eur. Phys. J. C",
	volume = "80",
	number = "4",
	pages = "347",
	year = "2020"
}

@article{Rui:2016opu,
	author = "Rui, Zhou and Li, Hong and Wang, Guang-xin and Xiao, Ying",
	title = "{Semileptonic decays of $B_c$ meson to S-wave charmonium states in the perturbative QCD approach}",
	eprint = "1602.08918",
	archivePrefix = "arXiv",
	primaryClass = "hep-ph",
	doi = "10.1140/epjc/s10052-016-4424-y",
	journal = "Eur. Phys. J. C",
	volume = "76",
	number = "10",
	pages = "564",
	year = "2016"
}

@article{Cooper:2021bkt,
	author = "Cooper, Laurence J. and Davies, Christine T. H. and Wingate, Matthew",
	collaboration = "HPQCD",
	title = "{Form factors for the processes $B^+_c \to D^0 \ell^+\nu_\ell$ and $B^+_c \to D^+_s \ell^+ \ell^+ (\nu \bar \nu)$ from lattice QCD}",
	eprint = "2108.11242",
	archivePrefix = "arXiv",
	primaryClass = "hep-lat",
	doi = "10.1103/PhysRevD.105.014503",
	journal = "Phys. Rev. D",
	volume = "105",
	number = "1",
	pages = "014503",
	year = "2022"
}

@article{Jenkins:1992nb,
	author = "Jenkins, Elizabeth Ellen and Luke, Michael E. and Manohar, Aneesh V. and Savage, Martin J.",
	title = "{Semileptonic B(c) decay and heavy quark spin symmetry}",
	eprint = "hep-ph/9204238",
	archivePrefix = "arXiv",
	reportNumber = "UCSD-PTH-92-13",
	doi = "10.1016/0550-3213(93)90464-Z",
	journal = "Nucl. Phys. B",
	volume = "390",
	pages = "463--473",
	year = "1993"
}

@article{Wang:2008xt,
	author = "Wang, Wei and Shen, Yue-Long and Lu, Cai-Dian",
	title = "{Covariant Light-Front Approach for B(c) transition form factors}",
	eprint = "0811.3748",
	archivePrefix = "arXiv",
	primaryClass = "hep-ph",
	doi = "10.1103/PhysRevD.79.054012",
	journal = "Phys. Rev. D",
	volume = "79",
	pages = "054012",
	year = "2009"
}

@article{Hu:2019qcn,
	author = "Hu, Xue-Qing and Jin, Su-Ping and Xiao, Zhen-Jun",
	title = "{Semileptonic decays $B_c \to (\eta_c,J/\psi) l \bar{\nu}_l $ in the ''PQCD + Lattice'' approach}",
	eprint = "1904.07530",
	archivePrefix = "arXiv",
	primaryClass = "hep-ph",
	doi = "10.1088/1674-1137/44/2/023104",
	journal = "Chin. Phys. C",
	volume = "44",
	number = "2",
	pages = "023104",
	year = "2020"
}

@article{Wang:2012lrc,
	author = "Wang, Wen-Fei and Fan, Ying-Ying and Xiao, Zhen-Jun",
	title = "{Semileptonic decays $B_c\to(\eta_c,J/\Psi)l\nu$ in the perturbative QCD approach}",
	eprint = "1212.5903",
	archivePrefix = "arXiv",
	primaryClass = "hep-ph",
	doi = "10.1088/1674-1137/37/9/093102",
	journal = "Chin. Phys. C",
	volume = "37",
	pages = "093102",
	year = "2013"
}

@article{Colangelo:2021dnv,
	author = "Colangelo, Pietro and De Fazio, Fulvia and Loparco, Francesco",
	title = "{Role of $B_c^+ \to B_{s,d}^{(*)} \, \bar \ell \, \nu_\ell$ in the Standard Model and in the search for BSM signals}",
	eprint = "2102.05365",
	archivePrefix = "arXiv",
	primaryClass = "hep-ph",
	reportNumber = "BARI-TH/21-726",
	doi = "10.1103/PhysRevD.103.075019",
	journal = "Phys. Rev. D",
	volume = "103",
	number = "7",
	pages = "075019",
	year = "2021"
}

@article{Ball:2004rg,
	author = "Ball, Patricia and Zwicky, Roman",
	title = "{$B_{d,s} \to  \rho, \omega, K^*, \phi$ decay form-factors from light-cone sum rules revisited}",
	eprint = "hep-ph/0412079",
	archivePrefix = "arXiv",
	reportNumber = "IPPP-04-74, DCPT-04-48, TPI-MINN-04-39",
	doi = "10.1103/PhysRevD.71.014029",
	journal = "Phys. Rev. D",
	volume = "71",
	pages = "014029",
	year = "2005"
}

@article{Ivanov:2024iat,
	author = "Ivanov, M. A. and Pandya, J. N. and Santorelli, P. and Soni, N. R.",
	title = "{Decay $B_c^+ \to D_{(s)}^{(*)+} \ell^+\ell^-$ within covariant confined quark model}",
	eprint = "2404.15085",
	archivePrefix = "arXiv",
	primaryClass = "hep-ph",
	doi = "10.1103/PhysRevD.110.096003",
	journal = "Phys. Rev. D",
	volume = "110",
	pages = "096003",
	year = "2024"
}

@article{Dey:2025xdx,
	author = "Dey, Utsab and Nandi, Soumitra",
	title = "{Correlated study on some B$_{c}${\textrightarrow} P and B$_{c}${\textrightarrow} S wave channels in light of new inputs}",
	eprint = "2503.01693",
	archivePrefix = "arXiv",
	primaryClass = "hep-ph",
	doi = "10.1007/JHEP07(2025)144",
	journal = "JHEP",
	volume = "07",
	pages = "144",
	year = "2025"
}

@article{Colangelo:2010bg,
	author = "Colangelo, Pietro and De Fazio, Fulvia and Wang, Wei",
	title = "{$B_s\to f_0(980)$ form factors and $B_s$ decays into $f_0(980)$}",
	eprint = "1002.2880",
	archivePrefix = "arXiv",
	primaryClass = "hep-ph",
	reportNumber = "BARI-TH-625-10",
	doi = "10.1103/PhysRevD.81.074001",
	journal = "Phys. Rev. D",
	volume = "81",
	pages = "074001",
	year = "2010"
}

@article{CLEO:1997rew,
	author = "Bartelt, John E. and others",
	collaboration = "CLEO",
	title = "{Observation of the radiative decay D*+ ---{\ensuremath{>}} D+ gamma}",
	eprint = "hep-ex/9711011",
	archivePrefix = "arXiv",
	reportNumber = "SLAC-PUB-9758, CLNS-97-1518, CLEO-97-25",
	doi = "10.1103/PhysRevLett.80.3919",
	journal = "Phys. Rev. Lett.",
	volume = "80",
	pages = "3919--3923",
	year = "1998"
}

@article{LHCb:2020gog,
	author = "Aaij, Roel and others",
	collaboration = "LHCb",
	title = "{Angular Analysis of the  $B^{+}\rightarrow K^{\ast+}\mu^{+}\mu^{-}$ Decay}",
	eprint = "2012.13241",
	archivePrefix = "arXiv",
	primaryClass = "hep-ex",
	reportNumber = "LHCb-PAPER-2020-041, CERN-EP-2020-239",
	doi = "10.1103/PhysRevLett.126.161802",
	journal = "Phys. Rev. Lett.",
	volume = "126",
	number = "16",
	pages = "161802",
	year = "2021"
}

@article{Aliev:2007gr,
    author = "Aliev, T. M. and Cornell, Alan S. and Gaur, Naveen",
    title = "{B ---{\ensuremath{>}} K(K*) missing energy in Unparticle physics}",
    eprint = "0705.4542",
    archivePrefix = "arXiv",
    primaryClass = "hep-ph",
    reportNumber = "KEK-TH-1153, LYCEN-2007-05",
    doi = "10.1088/1126-6708/2007/07/072",
    journal = "JHEP",
    volume = "07",
    pages = "072",
    year = "2007"
}

@article{Cheng:2014fwa,
    author = {Cheng, Shan and Fan, Ying-Ying and Yu, Xin and L{\"u}, Cai-Dian and Xiao, Zhen-Jun},
    title = "{The NLO twist-3 contributions to $B \to \pi$ form factors in $k_{T}$ factorization}",
    eprint = "1402.5501",
    archivePrefix = "arXiv",
    primaryClass = "hep-ph",
    doi = "10.1103/PhysRevD.89.094004",
    journal = "Phys. Rev. D",
    volume = "89",
    number = "9",
    pages = "094004",
    year = "2014"
}

@article{Mahajan:2004dx,
    author = "Mahajan, Namit",
    title = "{B ---{\ensuremath{>}} rho form-factors including higher twist contributions and reliability of pQCD approach}",
    eprint = "hep-ph/0405161",
    archivePrefix = "arXiv",
    month = "5",
    year = "2004"
}

@article{Liu:2020upy,
    author = "Liu, Xin and Li, Hsiang-nan and Xiao, Zhen-Jun",
    title = "{Next-to-leading-logarithm $k_T$ resummation for $B_c\to J/\psi$ decays}",
    eprint = "2006.12786",
    archivePrefix = "arXiv",
    primaryClass = "hep-ph",
    doi = "10.1016/j.physletb.2020.135892",
    journal = "Phys. Lett. B",
    volume = "811",
    pages = "135892",
    year = "2020"
}

@article{Keum:2000wi,
    author = "Keum, Y. Y. and Li, Hsiang-Nan and Sanda, A. I.",
    title = "{Penguin enhancement and $B \to K \pi$ decays in perturbative QCD}",
    eprint = "hep-ph/0004173",
    archivePrefix = "arXiv",
    reportNumber = "NCKU-HEP-00-01A, APCTP-00-05, DPNU-00-14",
    doi = "10.1103/PhysRevD.63.054008",
    journal = "Phys. Rev. D",
    volume = "63",
    pages = "054008",
    year = "2001"
}

\end{document}